\documentclass[a4paper,12pt]{article}
\usepackage{jheppub,esint,shuffle,psfrag}
 \usepackage[utf8]{inputenc}
 \usepackage{booktabs}
 
\usepackage{esint} 
\usepackage{breqn}

\def \tr {\mathop{\rm tr}\nolimits}

\def \res{\mathop{\rm res}\nolimits}
\def \e {\mathop {e}\nolimits}

\newcommand\lr[1]{{\left({#1}\right)}}

\newcommand \vev [1] {\langle{#1}\rangle}

\newcommand\re[1]{(\ref{#1})}

\def \qqqquad {\qquad\qquad}

\newcommand{\ft}[2]{{\textstyle\frac{#1}{#2}}}

\def\numberbysection{\@addtoreset{equation}{section}
                     \def\theequation{\thesection.\arabic{equation}}}
                     
\def\Xint#1{\mathchoice
   {\XXint\displaystyle\textstyle{#1}}%
   {\XXint\textstyle\scriptstyle{#1}}%
   {\XXint\scriptstyle\scriptscriptstyle{#1}}%
   {\XXint\scriptscriptstyle\scriptscriptstyle{#1}}%
   \!\int}
\def\XXint#1#2#3{{\setbox0=\hbox{$#1{#2#3}{\int}$}
     \vcenter{\hbox{$#2#3$}}\kern-.5\wd0}}

\def\dashint{\Xint-}                     

\def \N {{\cal N}}

\newcommand{\sa}{\mathsf{SA}}
\newcommand{\fa}{\mathsf{FA}}
\newcommand{\fatilde}{\widetilde{\fa}}
\newcommand{\M}{\mathsf{M}}

\def \ed {\end{document}}
\def \rf {\re}
\def\la{\label}
\def\l{\lambda}
\def \ov {\over}
\def \tr {{\rm tr}}
\def \ha {{1 \over 2}}

\def \ci {\cite}

\def \N  {{\cal N}}
 
\def \be {\begin{equation}}
\def \ee {\end{equation}}
\def \foot {\footnote}
\def \BB {{\rm B}}
\def \loc  {{\rm loc}}
\def \GG {{\rm G}}
\def \ha  {{1\ov 2}} 
\def \f {{\rm C}}

\def \rI {{I}}
\def \c {  c}

\def \l {\lambda}

\begin{document}

\vspace{-2cm }

\begin{flushleft}
 \hfill \parbox[c]{50mm}{
 IPhT--T22/040
 \\
 Imperial-TP-AT-2022-04}
\end{flushleft}

%\vspace{-1cm }

\author{M. Beccaria$^a$, G.P. Korchemsky$^{b,c}$ and A.A. Tseytlin$^{d,}$\footnote{Also on leave from  Institute for Theoretical and Mathematical Physics (ITMP)  and Lebedev Institute.}}

\affiliation{
$\null$
$^a$Universit\`a del Salento, Dipartimento di Matematica e Fisica \textit{Ennio De Giorgi},\\ 
\phantom{a} and I.N.F.N. - sezione di Lecce, Via Arnesano, I-73100 Lecce, Italy
\\		
$\null$
$^b${Institut de Physique Th\'eorique\footnote{Unit\'e Mixte de Recherche 3681 du CNRS}, Universit\'e Paris Saclay, CNRS, 91191 Gif-sur-Yvette, France}  
\\
$\null$
$^c${Institut des Hautes \'Etudes Scientifiques, 91440 Bures-sur-Yvette, France}  
\\
$\null$
$^d${Blackett Laboratory, Imperial College London,  SW7 2AZ, U.K.}
}

\title{Strong coupling expansion  \\ in  $\mathbf{\N=2}$   superconformal  theories 
\\  
and the Bessel kernel
}
 
 \abstract
{
We consider strong  't Hooft  coupling expansion in special four-dimensional $\mathcal N=2$ superconformal models
that are planar-equivalent to  $\mathcal N=4$ super Yang-Mills theory. Various observables in these models that admit localization matrix model representation 
can be  expressed at large $N$  in terms of a Fredholm determinant of a Bessel operator. The latter previously appeared in the study of level spacing 
distributions in matrix models and, more recently, in four-point correlation functions of infinitely heavy half-BPS operators in planar $\mathcal N=4$ SYM. 
We  use this relation  and a  suitably generalized Szeg\H{o}-Akhiezer-Kac formula to derive the  strong  't Hooft coupling expansion of the leading 
corrections to free energy,  half-BPS circular Wilson loop, and certain correlators of chiral primaries operators in  the  $\mathcal N=2$ models.
This substantially generalizes partial results in the literature and represents a challenge for dual string theory calculations in AdS/CFT 
 context. We also demonstrate that the resulting strong-coupling expansions suffer from Borel singularities and 
require adding non-perturbative, exponentially suppressed corrections. As a byproduct of our analysis, we determine 
the  non-perturbative correction to the above mentioned four-point correlator in planar $\mathcal N=4$ SYM.
 }
 
\maketitle
\flushbottom
\setcounter{footnote} 0

\newpage

\section{Introduction}

In this paper, we address the problem of computing a special class of observables in four-dimensional 
$\mathcal N=2$ and $\mathcal N=4$ superconformal Yang-Mills theories (SYM) for an arbitrary 't~Hooft coupling $\lambda=g_{\rm YM}^2 N$. 
A distinguished feature of these observables (denoted  as ${\mathcal F_\ell(g)}$) is that they can be expressed in terms of  determinants of a certain semi-infinite matrix  
\begin{align}\label{det-K}
\exp\big( {\mathcal F_\ell(g)} \big)  &= \det\Big(\delta_{nm}- K_{nm}(g, \ell)\Big)\Big|_{1\le n,m< \infty}\, , \qqqquad g= {\sqrt{\lambda}\over 4\pi} \,.
\end{align}
The semi-infinite matrix $K_{nm}$ carries information about the dynamics of the underlying gauge theory. 
In addition to  the  coupling $g$ it may depend on other (``kinematical'') parameters such as a non-negative integer $\ell$  to be specified below.

The derivation of the representation \re{det-K} is rather non-trivial and  relies on different techniques -- integrability \cite{Beisert:2010jr} (in the case of  four-point correlation function in planar $\mathcal N=4$ SYM) and localization \cite{Pestun:2007rz,Pestun:2016zxk} (in the case of leading non-planar correction to $\N=2$ free energy on $S^4$). 
 
In planar $\mathcal N=4$ SYM theory, the function $\mathcal F_\ell(g)$ defines the four-point correlation function of infinitely heavy half-BPS operators \cite{Coronado:2018ypq,Coronado:2018cxj,Kostov:2019stn,Bargheer:2019kxb,Kostov:2019auq,Belitsky:2019fan,Bargheer:2019exp,Belitsky:2020qrm,Belitsky:2020qir,Kostov:2021omc}.  
The determinant expression  \re{det-K} arises in  conjectured   representation 
 of this correlation function in terms of an effective two-dimensional integrable theory corresponding to  the  dual 
string world-sheet model   according to  the planar AdS/CFT correspondence \cite{Basso:2015zoa,Fleury:2016ykk,Eden:2016xvg,Bajnok:2017mdf}. 

In $\mathcal N=2$ superconformal  models that are planar-equivalent to $\N=4$ SYM (in particular,  in  some  $SU(N)$ 
models with  matter in the fundamental, rank-two symmetric  or   antisymmetric representations)
${\mathcal F_\ell(g)} $  stands for the  leading non-planar correction to  free energy  on $S^4$ 
  \cite{Beccaria:2020hgy,Beccaria:2021ksw,Beccaria:2021vuc,Beccaria:2021ism}. The localization technique allows one to express it in terms of a matrix model integral whose evaluation (to leading non-planar order) 
leads to the determinant  representation  \re{det-K} (see  Appendix~\ref{app:mat} for details).

There is a priori no reason why the semi-infinite matrices in \re{det-K} corresponding to these   different observables in the 
two different theories should be related to each other in a simple way. 
It is therefore surprising that in both cases 
the matrix $K_{nm}$  turns out to have   essentially the  same, universal form.
Its matrix elements are given by integrals involving the product of two Bessel functions  with indices of the same parity  
\begin{align}\label{K-def}
%v3
K_{nm}(g,\ell)=2(-1)^{n+m} \sqrt{(2n+\ell-1)(2m+\ell-1)}\int_0^\infty {dx\over x}\,    J_{2n+\ell-1}(x)\,  J_{2m+\ell-1}(x)\,  \chi\lr{ {x\ov 2g}}\, .
\end{align}
The dependence of $K_{nm}$ on the  coupling constant $g$ enters through a function $\chi$,  conventionally called the symbol of the matrix $K$. 
As we will explain below, the value of  a non-negative  integer number $\ell$ and the explicit form of the function $\chi(x)$ depend on a choice of a particular  observable. 

It is remarkable that the same matrix \re{K-def}  has previously appeared in a completely different area of mathematical physics and mathematical analysis. Namely, for $\ell=0$ and for the special choice of the function $\chi(x) = \theta(1-x)$,
the determinant \re{det-K} describes 
the level spacing distributions in the Laguerre ensemble in random matrix theory \cite{Forrester:1993vtx,Mehta}. More precisely,
the function $\exp(\mathcal F_{\ell=0}(g))$ gives the probability that there are no eigenvalues on the interval $[0,g^2]$. This function has a number of interesting properties and  can be found exactly in terms of a Painlev\'e transcendent \cite{Tracy:1993xj}.

It is convenient to think about the semi-infinite matrix $K_{nm}$ as representing a certain integral operator $\mathbf B_\ell$ on a space spanned by basis functions, $\mathbf B_\ell\, \psi_m(t)= K_{nm} \psi_n(t)$  
{(see Appendix~\ref{app:B} for details)}. In mathematical literature, this operator is known as a truncated (or finite temperature) Bessel operator \cite{BasorEhrhardt03}. The function \re{det-K}  is then  a Fredholm determinant of this operator, $\det (1-\mathbf B_\ell)$. Such an identification proves to be very useful because it allows us to study the observable \re{det-K} using powerful methods developed in the literature (for  reviews see \cite{Korepin:1993kvr,Bttcher2006AnalysisOT,Basor:2012,Krasovsky13}). 

\subsection*{Applications}

In  the application to superconformal gauge theories, the relations \re{det-K} and \re{K-def} provide a concise representation of 
the observable $\mathcal F_\ell (g)$ that is valid for an arbitrary 't~Hooft coupling. 
In this paper, we exploit this representation to evaluate $\mathcal F_\ell (g)$ at both  weak and  strong coupling and, then, study a transition between the two regimes. 
A detailed discussion  of $\mathcal F_\ell (g)$ in   \re{det-K} in $\mathcal N=2$ superconformal theories is presented below in Section~\ref{sect:N=2app}. 
Each observable \re{det-K} corresponds to a specific choice  of  the non-negative integer $\ell$ and the function $\chi(x)$. 
We will encounter three different  examples of the choices  of  $\ell$ and $\chi(x)$. 

The first example corresponds to  
\begin{align}\label{chi-BES}
\chi_{_{\text{BES}}}(x) = {2\over 1-\e^{x}}=1-\coth{(x/2) } \,.
\end{align}
For $\ell=0$ the resulting semi-infinite matrix \re{K-def} governs the cusp anomalous dimension in planar $\mathcal N=4$ SYM through the BES equation~\cite{Beisert:2006ez}.~\footnote{More precisely, in addition to  \re{K-def}    the 
BES equation  also involves    semi-infinite matrices of the same form as \re{K-def} but with  the product of Bessel functions that have indices of different parity. We do not consider these  latter  matrices in this paper.}

The second example comes from the study of four-point correlation functions of half-BPS operators  in planar $\mathcal N=4$ SYM.
%v3
%in Euclidean space.
 In the limit, when the $R$-charge of the operators becomes infinitely large, this correlation function factorizes into a product of two building blocks, the octagons~\cite{Coronado:2018ypq,Coronado:2018cxj}.
%v3
They depend on the  kinematical variables $y$ and $\xi$ which are expressed in terms of the  two cross-ratios built out of the coordinates of 
the four operators \foot{Explicitly,  $u= { x^2_{12} x^2_{34} \ov x^2_{13} x^2_{24}} =   e^{-2\xi}$ and $v= { x^2_{23} x^2_{41} \ov x^2_{13} x^2_{24}} = (1 +  e^{-y-\xi}) (1 + e^{y-\xi})$.} as well nonnegative integer $\ell$, the so-called bridge length. This parameter defines the smallest number of scalar propagators stretched between the four operators at zero coupling. Its value depends on the choice of polarizations of the 
operators. For the so-called ``simplest'' correlation function,  the scalar propagators connect the four operators   in a sequential (and not pairwise) manner. In this case,
the octagon is given by \re{det-K} and \re{K-def}   with  $\ell=0$
and
\begin{align} 
\chi_{\text{oct}}(x) = {\cosh y +\cosh\xi\over \cosh y + \cosh \sqrt{x^2+\xi^2}}\,.
\end{align}
%It depends on the  kinematical variables $y$ and $\xi$ which are expressed in terms of the  two cross-ratios built out of the coordinates of 
%the four operators.\foot{Explicitly,  $u= { x^2_{12} x^2_{34} \ov x^2_{13} x^2_{24}} =   e^{-2\xi}$ and $v= { x^2_{23} x^2_{41} \ov x^2_{13} x^2_{24}} = (1 +  e^{-y-\xi}) (1 + e^{y-\xi})$.}
 The properties of the octagon for arbitrary $y$ and $\xi$ have been studied in Refs.~\cite{Bargheer:2019kxb,Belitsky:2019fan,Bargheer:2019exp,Belitsky:2020qrm,Belitsky:2020qir}. For our purposes it will be sufficient to consider the special kinematical configuration $y=\xi=0$, in which case
\begin{align}\label{chi-oct}
\chi_{\text{oct}}(x) =  {2 \over 1 + \cosh x}\,.
\end{align}
The  choice of 
$y=\xi=0$  corresponds to the so-called bulk point singularity of four-point correlation function \cite{Maldacena:2015iua}. 
One  reason why this kinematical point is interesting is that, as we show in  Appendix~\ref{app:mat}, the corresponding expression for the octagon admits a matrix model representation similar to that  
coming from the localization.

%$\ell=0,1,2$
%It also  controls the asymptotic behaviour of 
 %the four-point correlation function of infinitely heavy half-BPS operators 
 %v3
 %in Minkowski space 
 %in the limit when four operators are located at the vertices of
 %a   light-like rectangle \cite{Belitsky:2019fan}.

The third example is related to a class of  special $\mathcal N=2$  superconformal gauge theories
that are planar-equivalent to $\N=4 $ SYM. 
 In these theories, the localization technique can be used to compute  some observables 
 (free energy, circular Wilson loop, two-point correlation functions  of  chiral operators)
 as functions of $\l$ and $N$ 
  in terms of  a  non-gaussian matrix model 
  (see~\cite{Pestun:2007rz,Fiol:2015mrp}). 
  It turns out that the  leading $1/N^2$ non-planar corrections in 
  the corresponding matrix integrals can be expressed in terms of the determinant \re{det-K} and \re{K-def} with  $\ell=1,2$ 
  %v3
  \foot{These  choices 
  of $\ell $ apply to cases of the $\mathbb Z_2$ orbifold of $\N=4$ SYM  and its orientifold -- the $SU(N)$ SA model discussed below. 
  They are not directly related to the number of nodes. Indeed, they also apply to certain two-point functions in the 
  $L$-node quivers with equal couplings ($\mathbb Z_L$ orbifolds of $\N=4$ SYM).
    }
      and
\begin{align}\label{chi-loc}
\chi_{\text{loc}}(x) = -{4\e^{x}\over (1-e^x)^2}= -{1\over \sinh^2({x / 2}) }\,.
\end{align}

\

Below  we first study the observable
 \re{det-K}   for a generic function $\chi(x)$ and, then, specify the resulting expressions 
   to the  cases of the symbols $\chi(x)$  in  \re{chi-BES}, \re{chi-oct} and \re{chi-loc} that are relevant for physics applications.

\subsection*{Weak and strong-coupling expansions}

Computing ${\mathcal F_\ell(g)}$  in  \re{det-K} and \re{K-def}  for an arbitrary value of  't~Hooft coupling   is a challenging problem. A significant simplification occurs in  the limits  of weak and strong coupling $g$.

At weak coupling, it is straightforward to expand the determinant \re{det-K} in powers of $g^2$. Changing the integration variable in \re{K-def} as $x\to xg$, we find that
the matrix elements scale as $K_{nm} = O(g^{n+m+\ell+1})$. As a consequence, to any finite order in $g^2$, an infinite-dimensional matrix $K_{nm}$ can be replaced  by its finite-dimensional minor and  the determinant \re{det-K} can be expanded in 
 powers of $\tr (K^r)$ (with $r=1,2,\dots$). 

At strong coupling, the study of asymptotic behaviour of the determinants like  \re{det-K} has a long history in mathematical analysis, see,  e.g.,   \cite{Bttcher2006AnalysisOT,Basor:2012,Krasovsky13}.
Relying on the strong Szeg\H{o} limit theorems \cite{Szego:1915,Szego:1952}, we expect that, for  sufficiently smooth function $\chi(x)$, the determinant \re{det-K} should admit a semiclassical expansion in the effective  $\hbar \sim 1/g$. This leads to the following asymptotic behaviour
\begin{align}\label{semi}
\mathcal F_\ell (g) = {-g A_0 +\frac12 A_1^2 \log g +B +  \Delta \mathcal F_\ell(g)}\,, 
\end{align}
where $\Delta \mathcal F_\ell(g)$ vanishes for $g\to\infty$.
Each term on the right-hand side of \re{semi}  depends on
the symbol $\chi(x)$ and  has a different origin.

The first three terms in \re{semi} give  the expression for $\mathcal F_\ell (g)$ which is known as the Szeg\H{o}-Akhiezer-Kac (SAK) formula \cite{Szego:1915,Szego:1952,Kac:1964,Akhiezer:1964}.  It involves the  coefficients $A_0$, $A_1$ and $B$, where  $B$ is called the  Widom-Dyson constant.  
For the matrix \re{K-def}, these coefficients are known 
in mathematical literature only for the 
unphysical values $\ell=\pm {1\ov 2}$, see \cite{BasorEhrhardt05}.  For arbitrary $\ell$, the expressions for these coefficients were conjectured in \cite{Belitsky:2020qir}. 

The appearance of $O(\log g)$ term in  \re{semi} can be attributed to Fisher-Hartwig singularity of the symbol $\chi(x)$, see \cite{Fisher68}. It corresponds to the behaviour 
\be 1-\chi(x) = O(x^{2\beta}) \ , \qquad \qquad   x\to 0 \ , \la{2} \ee
where  the  parameter $\beta$ defines the strength of the singularity.
The  coefficient $A_1$ in \re{semi} depends on $\beta$  and   vanishes  for $\beta=0$. 
For the symbols \re{chi-BES}, \re{chi-oct} and \re{chi-loc} the parameter
$\beta$ is different from zero and takes (half) integer values. Notice that $1-\chi(x)$ is an even (odd) function of $x$ for even (odd) $2\beta$. 

The last term in \re{semi} can be split into the sum of two terms of different kinds
\begin{align}\label{sum-f}
 \Delta \mathcal F_\ell(g) = f_\ell(g)+ \Delta f_\ell(g)\,.
\end{align}
Here $f_\ell(g)$ is given by a perturbative series in $1/g$
\begin{align}\label{f-PT}
f_\ell(g) = \sum_{k\ge 1} {A_{k+1}\over 2k(k+1)} g^{-k}\,,
\end{align}
where the expansion coefficients can be expressed in terms of the symbol $\chi(x)$~\cite{Belitsky:2020qrm,Belitsky:2020qir}. 
The function $\Delta f_\ell(g)$ in    \re{sum-f}  describes non-perturbative exponentially small  (in $\hbar^{-1}  \sim g$) corrections to $ \mathcal F_\ell(g)$.

 We show below that, for the symbols \re{chi-BES}--\re{chi-loc}, the properties of the perturbative series \re{f-PT} depend on the value of the parameter $\beta$, i.e. the strength of the Fisher-Hartwig singularity in \rf{2}  or, equivalently, on the  parity of the function $1-\chi(x)$.
   For half-integer $\beta$, the series \re{f-PT} can be resummed to all orders in $1/g$ to yield a well-defined function of $g$. \footnote{Moreover, as we show below (see Eq.~\re{B-guess}), in this case the function \re{det-K} can be found in a closed form.}
For integer $\beta$, the expansion coefficients in \re{f-PT} grow factorially.  Performing the Borel transform
\begin{align}\label{Bor}
f_\ell(g) = \int_0^\infty d\sigma \e^{-\sigma} \mathcal B_f(\sigma/g)\,,
\end{align} 
we find that the function $\mathcal B_f(\sigma/g)$ has poles at positive $\sigma$.  As a consequence, the function $f_\ell(g)$ is ill-defined and requires a regularization of the Borel singularities. 
 
\subsection*{Non-perturbative corrections}

The non-perturbative function $\Delta f_\ell(g)$ in \re{sum-f} takes the form
\begin{align}\label{f-nonPT}
\Delta f_\ell (g) = c_1\, g^{n_1} \e^{- 8\pi  g \,x_1}  \lr{1+O(1/g)}\,,
\end{align}
where the $O(1/g)$ term denotes a series in $1/g$. We show below that the parameters $c_1$, $n_1$ and $x_1$ depend on the choice of the function $\chi(x)$ and have a simple interpretation, e.g.,  $x=2\pi x_1$ is the root of $1-\chi(x)$ of degree $n_1$ closest to the origin.  In general, the expression on the right-hand side of \re{f-nonPT} contains an infinite sum of terms of the same form but with different parameters $c_i$, $n_i$ and $x_i$. The leading contribution to \re{f-nonPT} comes from the term with the minimal  value of $x_i$.

Despite the fact that the non-perturbative corrections \re{f-nonPT} are exponentially small at large $g$,
 they play an important role in understanding the properties of the function \re{semi} in
the transition region from strong to weak coupling. Determining the non-perturbative function $\Delta f_\ell(g)$ is an important open problem that we address in this paper. 

For the symbol $\chi(x)$ given by \re{chi-BES}, a closely related question
 of finding non-perturbative corrections to the cusp anomalous dimension in planar $\mathcal N=4$ SYM was studied 
in~\cite{Basso:2007wd,Basso:2009gh}. In what follows we employ the approach developed in these papers to determine $\Delta f_\ell (g)$ for a generic symbol $\chi(x)$ including the cases of \re{chi-oct} and \re{chi-loc}.

We show below that,  similarly  to the perturbative series \re{f-PT}, the non-perturbative function $\Delta f_\ell (g)$ has different properties for integer and half-integer $\beta$  in \rf{2}. For half-integer $\beta$, the  coefficients 
 in \re{f-nonPT} can 
be determined unambiguously in terms of the function $\chi(x)$.
For integer $\beta$, due to the presence of Borel singularities in the perturbative series \re{f-PT}, the functions $f_\ell (g)$ and $\Delta f_\ell (g)$ 
are not well-defined separately.  However,  all ambiguities related to a freedom in regularizing these singularities cancel  in their sum \re{sum-f}.
We determine the leading non-perturbative correction \re{f-nonPT} by specifying a regularization procedure of 
 the perturbative series \re{f-PT}. 
It amounts to a deformation of the integration contour in the Borel transform \re{Bor} in the vicinity of poles of
the integrand.

As a relevant example, illustrating different properties of \re{det-K} for  integer and half-integer $\beta$,  one 
 can consider an asymptotic expansion of a 
(properly normalized) modified Bessel function
\begin{align}\label{bes-exp}
{\rI_\beta(\sqrt\lambda)\over \lambda^{\beta/2}} = \e^{\sqrt\lambda} P(\lambda^{-1/2}) + \e^{-\sqrt\lambda} P(-\lambda^{-1/2})\,.
\end{align}
At large $\lambda$, the second term is exponentially small.   
For half-integer $\beta$, $P(x)$ is a polynomial in $x$ of degree $2\beta+1$. For integer $\beta$, $P(x)$ is given by the product of $\sqrt{x}$ and a Borel non-summable series in $x$. We show below that $\mathcal F_\ell(g)$ has similar properties for the symbols \re{chi-BES}--\re{chi-loc}. For $\beta=1$, the relation \re{bes-exp} yields the strong-coupling expansion of the   circular Wilson-Maldacena 
 loop in planar $\mathcal N=4$ SYM \cite{Erickson:2000af,Drukker:2000rr}  (see \re{W-planar} below). In this case,  the first and the second terms on the right-hand side of \re{bes-exp} define, respectively, the perturbative and the non-perturbative contribution to the circular Wilson loop at strong coupling.
 
The  rest of this  paper is organized as follows. In Section~\ref{sect:SAK} we present the relation of the observable \re{det-K}  to the
Fredholm determinant of the truncated Bessel operator and discuss its expansion at weak and strong coupling. We show that the strong-coupling expansion of \re{det-K}  has different properties for odd and even functions
$1-\chi(x)$. In the latter case, the perturbative series in $1/g$ is  not Borel summable and requires a regularization. In Section~\ref{sect:np-SAK}, we analyse 
 the resulting non-perturbative, exponentially small corrections that appear  in \re{det-K}. 
We apply the approach developed 
in~\cite{Basso:2009gh} to establish the relation \re{f-nonPT} and present the explicit expressions for the 
coefficients there. 
In Section~\ref{sect:N=2app} we use the general results of Sections  \ref{sect:SAK}   and  \ref{sect:np-SAK}
    to compute the strong-coupling expansion  of the 
 leading non-planar corrections to   observables in   special 
 $\mathcal N=2$ superconformal  $SU(N)$  models. We find, in particular, 
 the   analytic expressions for the leading strong-coupling coefficients   that   were previously   estimated 
  only numerically  (cf.  \cite{Beccaria:2021ksw,Beccaria:2021vuc,Beccaria:2021ism}). 
 We  summarize our  results and make some  concluding comments (in particular, on possible dual string theory interpretation) 
  in Section \ref{s5}.
 Some   technical details are presented in  Appendices~\ref{app:mat}--\ref{app:sc-orb}.

\section{Szeg\H{o}-Akhiezer-Kac  formula}\label{sect:SAK}
 
In this Section, we summarize the properties of the function $\mathcal F_\ell(g)$ defined in \re{det-K} and \re{K-def} and present the expressions for the  coefficients in  its strong-coupling expansion \re{semi}.

\subsection{Truncated Bessel operator}

Discussing the properties of \re{det-K}, it is convenient to switch from the semi-infinite matrix $K_{nm}$ to an integral  operator $\mathbf B_\ell(\chi)$
 defined as
\begin{align}\label{def-B}
\mathbf B_\ell(\chi) f(t) = \int_0^{2g} dt'\, \BB_\ell(t,t') f(t')\,,
\end{align}
where $f(t)$ is a test function and the kernel is given by
\begin{align}\label{ker-B}
\BB_\ell(t,t') = (tt')^{1/2} \int_0^\infty dx\, x J_\ell(t x)\,  \chi(x)\,  J_\ell(t' x) \,.
\end{align}
In mathematical literature, it is called truncated Bessel operator, see, e.g., \cite{BasorEhrhardt03}. The relation between the semi-infinite matrix $K_{nm}$ and the Bessel operator $\mathbf B_\ell(\chi)$ is discussed in Appendix~\ref{app:B}.

The function \re{det-K} admits a representation as a Fredholm determinant of the Bessel operator
\begin{align}\label{det-B}
\e^{\mathcal F_\ell (g)} = \det (1- \mathbf B_\ell(\chi))_{[0,2g]}\,,
\end{align}
where the  subscript  ${[0,2g]}$ indicates that the operator $ \mathbf B_\ell$ acts on the interval $[0,2g]$.
  
The equivalence of the two representations \re{det-K}, \re{K-def}  and \re{det-B} follows from the following  identity
for the trace of the product of $n$ copies of the matrices \re{K-def} (see \cite{Belitsky:2020qir})
\begin{align}\label{trB}
 \tr(K^n) = \int_0^{2g} dt_1 \dots \int_0^{2g} dt_n\,  \BB_\ell(t_1,t_2) \dots \BB_\ell(t_n,t_1) \equiv \tr ( \mathbf B_\ell^n)  \,.
\end{align} 
The representation \re{det-B} is very useful for deriving an expansion of the function $\mathcal F_\ell(g)$ at small $g$. As mentioned in the Introduction, 
it can be obtained by expanding the determinant \re{det-B}  in terms of  traces of  powers of the operator $\mathbf B_\ell$
\begin{align}\label{exp-tr} 
\mathcal F_\ell (g) = - \tr  (\mathbf B_\ell) -\frac12 \tr (\mathbf B_\ell^2) -\frac13\tr (\mathbf B_\ell^3) + \cdots\, .
\end{align}
According to its definition \re{trB}, $\tr (\mathbf B_\ell^n)$ is given by the $n$-fold integral \re{trB}. 
Changing the integration variables in \re{trB} as $t_i \to g t_i$ and taking into account \re{ker-B}, we find that $\tr (\mathbf B_\ell^n)=O(g^{2n(\ell+1)})$ and, therefore, the expansion in \re{exp-tr} runs in powers of $g^{2(\ell+1)}$. 

The leading term of the expansion looks as
\begin{align}\label{F-weak}
\mathcal F_\ell (g) =  \sum_{k\ge 0}  g^{2(\ell+k+1)} q_{\ell+k+1}  {(-1)^{k+1} (2\ell+2k)!\over k! (2\ell+k)! [(\ell+k+1)!]^2}
+ O(g^{4(\ell+1)})\,,
\end{align}
where  we introduced  
\begin{align}\label{qk}
q_k(\chi) = 2k\int_0^\infty dx\, x^{2k-1} \chi(x)\,.
\end{align}
The subleading terms in \re{F-weak} are given by multi-linear combinations of the coefficients $q_k$
(their expressions can be found in \cite{Belitsky:2020qrm,Belitsky:2020qir}). 
For the  three examples of the function $\chi$  defined above in \re{chi-BES}, \re{chi-oct} and \re{chi-loc} we have
\begin{align}\notag\label{q-weak}
& q^{\text{BES}}_k=-2 (2 k)! \zeta (2 k)\,,
\\[1.5mm] \notag
& q^{\text{oct}}_k=4 \left(1-4^{1-k}\right) (2 k)! \zeta (2 k-1)\,,
\\[1.5mm]
& q^{\text{loc}}_k=-4(2 k)! \zeta (2 k-1)\,,
\end{align}
where $\zeta(n)$ is the  Riemann zeta-function.

Notice that the function $\mathcal F_\ell (g)$ receives corrections only starting at order $O(g^{2(\ell+1)})$. This property does not depend on the explicit form of the function $\chi(x)$. In particular, it holds in both cases \re{chi-oct} and \re{chi-loc} mentioned above but its physical interpretation is different. For the octagon, the leading correction to $\mathcal F_\ell$ is associated with a scattering of elementary excitations (magnons) off a heavy state built of $\ell$ scalar particles in $\mathcal N=4$ SYM. At weak coupling, the corresponding amplitude behaves as $g^{2(\ell+1)}$. 
In $\mathcal N=2$ superconformal models  planar-equivalent to $\N=4$ SYM, the   partition function does not depend on
 the matter content of the theory to  first few orders in $g^2$ and, therefore, it  does not get corrections (coinciding  with the partition function in $\mathcal N=4$ SYM which is protected). 
  
\subsection{Specifying   the symbol}

The determinants \re{det-K} and \re{det-B}  depend on the  function $\chi(x)$ in a non-trivial way. In mathematical literature,
this function is called a symbol of the Bessel operator \re{def-B}.
It is assumed to be a smooth function on real semi-axis. 

For the strong-coupling expansion \re{semi} to be well-defined, the symbol $\chi(x)$ has to verify additional conditions. The expansion coefficients in \re{semi} and \re{f-PT} are given by multilinear combinations of the integrals  (see \re{As} below)
\begin{align}\label{In}
I_n (\chi) = {1\over (2n-1)!!} \int_0^\infty {dx\over\pi} (x^{-1}\partial_x)^n x \partial_x \log(1-\chi(x))\,,\qquad \ \ \ n=0,1, \dots \ . 
\end{align}
 For these integrals to be finite, the function $1-\chi(x)$ should be positive definite for $x>0$ and decrease at infinity faster than $1/x$. 

It is easy to see that the symbols  \re{chi-BES}, \re{chi-oct} and \re{chi-loc} satisfy these conditions.
Notice that these symbols are expressed in terms of the same function
and, as a consequence, they are related to each other as
\begin{align}\label{chi's}
1-\chi_{\text{loc}}(x) = {1\over 1-\chi_{\text{oct}}(x)} = [1-\chi_{_{\text{BES}}}(x)]^2=\coth^2(x/2)\,.
\end{align}
Substituting these expressions into \re{In} we find
\begin{align}\label{In1}
&I_n^{\text{loc}}= - I_{n} ^{\text{oct}}=2I_n^{\text{BES}} =(-1)^{n-1} (1-2^{2-2n}) {2 \zeta(2n-1)\over \pi^{2n-1}}\ , 
\end{align}
in particular, 
\begin{align} 
&I_0^{\text{loc}}=-\frac{\pi }{2} \ , \qquad   I_1^{\text{loc}}= \frac{2\log 2}{\pi } \ , \qquad 
I_2^{\text{loc}}=-\frac{3 \zeta (3)}{2 \pi ^3} \ , \qquad   I_3^{\text{loc}}= \frac{15 \zeta   (5)}{8 \pi ^5} \ , \ \  \dots \ . \la{3} 
\end{align}  
Generalizing the relations \re{chi-BES}, \re{chi-oct} and \re{chi-loc}, let us  choose the following ansatz for the function $\chi(x)$
\begin{align}\label{zeros} 
1-\chi(x) = b \, x^{2\beta}  \prod_{n\ge 1} {  1+x^2/(2\pi x_n)^2 \over  1+x^2/(2\pi y_n)^2}\,,
\end{align}
where the parameters $b$, $x_n$ and $y_n$ are real positive numbers.  
For this choice of $\chi(x)$, the integrals \re{In} become functions of $x_n$ and $y_n$, e.g.
\begin{align}\label{I1}
I_1={1\over 2\pi} \sum_{n\ge 1} \lr{{1\over x_n}-{1\over y_n}}\,.
\end{align}
We also assume that $\chi(x)$ is analytical at the origin, so that $2\beta$ takes integer values.
For integer (half-integer) $\beta$, the symbol \re{zeros} is an even (odd) function of $x$. 

As was mentioned in the Introduction, the parameter $\beta$ defined  in \rf{2}  plays an important role in our analysis.
 According to \re{zeros}, it controls behaviour of the symbol around the origin 
\begin{align}\label{small-x}
1-\chi(x) = b \, x^{2\beta} \big(1+O(x^2)\big)\,.
\end{align}
Substituting this relation into \re{K-def} and requiring the matrix elements $K_{nm}$ to be finite for $n,m\ge 0$, we find that $\ell$ and $\beta$ have to satisfy 
\begin{align}\label{lbeta}
\ell_\beta\equiv \ell+\beta>-1\,.
\end{align}
The relation \re{small-x} implies that the symbol \re{zeros} possesses the Fisher-Hartwig singularity~\cite{Fisher68}.
It is responsible for the appearance of the $O(\log g)$ term in the exponent of \re{semi}. The corresponding coefficient is given by \cite{Belitsky:2020qir}
\begin{align}\label{A1}
A_1^2=2\beta\ell+\beta^2\,.
\end{align}
It only depends on $\ell$ and $\beta$ and is insensitive to the values of $x_n$ and $y_n$ in \re{zeros}.

As follows from \re{zeros}, the function $1-\chi(x)$ has an infinite number of poles and zeros in the complex $x$-plane. They are located along the  imaginary axis at $x=\pm 2\pi i y_n$ and $x=\pm 2\pi i x_n$, respectively. It proves convenient to decompose the function $1-\chi(x)$ into a product of functions analytical in upper and lower half-planes
(the Wiener-Hopf decomposition)
\begin{align}\notag\label{Phi}
& 1-\chi(x) =b \, x^{2\beta} \Phi(x) \Phi(-x) \,,
\\[1.5mm]
& \Phi(x) = \prod_{n\ge 1} {1- ix/(2\pi x_n)\over 1 -ix/(2\pi y_n)}\,,
\end{align}
where $x_n$ and $y_n$ are positive. The function $\Phi(x)$ has poles and zeros located in the lower half-plane. 
By definition, it satisfies the normalization condition $\Phi(0)=1$. Some of the parameters $x_n$ and $y_n$ may take infinite value so that the number of poles and zeros can be different. In addition, some of $x$'s and $y$'s can coincide so that the roots and poles can be double, triple,   etc. 
Recall that the symbol $\chi(x)$ has to vanish at infinity. Then, it follows from \re{Phi} that  for $x\to\infty$
\begin{align}\label{Phi-inf}
\Phi(x) \sim b^{-1/2} (-ix)^{-\beta} \,. 
\end{align}
This relation imposes non-trivial conditions on large $n$ behaviour of $x_n$ and $y_n$ in \re{Phi}.

Let us examine the Wiener-Hopf decomposition \re{Phi}  of the symbols \re{chi's}. 
For the symbol $1-\chi_{_{\text{BES}}}(x)$ it  gives 
\begin{align}\label{Phi-BES}
& b_{_{\text{BES}}} = 2 \,,&& 
\beta_{_{\text{BES}}} = -\frac12 \,,&& 
\Phi_{_{\text{BES}}}(x) = \sqrt{\pi}\frac{\Gamma \left(1-\frac{i x}{2\pi}\right)}{\Gamma
   \left(\frac{1}{2}-\frac{i x}{2\pi}\right) }\,.
\end{align}
Matching this relation to \re{Phi}, we identify the values of roots $x_n=n-1/2$ and poles $y_n=n$ (with $n\ge 1$).
For the two remaining symbols in \re{chi's} we get
\begin{align}\notag\label{Phi-oct}
& b_{\text{oct}} = \frac14 \,,&& 
\beta_{\text{oct}} = 1 \,,&& 
\Phi_{\text{oct}}(x) = \left[ \Phi_{_{\text{BES}}}(x)\right]^{-2},
\\[1.2mm]
& b_{\text{loc}} = 4 \,,&& 
\beta_{\text{loc}} = -1 \,,&& 
\Phi_{\text{loc}}(x) =\left[ \Phi_{_{\text{BES}}}(x)\right]^2
   \,.
\end{align}
It is straightforward to check that the functions $\Phi(x)$ in \re{Phi-BES} and \re{Phi-oct} verify the relation \re{Phi-inf}.

Notice that the poles and roots of $\Phi_{\text{oct}}(x)$ and $\Phi_{\text{loc}}(x)$ are double degenerate. We will show below that this has important consequences for the properties of non-perturbative corrections at strong coupling. Recall that $\ell$ and $\beta$ have to satisfy the condition \re{lbeta}. For the symbols \re{chi's} this leads to 
$\ell_{_{\text{BES}}}\ge 0$, $\ell_{\text{oct}}\ge 0$ and  $\ell_{\text{loc}}\ge 1$.

\subsection{Widom-Dyson constant} 
 
For the symbol \re{zeros} the Widom-Dyson constant is given by \cite{Belitsky:2020qir}
\begin{align}\label{B}
B_\ell =\frac12 \int_0^\infty dk \left[ k (\tilde\psi(k))^2-\beta^2 {1-\e^{-k}\over k}\right]+{\beta\over 2}\log(2\pi) -{\ell\over 2} \log b+\log {G(1+\ell)\over G(1+\ell+\beta)}\,,
\end{align}
where the subscript on the left-hand side was introduced to indicate its  dependence 
 on $\ell$. Here $G(x)$ is the Barnes function satisfying $G(x+1)=G(x)\, \Gamma(x)$ and $\widetilde \psi(k)$ is given by a Fourier transform of $\log(1-\chi(x))$
\begin{align}\label{psi}
\widetilde \psi(k) = \int_0^\infty {dx\over\pi} \cos(kx) \log(1-\chi(x))\,.
\end{align}
The relation \re{small-x} translates to $\widetilde \psi(k) \sim -\beta/k$ at large $k$. The second term inside the brackets in the first term in  \re{B} ensures that the integral converges at large $k$.

Replacing $\chi$  in \re{B} and \re{psi}  with one of  in  \re{chi's} 
we find after some algebra~\footnote{We use \re{psi} and \re{chi's} to get
$\widetilde \psi_{\text{loc}}(k) =-\widetilde \psi_{\text{oct}}(k) = 2\widetilde \psi_{_{\text{BES}}}(k) ={1\over k} \tanh (k\pi/2)$.
}
 \begin{align}\notag\label{Bs}
 & B_\ell^{\text{BES}}=\log \Big(\frac{\pi^{\frac18}\,  G(\frac{1}{2}) \,   G(\ell +1)}{2^{\frac{\ell
   }{2}+\frac{1}{8}} G\left(\ell +\frac{1}{2}\right)}\Big),
 \\\notag
& B_\ell^ {\text{oct}}= \log \Big(\frac{\pi ^2 \, G^4(\frac{1}{2}) \, 2^{\ell +1}}{\Gamma (\ell
   +1)}\Big),
\\[2.2mm] 
& B_\ell^ {\text{loc}}=   \log \Big(\pi G^4(\ft{1}{2})\,  \Gamma (\ell ) \, 2^{-\ell }\Big)\,.
\end{align}
Here the Barnes function $G({1\ov 2} ) = 2^{1/24} e^{1/8} \pi^{-1/4} \mathsf{A}^{-3/2}$ can be expressed in terms of Glaisher's constant  $\mathsf{A}$. 
 
\subsection{Perturbative corrections at strong coupling} 

With $A_1$ given by \rf{A1} 
the expressions for the remaining expansion coefficients in \re{semi} and \re{f-PT} are \cite{Belitsky:2020qir}
\begin{align}\notag\label{As}
& A_0=2I_0\,,
\\\notag
& A_2=-\frac14(4\ell_\beta^2-1) I_1\,, 
\\\notag
& A_3=-\frac3{16}(4\ell_\beta^2-1) I_1^2\,, 
\\
&A_4=-\frac1{128}(4\ell_\beta^2-1)\big( (4\ell_\beta^2-9)I_2+16 I_1^3\big)\,, \ \ \dots
\end{align}
where $\ell_\beta=\ell+\beta$ and  $I_n$ are defined in \re{In} and \rf{In1}. 

Replacing the coefficients in \re{f-PT} with their explicit expressions \re{As}, we obtain the function $f_\ell(g)$ in \rf{sum-f}  which describes the subleading corrections to 
\re{semi} suppressed by powers of $1/g$. This function has the following interesting properties. 

According to \re{As}, the expansion coefficients involve powers of $I_1$ in  \re{In}. 
All such terms can be eliminated at once by shifting the coupling constant as 
\be \la{4}
g'\equiv g-  {1\ov 2}  I_1 \ , \ee 
thus getting 
\begin{align}\notag
&f_{\ell} (g)  =  \frac18 (4\ell_\beta^2-1) \log(g'/g) 
-\frac{(4 \ell_\beta^2-1)(4 \ell_\beta^2-9)}{3072 {g'}^3}I_2
 \\[2mm]
  & 
 -\frac{ (4 \ell_\beta^2-1)(4 \ell_\beta^2-9)(4 \ell_\beta^2-25)}{163840 {g'}^5}I_3
  -\frac{(4 \ell_\beta^2-1)
   (4 \ell_\beta^2-9) (4 \ell_\beta^2-21)}{196608
   {g'}^6}I_2^2+O\Big({1\ov {g'}^7}\Big) 
   \, .\label{f-imp}
\end{align}
Because the  coefficients  $I_n$ are independent of $\ell$, it is obvious from this relation that $f_\ell (g)$ depends on $\ell$ only through $\ell_\beta=\ell+\beta$.  We show below that the same is true for the non-perturbative function $\Delta f_\ell (g)$ in \re{semi}. 
We can use this observation to show that  the function $f_\ell(g) \equiv f_{\ell_\beta}(g)$ has different properties for integer and half-integer $\beta$.

For half-integer $\beta$, or,  equivalently,  half-integer $\ell_\beta=\ell+\beta$, the series \re{f-imp} simplifies dramatically. For instance, 
for $\ell_\beta=\frac12$ and $\ell_\beta=\frac32$ all but the first few terms of the expansion \re{f-imp} vanish. For an arbitrary 
half-integer $\ell_\beta=n+\ha $, or equivalently $\beta=n+\ha -\ell$, the series \re{f-imp} can be resummed to all orders in $1/g$, see \cite{Belitsky:2020qrm,Belitsky:2020qir}. The resulting expression for $f_{\ell}(g)$ looks as
\begin{align}\label{f-well}
f_\ell(g) = \frac{n(n+1)}{2} \log(g'/g) + \log P_{n(n+1)\ov 2}(1/g')\,,
\end{align}
where $n=\beta+\ell-\ha$ is non-negative integer and $g'=g-\ha I_1$. Here $P_{n(n+1)\ov 2}(x)$ is a polynomial in $x$ of degree $n(n+1)\ov 2$ with the expansion coefficients given by multilinear combination of $I_k$ with $k\ge 2$. For instance, for $n=0,1,2,3$ we have
\begin{align}\notag\label{P-pol}
& P_0(x)=P_1(x)=1\,,  
\\%[2mm]
 \notag
& P_3(x)=1-\frac18 I_2 x^3 \,, 
\\
& P_6(x)=1-\frac{5}{8} I_2 x^3-\frac{9}{32} I_3 x^5-\frac{5}{64} I_2^2 x^6\,.
\end{align}

For integer $\beta$, or equivalently integer $\ell_\beta$, the situation is different. For a generic symbol $\chi(x)$ the expansion coefficients \re{As} are different from zero and the function $f_\ell(g)$ is given by an asymptotic series in $1/g$ with factorially growing coefficients. 
Moreover, as we will show below, this series is not Borel summable and its Borel transform \re{Bor} develops a pole at positive $\sigma$
\begin{align}
\mathcal B_f(\sigma) \sim {1\over (\sigma- 8\pi x_1)^{n_1}}\,,
\end{align}
where $x_1$ is the smallest root of the symbol \re{zeros} of degree $n_1$. As a consequence, the series \re{f-imp} approximates the function $f_\ell(g)$ up to an exponentially small correction proportional to the residue of \re{Bor} at the pole $\sigma=8\pi g x_1$. The latter takes the same form as the non-perturbative correction \re{f-nonPT}. To define $f_\ell(g)$ unambiguously, one has to specify the prescription for deforming the integration contour in \re{Bor} in the vicinity of the Borel pole. We return to this question in Section~\ref{sect:bor} below.
 
\subsection{Physics applications}
\label{sec:phys-app}

In this subsection, we combine together the above relations and present the results for the strong-coupling expansion \re{semi} for the choice of the  symbols in  \re{chi-BES}, \re{chi-oct} and \re{chi-loc} that appear in different gauge-theory observables. For the time being, we shall neglect the non-perturbative correction $\Delta f_\ell(g)$ in \re{semi}.
 
We start with the BES  symbol \re{chi-BES}. Taking into account the relations \re{A1}, \re{As}, \re{f-imp} and \re{Bs}, we obtain the strong-coupling expansion of the corresponding observable $\mathcal F_{\ell} ^{\text{BES}}(g)$ for $\ell=0,1,2$
\begin{align}\notag\label{bes-ex}
&  \mathcal F_{\ell=0} ^{\text{BES}}= {\frac{\pi  g}{2}+\frac{1}{8} \log( \frac{\pi  g}{2})+\dots}\,,
\\\notag
&   \mathcal F_{\ell=1} ^{\text{BES}}= {\frac{\pi  g}{2}-\frac{3}{8} \log (\pi  g)-\frac{5 \log 2}{8} +\dots}\,,
\\
&   \mathcal F_{\ell=2} ^{\text{BES}}= {\frac{\pi  g}{2}-\frac{7}{8} \log (\pi  g)-\frac{\log 2}{8} 
+ \log\Big({1-{\log 2 \over 2\pi g}}\Big)+\dots} \,,
\end{align}
where dots stand only for non-perturbative corrections of the form \re{f-nonPT}. 

Indeed, 
according to \re{Phi-BES}, the parameter $\beta$ is half-integer in this case and, therefore, the perturbative $1/g^n$  corrections to  $\mathcal F_{\ell} ^{\text{BES}}(g)$ are expected to  be  very simple.  Namely, the first two relations in \re{bes-ex} do not receive perturbative corrections in $1/g$ at all. In the last relation, they all  come only  from 
the expansion of  $\log({1-{\log 2\ov  2\pi g}})$. These properties are in a perfect agreement with the relations \re{f-well} and \re{P-pol} for $n=\ell-1$ and $\ell=0,1,2$.

Yet another remarkable feature of the symbol \re{chi-BES} is that the functions \re{bes-ex} can be found  exactly for arbitrary $g$, see \cite{Belitsky:2019fan,Basso:2020xts,Belitsky}
\begin{align}\notag\label{B-guess}
&  \mathcal F_{\ell=0}^{\text{BES}}  = \frac38 \log \cosh(2\pi g) -\frac18 \log { \sinh(2\pi g)\over 2\pi g}\,,
\\\notag
&  \mathcal F_{\ell=1}^{\text{BES}}  = - \frac18 \log \cosh(2\pi g) +\frac38 \log { \sinh(2\pi g)\over 2\pi g}\,,
\\
&  \mathcal F_{\ell=2}^{\text{BES}} = \frac38 \log \cosh(2\pi g) -\frac18 \log { \sinh(2\pi g)\over 2\pi g}
+ \log {\log \cosh(2\pi g) \over 2\pi^2 g^2}\,.
\end{align} 
It is easy to check that, at strong coupling, these relations reproduce \re{bes-ex}. In addition, they allow us to identify non-perturbative corrections to \re{bes-ex}. We discuss them in Section~\ref{sect:app} below.

We can use \re{B-guess} to define the following functions
\begin{align}\notag \la{5}
& \Gamma_{\rm oct}(g) =2\pi^2 g^2\exp\big({\mathcal F_{\ell=2}^{\text{BES}}-\mathcal F_{\ell=0}^{\text{BES}}} \big)= \log \cosh(2\pi g) \,,
\\[2mm]
& C_{\rm oct}(g) =- {4\big(\mathcal F_{\ell=1}^{\text{BES}}+\mathcal F_{\ell=0}^{\text{BES}}\big)} = -\log  { \sinh(4\pi g)\over 4\pi g}\,. 
\end{align}
As was shown in \cite{Belitsky:2019fan,Belitsky:2020qzm}, these  ``octagon anomalous dimensions''  determine the asymptotic behaviour of the four-point correlation function of 
infinitely heavy half-BPS operators in the limit when the  four operators are located in the vertices of a light-like rectangle.~\footnote{It is interesting to note that the same functions appear in the analysis of the six-gluon amplitudes in planar $\mathcal N=4$ SYM in a special kinematical limit \cite{Basso:2020xts}. }

In the cases of  the octagon  \re{chi-oct} and localization \re{chi-loc} symbols we
similarly  find (cf. \re{semi} and \re{sum-f})
\begin{align}\la{7}
&  \mathcal F_{\ell} ^{\text{oct}}(g)= {-\pi g + \lr{\ell+\ft12}\log g + B_\ell^ {\text{oct}}  + f^{\text{oct}}_{\ell}(g) + \dots}\,,
\\[2mm]
&  \mathcal F_{\ell} ^{\text{loc}}(g)= {\pi g - \lr{\ell-\ft12}\log g +  B_\ell^ {\text{loc}}+ f^{\text{loc}}_{\ell}(g) + \dots}\,,\la{8}
\end{align}
where $ f_{\ell}(g) $ stand  for $1/g^n$   corrections \rf{f-PT} and  dots denote non-perturbative corrections \re{f-nonPT}. Notice that the leading $O(g)$ term has an opposite sign in \rf{7} and \rf{8}.  
The Widom-Dyson constants $B_\ell^ {\text{oct}} $ and $B_\ell^ {\text{loc}} $ are given by \re{Bs}. The $O(\log g)$ term in both expressions is generated by the Fisher-Hartwig singularity of the symbol.
According to \re{Phi-oct}, the corresponding parameters $\beta_{\rm oct}$ and $\beta_{\rm loc}$ are integer and, therefore, the perturbative functions $f^{\text{oct}}_{\ell}(g)$ and $f^{\text{loc}}_{\ell}(g)$ are given by Borel non-summable series. 

For instance, for $\ell=2$ and $\beta=-1$ we find from \re{f-PT}, \re{As} and \re{In1}
\begin{align} \label{f-oct-weak}\notag
& f^{\text{loc}}_{\ell=2}(g)= - {3\log 2\over 8\pi}g^{-1} - {3\log^22\over 16\pi^2}g^{-2} - \lr{\frac{15 \zeta (3)}{2048 \pi ^3}+\frac{\log ^32}{8
   \pi ^3}} g^{-3}
\\[2mm]   
& -\lr{\frac{45 \zeta (3) \log 2}{2048 \pi ^4}+\frac{3 \log ^42}{32
   \pi ^4}}g^{-4}-\lr{\frac{945 \zeta (5)}{262144 \pi ^5}+\frac{45 \zeta (3) \log
   ^22}{1024 \pi ^5}+\frac{3 \log ^52}{40 \pi ^5}}g^{-5}
%   -\frac{\frac{765 \zeta
%%   (5) \log (2)}{262144 \pi ^6}+\frac{\log ^6(2)}{16 \pi ^6}}{g^6}
+\dots
\end{align}
One can verify that, in agreement with \re{f-imp}, \rf{4} and \rf{3}, all terms in this expression involving $\log 2\ov \pi$ can be eliminated by changing the expansion parameter to
$g'=g+{\log 2\ov \pi}$.  

There exists an interesting relation between the  two different perturbative functions $f^{\text{oct}}_{\ell}(g)$ and $f^{\text{loc}}_{\ell}(g)$. 
It follows from the identity $I_n^{\text{loc}}= - I_{n} ^{\text{oct}}$  in \re{In1}. A close examination of \re{f-imp} shows that the function $f_{\ell}(g)$ is formally invariant under transformation $g\to -g$ and $I_n\to -I_n$  (with $n=0,1,2,\dots$). This leads to
\begin{align}\label{-g}
f^{\text{oct}}_{\ell}(g)=f^{\text{loc}}_{\ell+2}(-g)\,.
\end{align}
We recall that both functions depend on $\ell_\beta=\ell+\beta$ and 
the shift $\ell\to \ell+2$ on the right-hand side is needed to compensate the difference $\beta_{\rm oct}-\beta_{\rm loc}=2$.
Because the  functions on both sides of \re{-g} suffer from Borel singularities, this 
relation
 is rather formal and it 
should be understood as 
an equality 
between the expansion coefficients in  the two series. Equivalently, upon the Borel transform \re{Bor},  Eq. \rf{-g}  leads to the relation 
\begin{align}
\mathcal B_\ell ^{\rm oct}(\sigma)=\mathcal B_{\ell+2}^{\rm loc}(-\sigma)\,,
\end{align}
which maps the Borel singularities of the two functions into each other. In particular, the leading Borel pole of $\mathcal B_\ell ^{\rm oct}(\sigma)$ for $\sigma>0$ is in one-to-one correspondence with the pole of $\mathcal B_{\ell+2}^{\rm loc}(\sigma)$ for $\sigma<0$ closest to the origin.
 
\section{Non-perturbative corrections to  SAK formula}\label{sect:np-SAK}
 
Let us   now   employ the method developed in \cite{Basso:2009gh} to compute non-perturbative corrections to $\mathcal F_\ell(g)$ in  \re{semi} at strong coupling. 
They are described by the function $\Delta f_\ell(g)$ 
which is expected to have a general form \re{f-nonPT}. 

To find the dependence of  $\Delta f_\ell(g)$ on the coupling constant $g$ and non-negative integer $\ell$, we shall examine two functions related to 
$\mathcal F_\ell(g)$. The first one is 
\begin{align}\label{deq}
g\partial_g   \mathcal F_\ell(g) = -\tr \Big[g\partial_g K {1\over 1-K}\Big],
\end{align}
where we used  the determinant representation \re{det-K} of 
 $\mathcal F_\ell(g)$ in terms of matrix $K$.

The second one is $\mathcal F_{\ell+2}(g)- \mathcal F_\ell(g)$, i.e.   the difference of  functions with shifted indices.
It follows from the definition \re{det-K} and \re{K-def} that $\exp(\mathcal F_{\ell+2})$ is given by the determinant of the matrix $(1-K)$ with the first row and column removed, $\exp(\mathcal F_{\ell+2})=\det(1- K)\big|_{  n,m\ge 2}$. Then, the ratio of the determinants 
$\exp(\mathcal F_{\ell+2})$ and $\exp(\mathcal F_{\ell})$
can be evaluated using   Cramer's rule  as  
\begin{align} \label{feq}
D_\ell(g) \equiv \e^{\mathcal F_{\ell+2}(g)- \mathcal F_\ell(g)}  =\Big({1\over 1-K}\Big)_{11}\,.
\end{align}
Having computed \re{deq} and \re{feq} at strong coupling, we can obtain relations for $g\partial_g \Delta f_\ell(g)$ and $\Delta f_{\ell+2}(g)-\Delta f_\ell(g)$. 

It proves convenient to introduce an auxiliary function  $\Gamma(x,y)$ \cite{Basso:2009gh,Basso:2020xts}
\begin{align} \label{Gamma}
& \Gamma(x,y) = {1\over y}\Big[1-\chi\Big({x\over 2g}\Big)\Big]\gamma(x,y) \,,
\\\notag
& \gamma(x,y) = 2\sum_{n,m\ge 1}(-1)^{n+m} \sqrt{(2n+\ell-1)(2m+\ell-1)}J_{2n+\ell-1}(x) J_{2m+\ell-1}(y)\Big({1\over 1-K}\Big)_{mn} \,.
\end{align}
The rationale for defining this function is that both quantities \re{deq} and \re{feq} can be expressed in terms of $\Gamma(x,y)$ in a simple way. 

Indeed, taking into account \re{K-def}, we can rewrite \re{deq} as
\begin{align}\notag\label{meq1}
g\partial_g \mathcal F_\ell(g) &=  \int_0^\infty dx\, \Gamma(x,x)\, g\partial_g\log(1-\chi(x/(2g)))
\\
&=  -2g\int_0^\infty dx\, \Gamma(2gx,2gx)\, x\partial_x\log(1-\chi(x))\,.
\end{align}
To find the ratio \re{feq}, we examine asymptotic behaviour of  the function $\Gamma(x,y)$ for small $x$ and/or $y$. In both cases, the leading contribution to \re{Gamma} comes from the Bessel functions with the minimal index
\begin{align}\label{small-x1}
\Gamma(x,y) \stackrel{x\to 0}{\sim} x^{2\beta+\ell+1}\,,\qqqquad
\Gamma(x,y) \stackrel{y\to 0}{\sim} y^{\ell}\,,
\end{align}
where in the first relation we applied \re{small-x} and replaced $J_{\ell+1}(x) \sim x^{\ell+1}$.
Combining together the two limits, $x\to 0$ and $y\to 0$, we find from \re{Gamma}
\begin{align}\label{G-small} 
\Gamma(2gx,2gy)=x^{2\beta+\ell+1} y^\ell \left[  {b\, g^{2\ell+1} D_\ell  \over \Gamma(\ell+1)\Gamma(\ell+2)}
+ \dots\right],
\end{align}
where dots denote terms suppressed by powers of $x$ and $y$. Thus, the ratio \re{feq} can be obtained from the leading behaviour of $\Gamma(2gx,2gy)$ at small $x$ and $y$.

In what follows, we first determine the function $\Gamma(x,y)$ and, then, apply \re{meq1} and \re{G-small} to compute the two quantities defined in \re{deq} and \re{feq}.
We use \re{K-def} together with \re{Gamma} to find that $\Gamma(x,y)$ verifies the (infinite) system of integral equations  
\begin{align}\label{meq}
\int_0^\infty {dx\over x} J_{2n+\ell -1}(x)\,  \Gamma(x,y) = {J_{2n+\ell -1}(y)\over y}\,, 
\qquad  \ \    n\ge 1\ .  
\end{align}
The function $\Gamma(x,y)$ has to satisfy the additional conditions that follow from its definition. According to \re{Gamma}, it is given by the product of the entire function $\gamma(x,y)$ and the meromorphic function $ \left(1-\chi\left({x/ (2g)}\right)\right)/y$. It follows from \re{zeros} that the latter function vanishes at $x=\pm 4\pi i g x_n$ and has poles at $x=\pm 4\pi i g y_n$ (with $n=1,2,\dots$). The function $\Gamma(x,y)$ inherits 
these properties, e.g., 
\begin{align}\label{zero-G}
\Gamma(\pm 4\pi i g x_n,y)=0\,.
\end{align}
We show below that the integral equation \re{meq}, supplemented with the information about analytical properties of $\Gamma(x,y)$, is sufficient to construct its solution.

As was explained in  Section \ref{sect:SAK}, the perturbative part of the strong-coupling expansion \re{semi} takes a different form for integer and half-integer $\beta$. This property can also be seen from \re{meq}. 
We verify using \re{Gamma} and \re{Phi} that the function $\Gamma(x,y)$ satisfies 
\begin{align}\label{par}
& \Gamma(x,y)=(-1)^{2\beta+\ell+1}\Gamma(-x,y)= (-1)^\ell \Gamma(x,-y)\,,  
\end{align}
where we took into account the Bessel function  identity $J_k(-x)=(-1)^k J_k(x)$. As a result, the integrand on the left-hand side of \re{meq} is an odd/even function of $x$ for even/odd 
$2\beta$. For half-integer $\beta$, this allows us to extend the integration in \re{meq} to the whole real $x$-axis and evaluate the integral by residues at the poles of $\Gamma(x,y)$. For integer $\beta$, we show below that the relation \re{meq} leads to a Riemann-Hilbert problem for $\Gamma(x,y)$.

Let us  consider separately the cases of  half-integer and integer  $\beta$.  

\subsection{`Easy' case: half-integer $\beta$}

It follows from \re{par} that $\Gamma(-x,y) = (-1)^{\ell} \, \Gamma(x,y)$ and, therefore, $\Gamma(x,y)$ is an even (odd) function of $x$ for $\ell$ even (odd). This allows us to rewrite the relation \re{meq} as
\begin{align}\label{eq-simp}
\int_{-\infty}^\infty {dx\over 2x} J_{2n+\ell -1}(x)\,  \Gamma(x,y) = {J_{2n+\ell -1}(y)\over y}\,, \qquad \ \ n\ge 1\ .
\end{align}
Solving the integral equation \re{eq-simp}, we follow the same steps as in \cite{Basso:2009gh}.  Details of the calculation can be found in Appendix~\ref{app:int-eq}.
We start with performing a Fourier transform of $\Gamma(x,y)$ with respect to $x$
\begin{align}\label{Four}
\Gamma (x,y) = \int_{-\infty}^\infty dk\, \e^{ikx} \tilde \Gamma(k,y)\,.
\end{align} 
Substituting \re{Four} into \re{eq-simp} and exchanging the order of integration, we find that the integral over $x$ vanishes for $k^2>1$. Then, solving the relation \re{eq-simp} we can determine the function $\tilde \Gamma(k,y)$  for $k^2\le 1$. 
To find the function $\tilde \Gamma(k,y)$ for $k^2>1$, we invert the relation \re{Four} and replace $\Gamma(x,y)$ with its representation \re{Gamma} as a product of $1-\chi(x/(2g))$ and an entire function $\gamma(x,y)/y$. Computing the integral over $x$ by residues, 
we obtain the representation for $\tilde \Gamma(k,y)$ at $k^2>1$ as a sum over poles of the function $1-\chi(x/(2g))$. 

Finally, 
splitting the integration region in \re{Four} into $k^2\le 1$ and $k^2>1$, we replace $\tilde \Gamma(k,y)$ with its expressions in each of these regions and arrive at the following result for $\Gamma(x,y)$ (see Appendix~\ref{app:int-eq})
\begin{align}\notag\label{Ge}
\Gamma(x,y) =& {1\over\pi}\left[ \frac{\sin (x-y)}{x-y}+\frac{\sin (x+y)}{x+y}\right] 
\\ \notag &
+ {1\over \pi} \sum_{n=0}^{\ell/2-1} a_n(y) 
 {\e^{ix}p_{2n} (ix)  - \e^{-ix}p_{2n} (-ix) \over (ix)^{2n+1}} 
 \\ &
 + \sum_{j\ge 1} c_j (y) \left[{\e^{-ix}\over  4\pi gy_j+ix} + {\e^{ix}\over 4\pi gy_j-ix}  \right]
\end{align}
for even $\ell$, and 
\begin{align}\notag\label{Go}
\Gamma(x,y) =&{1\over\pi}\left[ \frac{\sin (x-y)}{x-y}-\frac{\sin (x+y)}{x+y}\right] 
\\ \notag &
+{1\over \pi}\sum_{n=1}^{(\ell-1)/2} a_n(y) {\e^{ix}p_{2n-1} (ix)  - \e^{-ix}p_{2n-1} (-ix) \over i (ix)^{2n}} 
\\
&
+i  \sum_{j\ge 1} c_j (y) \left[{\e^{-ix}\over  4\pi gy_j+ix}- {\e^{ix}\over 4\pi gy_j-ix}  \right]
\end{align}
for odd $\ell$.  

The relations \re{Ge} and \re{Go} are valid for $\beta\ge -(\ell+1)/2$ in which case the function $\Gamma(x,y)$ remains regular at small $x$,
see \re{small-x1}.~\footnote{Having determined the function $\Gamma(x,y)$ under this condition, we can 
obtain the same function with $\beta\to \beta-1$ by rescaling $b\to b/(2\pi y_1)^2$ and taking the limit $y_1\to 0$ in \re{zeros}. In the similar manner, replacing $b\to b (2\pi x_1)^2$ in \re{zeros} and going to the limit $x_1\to 0$, one produces the transformation $\beta\to\beta+1$. \label{foot}}
The first two lines in \re{Ge} and \re{Go} come from integration over $k^2\le 1$ in \re{Four} and the last line from $k^2>1$. Both expressions 
involve the polynomials
\begin{align}\label{p-pol}
p_n(x) =\sum_{p=0}^n x^{n-p}  \frac{(-1)^p \ \Gamma (n+p+2)}{(2 p+1)\text{!!}\  \Gamma (n-p+1)}
\end{align}
as well as some functions $a_n(y)$ and $c_j(y)$. A distinguished property of $p_n(x)$ is that  the $x$-dependent coefficients in front of $a_n(y)$ in \re{Ge} and \re{Go} are regular for $x\to 0$.

The functions  $a_n(y)$ describe the contribution of zero modes of the integral equation 
\re{eq-simp} whereas $c_j(y)$ define the residue of $\Gamma(x,y)$ at the poles $x=\pm 4\pi i g y_j$.  
Both sets of functions can be found by requiring the functions \re{Ge} and \re{Go} to satisfy the relations \re{small-x1} and \re{zero-G}. To illustrate this, we take $\beta=-1/2$ and $\ell=0,1$.
 
\subsection*{Special solutions}
 
For $\beta=-1/2$ and $\ell=0$ the relation \re{Ge} simplifies as
\begin{align}\label{ini1}
\Gamma(x,y) &={1\over\pi}\left[ \frac{\sin (x-y)}{x-y}+\frac{\sin (x+y)}{x+y}\right] 
+ \sum_{j\ge 1} c_j (y) \left[{\e^{-ix}\over  4\pi gy_j+ix} + {\e^{ix}\over 4\pi gy_j-ix}  \right].
\end{align}
It does not involve the zero modes contribution and satisfies \re{small-x1}. 

To find the functions $c_j(y)$ we impose the condition \re{zero-G}.
Because \re{ini1} is an even function of $x$, it is sufficient to require that $\Gamma(x,y) $ vanishes for
$x=4\pi i g x_n$. This leads to the following (infinite) system of equations
\begin{align}\notag\label{inf-sys}
{}& \sum_{j\ge 1} {c_j(y)\over x_n-y_j} - {1\over 2\pi}\left(
\frac{ \e^{-i y}}{ x_n-iy/(4 \pi  g )}+\frac{\e^{i y}}{x_n+iy/(4 \pi  g)} \right) 
\\
{}& =
\e^{-8\pi g x_n} \left[\sum_{j\ge 1} {c_j(y)\over x_n+y_j} - {1\over 2\pi}\left(
\frac{ \e^{-i y}}{ x_n+iy/(4 \pi  g )}+\frac{\e^{i y}}{x_n-iy/(4 \pi  g)} \right)\right],
\end{align}
where $n\ge 1$ and $x_n$ are positive. At strong coupling, the expression on the second line is 
exponentially suppressed. This suggests to look for a solution to \re{inf-sys} in the form of expansion in powers of  $\e^{-8\pi g x_n}$. 

Arranging the roots $x_n$  in ascending order, $0<x_1<x_2 < \dots$, we find that 
the leading contribution to $c_j(y)$ comes from the smallest root
\begin{align}\label{c-exp}
c_j(y)= c_j^{(0)} (y)+ \e^{-8\pi g x_1}c_j^{(1)}(y) +\dots \,,
\end{align}
where dots denote subleading corrections of the form $\e^{-8\pi g (m_1 x_1+m_2 x_2 +\dots) }$
with $m_i$ non-negative integer. If all roots are multiples of the smallest one, e.g., $x_n/x_1$ are positive integer, the expansion 
in \re{c-exp} runs in powers of $\e^{-8\pi g x_1}$. It is easy to see that this is indeed the case for the symbols in \re{chi's}.
Combining together \re{ini1} and \re{c-exp} we obtain 
\begin{align}\label{Gamma-sum}
\Gamma(x,y)=\Gamma^{(0)}(x,y)+ \e^{-8\pi g x_1}\Gamma^{(1)}(x,y) +\dots \,,
\end{align}
where $\Gamma^{(0)}(x,y)$ is given by \re{ini1} with $c_j(y)$ replaced by the leading result $c_j^{(0)}(y)$.
In a similar manner, $\Gamma^{1)}(x,y)$ is given by the sum in \re{ini1} with $c_j(y)$ replaced by $c_j^{(1)}(y)$.
The dots in \re{Gamma-sum} denote subleading exponentially small corrections.

Substituting \re{c-exp} into \re{inf-sys} and matching the terms on both sides of the relation we find
\begin{align}\label{rc}\notag 
& \sum_{j\ge 1} {c_j^{(0)} (y)  \over x_n-y_j} ={1\over 2\pi}\left[ 
\frac{ \e^{-i y}}{ x_n-iy/(4 \pi  g )}+\frac{\e^{i y}}{x_n+iy/(4 \pi  g)} \right],
 \\  
& \sum_{j\ge 1} {c_j^{(1)} (y)  \over x_n-y_j} = \delta_{n1} \left[\sum_{j\ge 1}{c_j^{(0)} (y) \over x_1+y_j}  -\frac{ \e^{-i y}}{2 \pi  \left(x_1+iy/(4 \pi  g )\right)}-\frac{\e^{i y}}{2 \pi 
   \left(x_1-iy/(4 \pi  g)\right)} \right],
\end{align}
where $n\ge 1$. The left-hand side of both relations involve a Cauchy matrix $1/ (x_n-y_j)$. Inverting this matrix, we can determine the functions $c_j^{(0)} (y)$ and $c_j^{(1)} (y)$ (see Appendix~\ref{app:res} for details).  

The above relations were obtained for $\beta=-1/2$ and $\ell=0$. The same analysis can be carried  out for $\beta=-1/2$ and $\ell=1$. We can start with \re{Go} and go through the same steps to find that the function 
$\Gamma(x,y)$ takes the same form \re{Gamma-sum}. The only difference as compared to the previous case is that the coefficients $c_j^{(0)} (y)$
and $c_j^{(1)}(y)$ satisfy the system of equations that differ from \re{rc} by signs in front of various terms on the right-hand side.

Using the obtained expressions for $c_j^{(0)}(y)$, it is possible 
to express 
$\Gamma^{(0)}(x,y)$ for $\beta=-1/2$ and $\ell=0,1$ in terms of the function $\Phi(x)$ defined in \re{Phi}, see Appendix~\ref{app:res}. This leads to
\begin{align}\notag\label{rr2}
\Gamma^{(0)}(2gx,2gy) 
  = &{} {i\over 4g\pi} \Big[
  {\e^{-2ig (x-y)}\over x-y}{\Phi(-x)\over\Phi(-y)}- {\e^{2ig (x-y)}\over x-y}{\Phi(x)\over\Phi(y)} 
\\
&  
+   (-1)^\ell \left(  {\e^{-2ig (x+y)}\over x+y}{\Phi(-x)\over\Phi(y)} 
 - {\e^{2ig (x+y)}\over x+y}{\Phi(x)\over\Phi(-y)}\right) \Big].
\end{align}
We verify that this function satisfies the relation \re{zero-G} for $g\gg 1$. For $x=\pm 2\pi i x_n$ each term on the right-hand side of \re{rr2} is either exponentially small at strong coupling, or proportional to $\Phi(-2\pi i x_n)=0$. Note that the function \re{rr2} is regular for $x=\pm y$.

Similarly, the function $\Gamma^{(1)}(2gx,2gy)$ admits the following representation
\begin{align}\notag\label{gam2}
\Gamma^{(1)}(2gx,2gy) = {}& {1 \over 4g\pi^2 }  {F(2\pi i x_1)\over  F(-2\pi i x_1)}
\left[\e^{2igx} F(x) +(-1)^\ell \e^{-2igx}F(-x) \right]
\\[2mm] 
{}& \times 
{x_1\over x_1^2+y^2/(2\pi )^2}
\left[{\e^{2ig y}\over F(-y)}+(-1)^\ell{\e^{-2ig y}\over F(y)}\right],
\end{align}
where we introduced the notation 
\begin{align}\label{F}
F(x)\equiv  {\Phi(x)\over 1-ix/(2\pi x_1)}\,.
\end{align}
The additional factor in the denominator ensures that $F(-2\pi i x_1)$ is different from zero.
As above, we verify that $\Gamma^{(1)}(2gx,2gy)$ vanishes for $x=\pm 2\pi i x_n$ and $n\ge 2$. For $x=\pm 2\pi ix_1$ the function \re{gam2} is different from zero and scales as $\e^{4\pi gx_1}$. Going back to 
\re{Gamma-sum} we observe that its contribution to $\Gamma(\pm 4\pi ig x_1,2gy)$ cancels against the $O(\e^{-4\pi gx_1})$ contribution coming from $\Gamma^{(0)}(\pm 4\pi ig x_1 ,2gy)$.

We would like to emphasize that the relations \re{rr2} and \re{gam2}  hold only  for $\beta=-1/2$ and $\ell=0,1$.
In particular, they automatically satisfy the condition \re{small-x1}. For $\beta> -\ell/2$ the situation is different.
For arbitrary $a_n(y)$ and $c_j(y)$ the functions \re{Ge} and \re{Go} scale at small $x$ as $O(x^0)$ and $O(x)$, respectively. The relation  \re{small-x1} implies that the first $2\beta+\ell+1$ terms of the small $x$ expansion of both functions have to vanish. Imposing this condition on  \re{Ge} and \re{Go} allows us to express the zero modes $a_n(y)$ in terms of functions $c_j(y)$ (with $j\ge 1$) and, in addition, obtain non-trivial relations for infinite sums $\sum_j c_j(y)/y_j^m$ with $m=1,\dots,\beta+1/2$. Requiring the resulting expression for $\Gamma(x,y)$ to 
verity \re{zero-G} and going through the same steps as above, we find that at strong coupling $c_j(y)$ and  $\Gamma(x,y)$ have the same general form as before, Eqs.~\re{c-exp} and \re{Gamma-sum}. Important difference as compared with the previous case is that for $\beta=-1/2+p$ the expansion coefficients $c_j^{(0)} (y)$ and $c_j^{(1)} (y)$ in \re{c-exp} scale at strong coupling as $O(g^p)$. They satisfy a system of linear relations analogous to \re{rc} supplemented with the additional relations for the sums $\sum_j c_j(y)/y_j^m$ (with $m=1,\dots,p$) mentioned above. Solving these relations we can obtain the expressions for $\Gamma^{(0)}(2gx,2gy)$ and $\Gamma^{(1)}(2gx,2gy)$ that are valid for arbitrary half-integer $\beta\ge -1/2$ and non-negative $\ell$. 

\subsection*{Strong coupling expansion}\label{sect:str}

Taking into account \re{semi}, \re{f-nonPT}, \re{A1} and \re{As}, we expect that the strong-coupling expansion of $g\partial_g \mathcal F_\ell$ looks like
\begin{align}\label{logder-exp}
g\partial_g \mathcal F_\ell=-2g I_0 +\frac12(2\beta\ell+\beta^2)- 8\pi c_1 x_1\, g^{n_1+1} \e^{- 8\pi  g \,x_1} +\dots\,,
\end{align}
where dots denote terms suppressed by powers of $1/g$ as well as subleading exponentially small corrections.

Using the obtained expressions for the function $\Gamma(x,y)$ in  \re{Gamma-sum},
 we can apply \re{meq1} to compute \re{logder-exp}. In this way, we should be able to reproduce the first two terms on the right-hand side of \re{logder-exp} and, most importantly, identify the values of the parameters $c_1$, $n_1$ and $x_1$ defining the leading non-perturbative correction. 

Let us  start again with $\beta=-1/2$ and $\ell=0,1$. We use \re{rr2}  to get
\begin{align}  \notag\label{G0}
 \Gamma^{(0)}(2gx,2gx) =&{} {1\over\pi}  -{i\over 4\pi g}  \partial_x \log {\Phi(x)\over \Phi(-x)}
 \\
 &{}
 + (-1)^\ell { i\over 8\pi g x } \left[ \e^{-4gi x} {\Phi(-x)\over\Phi(x)} - \e^{4gi x} {\Phi(x)\over\Phi(-x)} \right] . 
\end{align}
Substituting this expression into \re{meq1} and taking into account \re{In},  we find that the first term on the right-hand side of \re{G0} gives rise to 
$(-2gI_0)$ term in \re{logder-exp}. The second term in \re{G0} yields the $O(g^0)$ correction to \re{logder-exp} 
\begin{align}\label{arc}
-\int_0^\infty {dx\over 2\pi i}\, x\partial_x\log(1-\chi(x))\,\partial_x \log {\Phi(x)\over \Phi(-x)}
=\int_{-\infty}^\infty {dx\over 2\pi i} \lr{\phi(x)-x\phi^2(x)} =-\frac18\,.
\end{align}
Here in the first relation we replaced the symbol $(1-\chi(x))$ with its expression \re{Phi}, introduced notation for $\phi(x) = \partial_x\log \Phi(x)$
 and extended the integration to the whole real axis. In the second relation, we took into account that $\phi(x)$ has poles in the lower half-plane and deformed the integration contour
to the upper half-plane to become an arc of infinite radius, $x=R\e^{i\alpha}$ with $R\to\infty$ and $0 \le \alpha\le \pi$. It follows from \re{Phi-inf} that $\phi(x)\sim -\beta/x$ on this arc and the integral \re{arc} can be easily evaluated at $\beta=-1/2$.

Similarly, the contribution of the last term in \re{G0} to \re{meq1} can be written as
\begin{align}\label{aux-int}
\frac{(-1)^\ell}4\int_{-\infty}^\infty {dx\over 2\pi i}\,  x \partial_x\log(1-\chi(x))  \e^{-4gi x} {\Phi(-x)\over\Phi(x)}  \lr{{1\over x+i0} +{1\over x-i0} }\,.
\end{align}
The last factor arises because the sum of the  two terms inside the brackets on the second line of \re{G0} is regular for $x\to 0$ but each term separately has a pole $1/x$. The integral \re{aux-int} can be evaluated by closing the integration contour to the lower half-plane and by picking up residue at the poles. Replacing the functions $\chi(x)$ and $\Phi(x)$ with their expressions \re{zeros} and \re{Phi}, we find that the integrand has simple poles at $x=0$, $x=-4\pi i y_n$ and double poles at $x=-4\pi i x_n$. The residue at the pole $x=0$ yields a constant $(-1)^\ell /4$, whereas the contribution of the two remaining sets of poles is exponentially small at strong coupling.  The residue at the double pole is enhanced by the factor of $g$ and the leading contribution comes from the double pole closest to the origin, $x=-4\pi i x_1$. As a result, the integral \re{aux-int} is given by
\begin{align}\label{bra}
&(-1)^\ell\left[{ 1 \over 4}- 8g\pi x_1 \e^{-8\pi g x_1} \Lambda(x_1) (1+O(1/g)) 
\right],
\end{align}
where the notation was introduced for
\begin{align}
\label{La}
&\Lambda(x_1) \equiv  {F(2i\pi x_1)\over F(-2i\pi x_1)}  = \prod_{n\ge 2} {x_n+x_1\over x_n -x_1}
\prod_{n\ge 1} {y_n -x_1\over y_n +x_1}\,. 
\end{align}
Here  the function $F(x)$ is given by \re{F}.

The first term inside the brackets \re{bra} contributes to the $O(g^0)$ term in  \re{logder-exp} whereas 
the
second term   defines a non-perturbative correction to $g\partial_g  \mathcal F_\ell$. Similar contribution also comes from the second term in \re{Gamma-sum} defined in
\re{gam2}. 
Substituting $\e^{-8\pi g x_1}\Gamma^{(1)}(2gx,2gx)$ into \re{meq1} and carrying out the integration, we find that its contribution to $g\partial_g  \mathcal F_\ell$ scales at strong coupling  as $\e^{-8\pi g x_1}$ and, therefore,  is subleading compared to \re{bra}.

Finally, combining together \re{arc} and \re{bra} we arrive at
\begin{align}\label{der-Lambda}
g\partial_g \mathcal F_\ell(g) = -2gI_0+\frac18\big(2 (-1)^\ell-1\big)- 8(-1)^\ell g\pi x_1 \e^{-8\pi g x_1} \Lambda(x_1)+\dots\ , 
\end{align}
where dots stand for subleading exponentially suppressed corrections. We recall that this relation holds for $\beta=-1/2$ and $\ell=0,1$. It is easy to see that for these values of the parameters the first two terms on the right-hand side of \re{logder-exp} coincide with the analogous terms in \re{der-Lambda}. 

Matching \re{der-Lambda} to \re{semi} we identify the leading non-perturbative correction to $\mathcal F_\ell(g)$ 
for $\beta=-1/2$ and $\ell=0,1$
\begin{align}\label{Delta-f}
\Delta f_\ell\Big|_{\beta=-1/2} = (-1)^\ell  \Lambda(x_1) \e^{-8\pi g x_1}(1+O(1/g))\,.
\end{align}
It is straightforward to generalize the relation \re{Delta-f} to arbitrary non-negative $\ell$ and half-integer $\beta$. As we will see in a moment, the non-perturbative correction to  the difference of functions $\Delta f_{\ell+2}-\Delta f_\ell$ scales as 
$O(g^{-1} \e^{-8\pi g x_1})$ and, therefore, it is suppressed by the factor of $g$ as compared with \re{Delta-f}. This means that the leading non-perturbative correction cancels in the difference $\Delta f_{\ell+2}-\Delta f_\ell$ and, as a consequence, the relation \re{Delta-f} holds for an arbitrary $\ell$. 

To restore the $\beta$-dependence of \re{Delta-f}, we recall (see footnote \ref{foot}) that sending one of the roots of the function \re{Phi} to zero, say $x_i\to 0$, generates the shift $\beta\to\beta+1$. It is easy to see from \re{La} that  under this transformation $\Lambda(x_1) \to -\Lambda(x_1)$. Thus, in order to restore the $\beta$-dependence of $\Delta f_\ell$, it is sufficient to insert $ (-1)^{\beta+1/2}$ on the right-hand side of \re{Delta-f}
\begin{align}\label{np-scale}
\Delta f_\ell  = (-1)^{\ell+\beta+1/2}\Lambda(x_1) \e^{-8\pi g x_1}(1+O(1/g))\,.
\end{align}
This relation defines the leading non-perturbative correction to $\mathcal F_\ell(g)$ for half-integer $\beta$.

Let us now examine the ratio of functions \re{feq}. Applying \re{semi}, \re{A1}, \re{B} and \re{As}, we find that it has the following form at strong coupling
\begin{align}\label{ratio-gen}
D_\ell  = {g^{2 \beta } \over b}
\frac{G(\ell+3)  G(\ell+\beta +1)  }{G(\ell+1) G(\ell+\beta +3)}
   \left[ 1- (\ell+\beta
   +1) {I_1\over g}  + \Delta f_{\ell+2}-\Delta f_{\ell} + O(1/g^2) \right] \,,
\end{align}
where $I_1$ is given by \re{I1}.

We can compute $D_\ell$ using \re{G-small}, by examining the leading behaviour of $\Gamma(x,y)$ at small $x$ and $y$. 
For $\beta=-1/2$ and $\ell=0,1$ we find from \re{Gamma-sum}, \re{rr2} and \re{gam2} for $x, y\to 0$
\begin{align}\notag\label{small-as}
& \Gamma(2gx,2gy) \Big|_{\ell=0} = \frac{2}{\pi}
\left[1 -{1\over 2g} \lr{i \Phi '(0)-{F(2\pi i x_1)\over  F(-2\pi i x_1)} \frac{ \e^{-8 \pi  g x_1}}{\pi x_1} } +O(1/g^2)\right]  \,,
\\[2mm]
& \Gamma(2gx,2gy) \Big|_{\ell=1} ={8\over 3\pi} g^2 xy\left[1-{3\over 2g} \lr{i \Phi '(0)+ {F(2\pi i x_1)\over  F(-2\pi i x_1)} \frac{ \e^{-8 \pi  g x_1}}{\pi x_1} } +O(1/g^2) \right] \,,
\end{align}
where the functions $\Phi(x)$ and $F(x)$ are defined in \re{Phi} and \re{F}, respectively.
Matching the relations \re{small-as} to \re{G-small} we obtain
\begin{align}\label{ratio-spec} 
{D_\ell} = {k_\ell \over g b}  \left[ 1- {1\over 2g}\left(2\ell
   +1\right)  \lr{I_1-(-1)^\ell  {\Lambda(x_1) \over \pi x_1}\e^{-8 \pi  g x_1}} + \dots \right],
\end{align}
where $k_0=2/\pi$, $k_1=8/(3\pi)$ and $\Lambda(x_1)$ is given by \re{La}. 

We verify that  \re{ratio-spec} agrees with \re{ratio-gen} for $\beta=-1/2$ and $\ell=0,1$ and obtain the 
following relation for the non-perturbative corrections
\begin{align}\label{delta-np}
(\Delta f_{\ell+2}-\Delta f_{\ell})\Big|_{\beta=-1/2} =(-1)^{\ell}\left(2\ell+1\right)\Lambda(x_1)  {\e^{-8 \pi  g x_1}\over 2g\pi x_1}\,.
\end{align}
As expected, the expression on the right-hand side is suppressed by the factor of $g$ as compared to \re{Delta-f}. The relation \re{delta-np} holds for $\beta=-1/2$ and $\ell=0,1$.

To find a general expression for $\Delta f_{\ell+2}-\Delta f_{\ell}$, we repeated the calculation of $\mathcal F_{\ell+2}-\mathcal F_\ell$ for $\ell=2,3$ and $\beta=p-1/2$ with $p=1,2,3$. In this way, we reproduced the first two terms inside the brackets on the right-hand side of \re{ratio-gen} and identified the leading non-perturbative correction to $\Delta f_{\ell+2}-\Delta f_{\ell}$. We found that this correction can be obtained by applying the following transformation to the $O(1/g)$ perturbative term in \re{ratio-gen}
\begin{align}\label{patt}
I_1 \to I_1 -(-1)^{\ell+\beta+1/2}  {\Lambda(x_1) \over \pi x_1}\e^{-8 \pi  g x_1}\,.
\end{align}
The resulting relation for the leading non-perturbative correction is then 
\begin{align}\label{fde0}
\Delta f_{\ell+2}-\Delta f_{\ell} = (-1)^{\ell+\beta+1/2}(\ell+\beta
   +1)\Lambda(x_1)  {\e^{-8 \pi  g x_1}\over g\pi x_1}\,.
\end{align}
Notice that the expression on the right-hand side depends on the sum $\ell+\beta$. Recall that the perturbative function \re{f-imp} has the same property.

\subsection{`Hard' case: integer $\beta$}

The main difference as compared to the previous case is that,
as follows from \re{par},  $ \Gamma(-x,y) = -(-1)^{\ell} \, \Gamma(x,y)$ and, therefore, $\Gamma(x,y)$ is an odd (even) function of $x$ for $\ell$ even (odd). As will see below, this entails a change of the properties of both perturbative and non-perturbative expansions of $\mathcal F_\ell$. 

As we  have seen in the previous subsection,  for half-integer $\beta$ the non-perturbative corrections depend on the sum $\ell+\beta$. We assume below that the same is true also for integer $\beta$. Indeed, this property holds for the perturbative function $f_\ell(g)$ that is given by a Borel non-summable series for integer $\beta$. The Borel singularities induce an exponentially small ambiguous contribution to the perturbative function $f_\ell (g)$ which is canceled 
 in the  sum of  $f_\ell (g)$   with $\Delta f_\ell (g)$. Because the former depends on $\ell+\beta$, the same should be true for $\Delta f_\ell(g)$. Taking the advantage of this property, we can restrict our analysis to $\ell=0,1$ and an arbitrary integer $\beta$.

Let us start with  $\beta=0$ and  then  extend consideration to arbitrary integer $\beta>0$. For $\ell=0,1$ an infinite system of equations \re{meq} can be cast into a compact form
\begin{align}\label{bes}\notag
 \int_0^\infty dx\, \cos(x u)\ \Gamma_{\ell=0}(x,y) = \cos(yu)\,, 
 \\ 
 \int_0^\infty dx\, \sin(x u)\ \Gamma_{\ell=1}(x,y) = \sin(yu)\,.
\end{align}
To recover the relations \re{meq}, it is sufficient to replace the  trigonometric functions with their Bessel series expansion and match the coefficients on both sides.  It is important to emphasize that the relations \re{bes} only hold for $-1<u<1$.

Applying the Fourier transform \re{Four} we get from \re{bes}
\begin{align}\label{RH}
 \dashint_{-\infty}^\infty {dk \,\tilde \Gamma(k,y)\over k-u} = (-i)^{1+\ell}\cos\lr{yu-\ell {\pi\over 2} }\,,\qquad  \ \ \ -1 < u< 1 \,,
\end{align}
where the integral is defined using the principal value prescription. The function $\tilde \Gamma(k,y)$ has a definite parity,
\begin{align}
\tilde \Gamma(-k,y)=-(-1)^\ell \, \tilde \Gamma(k,y)\,,
\end{align}
and its analytical properties are in one-to-one correspondence with analogous properties of the function $\Gamma(x,y)$ described at  the beginning of this Section. In particular, we can show, following~\cite{Basso:2009gh}, that because
$\Gamma(x,y)$ has an infinite sequence of simple poles at $x=\pm 4\pi i y_j$, the function $\tilde \Gamma(k,y)$ has the following form for $k>1$ 
\begin{align}\label{k>1}
\tilde\Gamma(k,y) = \sum_{j\ge 1}c_{j}(y) \e^{-4\pi g(k-1)y_{j}}  \,.
\end{align}
Here the functions $c_j(y)$ define the residue $\Gamma(x,y)$ at $x=\pm 4\pi i y_j$.

The resulting Riemann-Hilbert problem, Eqs.~\re{RH} -- \re{k>1}, is similar to that discussed in~\cite{Basso:2009gh}. Going along the same lines as in that paper we find 
\begin{align}\notag\label{k<1} 
 \tilde\Gamma(k,y)   =& {(-i)^{1+\ell}\over \pi \sqrt{1-k^2}}\,\dashint_{-1}^{1}\frac{dp}{\pi}\frac{\sqrt{1-p^2}}{k-p}  \cos\lr{py-\ell {\pi\over 2} }
\\
&+ {1\over \pi \sqrt{1-k^2}} \int_1^\infty  dp\, \tilde \Gamma(p,y) \sqrt{p^2-1} \lr{{1\over k-p} + {(-1)^\ell \over k+p}}\,.
\end{align}
These relations are valid for $|k|\le 1$. They allow us to express the function $\tilde\Gamma(k,y)$ for $k^2<1$ in terms of the same function $\tilde \Gamma(p,y)$ defined for $p^2>1$. 

Replacing $\tilde \Gamma(p,y)$ in the last relation with \re{k>1}, we obtain a representation for $\tilde\Gamma(k,y)$ in terms of an  infinite set of functions $c_j(y)$ (with $j\ge 1$). As in the previous subsection, we can determine these functions from the requirement of $\Gamma(x,y)$ to satisfy the additional conditions \re{G-small} and \re{zero-G}.
Substituting \re{k>1} and \re{k<1} into \re{Four} we find after some algebra  
\begin{align}\notag\label{G-l=0}
\Gamma_{\ell=0}(x,y) = & \frac{x}{x^{2}-y^{2}}[x J_{1}(x)J_{0}(y)-y J_{0}(x)J_{1}( y)] 
\\[2mm]\notag
& +\sum_{j\ge 1}c_{j}(y)\,\frac{x e^{4\pi g y_{j}}}{4\pi g y_{j}-ix}[\rI_{0}(ix) K_{1}(4\pi g y_{j})+\rI_{1}(ix)K_{0}(4\pi g y_{j})] 
 \\
& + \sum_{j\ge 1}c_{j}(y)\,\frac{x e^{4\pi g y_{j}}}{4\pi g y_{j}+ix}[\rI_{0}(ix) K_{1}(4\pi g y_{j})-\rI_{1}(ix)K_{0}(4\pi g y_{j})]\,,
 \end{align}
 \begin{align}\notag
\notag
\Gamma_{\ell=1}(x,y) = & \frac{x}{x^{2}-y^{2}}[x J_{0}(x)J_{1}(y)-y J_{1}(x)J_{0}( y)] 
\\[2mm]\notag
& +i\sum_{j\ge 1}c_{j}(y)\,\frac{x e^{4\pi g y_{j}}}{4\pi g y_{j}-ix}[\rI_{0}(ix) K_{1}(4\pi g y_{j})+\rI_{1}(ix)K_{0}(4\pi g y_{j})] 
 \\
& -i \sum_{j\ge 1}c_{j}(y)\,\frac{x e^{4\pi g y_{j}}}{4\pi g y_{j}+ix}[\rI_{0}(ix) K_{1}(4\pi g y_{j})-\rI_{1}(ix)K_{0}(4\pi g y_{j})]\,,  \label{G-l=1}
\end{align}
where $J_n$, $\rI_n$ and $K_n$ are  Bessel functions.
We verify that at small $x$ and $y$ these expressions satisfy the relation \re{small-x1} for $\beta=0$. For positive integer $\beta$, the condition $\Gamma(x,y) \sim x^{2\beta+\ell+1}$ as $x\to 0$, leads to the additional relations for the coefficient
functions $c_j(y)$.

For $\beta=0$ and $\ell=0$, we substitute \re{G-l=0} into \re{zero-G} to obtain the system of linear relations for the coefficient functions $c_j(y)$
\begin{align}\notag\label{sys1}
0=& \frac{4\pi  g x_{n} J_{0}(y)+r_{n} y J_{1}(y)}{16\pi^{2}g^{2}x_{n}^{2}+y^{2}}
\\\notag
& +\frac{1}{4\pi g}\sum_{j\ge 1} \,\frac{\tilde c_{j}(y)}{ y_{j}+ x_{n}}[r_{n} -K_{0}(4\pi g y_{j})/K_{1}(4\pi g y_{j})] 
\\
&
+\frac{1}{4\pi g}\sum_{j\ge 1}\,\frac{ \tilde c_{j}(y)}{y_{j}- x_{n}}[r_{n} +K_{0}(4\pi g y_{j})/K_{1}(4\pi g y_{j})] \,,
\end{align}
where $n\ge1$ and  we defined 
\begin{align}\label{r-def}
\tilde c_{j}(y) \equiv  c_{j}(y) e^{4\pi g y_{j}}K_{1}(4\pi g y_{j}) \,,\qqqquad
r_n\equiv {\rI_0(4\pi g x_n)\over \rI_1(4\pi g x_n)} \ . 
\end{align}
For $\beta=0$ and $\ell=1$, combining together \re{G-l=1} and \re{zero-G}, we find that $\tilde c_j(y)$ satisfy  similar relations. 

The relation \re{sys1} can be further simplified at strong coupling. 
For $g\gg 1$, the ratio $K_{0}(4\pi g y_{j})/K_{1}(4\pi g y_{j})$
is given by an asymptotic sign-alternating series in $1/g$. Replacing it by the leading term we find (for $n\ge 1$)
\begin{align}\notag\label{tc}
 &   \sum_{j\ge 1}\frac{\tilde c_{j}(y)\,}{x_{n}-y_j} -\frac14\lr{{J_0(y)-iJ_1(y)\over x_n-iy/(2\pi g)}+{J_0(y)+ iJ_1(y)\over x_n+ iy/(2\pi g)}}  
\\
 &={r_{n}-1\over r_{n}+1} \left[ \sum_{j\ge 1}\frac{\tilde c_{j}(y)\,}{x_{n}+y_{j}} -\frac14\lr{{J_0(y)-iJ_1(y)\over x_n+iy/(2\pi g)}+{J_0(y)+ iJ_1(y)\over x_n- iy/(2\pi g)}} \right]\,.
\end{align}
It is instructive to compare the last relation with \re{inf-sys}. We note that the exponential functions are replaced by  a linear combination of the Bessel functions and an exponentially small factor of $\exp(-8\pi g x_n)$ with
\begin{align}\label{rr}
{r_{n}-1\over r_{n}+1} = i \e^{-8\pi g x_n}\lr{1+O(1/g)}+ {1\over 16\pi g x_n} \lr{1+O(1/g)}+ \dots \,.
\end{align}
Here both terms are accompanied by series in $1/g$ and dots denote terms suppressed by powers of $\e^{-8\pi g x_n}$. 

Because the left-hand side of \re{rr} is a well-defined real function of the coupling constant $g$, the appearance of the factor `$i$' on the right-hand side of \re{rr}  may be  surprising. 
It
 is related to the fact that the series entering the second term in \re{rr} suffer from Borel singularities. To make it well-defined, we can move the coupling constant slightly above the real axis  by $g\to g+i0$. This transformation amounts to deforming the integration contour in the Borel plane in the vicinity of the singularities (see \re{Bor}). It induces an imaginary contribution to the second term in \re{rr} which cancels against a similar contribution from the first term in \re{rr} in such a way that their sum remains real and well-defined (see Appendix~\ref{app:bor} for details).

In a close analogy with \re{c-exp}, we look for solutions to \re{tc} in the form
\begin{align}\label{t-c}
\tilde c_j(y)= \tilde c_j^{(0)} (y)+ i \e^{-8\pi g x_1}\tilde c_j^{(1)}(y) +\dots \,,
\end{align}
where dots denote corrections suppressed by powers of $1/g$. 
Substituting \re{t-c} into \re{tc} and taking into account \re{rr}, we find that 
the coefficient functions $\tilde c_j^{(0)} (y)$ and $\tilde c_j^{(1)} (y)$ satisfy (for $n\ge 1$)
\begin{align}\notag\label{rc1}
 & \sum_{j\ge 1}\frac{\tilde c_{j}^{(0)} (y)\,}{x_{n}-y_j} = \frac14\lr{{J_0(y)-iJ_1(y)\over x_n-iy/(2\pi g)}+{J_0(y)+ iJ_1(y)\over x_n+ iy/(2\pi g)}} \,,
 \\
& \sum_{j\ge 1} {\tilde c_j^{(1)} (y)  \over x_n-y_j} = \delta_{n1} \left[\sum_{j\ge 1}{\tilde c_j^{(0)} (y) \over x_1+y_j}  -\frac{J_0(y)-iJ_1(y)}{4 \left(x_1+iy/(4 \pi  g )\right)}-\frac{J_0(y)+ iJ_1(y)}{4
   \left(x_1-iy/(4 \pi  g)\right)} \right].
\end{align}
These relations are similar to \re{rc}. As mentioned above, they can be obtained from \re{rc}  by replacing exponential functions with a linear combination of the Bessel functions
\begin{align}\label{transf}
\e^{\pm iy} \to J_\pm(y)\equiv  J_0(y)\pm iJ_1(y)\,.
\end{align}
It is therefore not unexpected  that the solutions to \re{rc} and \re{rc1} are related to each other by the same transformation. 

Solving the system of equations \re{rc1} as well as the analogous system for $\beta=0$ and $\ell=1$, we 
can find the functions \re{G-l=0} and \re{G-l=1} at strong coupling
\begin{align}\label{Gamma-sum1}
\Gamma(x,y)=\Gamma^{(0)}(x,y)+ i \e^{-8\pi g x_1}\Gamma^{(1)}(x,y) +\dots \,,
\end{align}
where dots  stand for  terms suppressed by powers of $1/g$ and by exponentially small factors $\e^{-8\pi g x_n}$ (with $n\ge 1$).  Here the functions $\Gamma^{(0)}(x,y)$ and $\Gamma^{(1)}(x,y)$ are obtained from \re{rr2} and \re{gam2} by applying 
the transformation \re{transf}
\begin{align}\notag\label{hard} 
\Gamma^{(0)}(2gx,2gy) 
  = &{} {ix \over 4} \Big[ \lr{
  {J_-(2g x) J_+(2g y) \over x-y}{\Phi(-x)\over\Phi(-y)}- {J_+(2g x) J_-(2g y) \over x-y}{\Phi(x)\over\Phi(y)} }
\\\notag
&  
+ (-1)^\ell\lr{ {J_-(2g x) J_-(2g y) \over x+y}{\Phi(-x)\over\Phi(y)} 
 - {J_+(2g x) J_+(2g y)  \over x+y}{\Phi(x)\over\Phi(-y)}}  \Big]\,,
\\[2mm]
\notag 
\Gamma^{(1)}(2gx,2gy) = {}& {x \over 4\pi}  {F(2\pi i x_1)\over  F(-2\pi i x_1)}
\left[J_+(2g x) F(x) +(-1)^\ell J_-(2g x) F(-x) \right]
\\[2mm] 
{}& \times 
{x_1\over x_1^2+y^2/(2\pi )^2}
\left[{J_+(2g y)\over F(-y)}+(-1)^\ell{J_-(2g y)\over F(y)}\right],
\end{align}
where the functions $\Phi$, $F$ and $J_\pm$ were introduced in \re{Phi}, \re{F} and \re{transf}, respectively.

The important difference between \re{Gamma-sum} and \re{Gamma-sum1} is that, in virtue of \re{rr},  the perturbative series in $1/g$ on the right-hand side of \re{Gamma-sum1}  contains Borel singularities. As discussed above, these singularities are regularized by replacing $g\to g+i0$.

We can now apply \re{hard} to determine the strong-coupling expansion of the observable $\mathcal F_\ell$. Following the same steps as in Section~\ref{sect:str}, we examine \re{hard} at the coinciding point $y=x$. Using the properties of Bessel functions  \re{transf}, we find from \re{hard} for $g\gg 1$
\begin{align}\notag\label{G0-h}
 \Gamma^{(0)}(2gx,2gx) &{} ={1\over\pi}  -{i\over 4\pi g}  \partial_x \log {\Phi(x)\over \Phi(-x)}
 \\
 &{}
 - (-1)^\ell { 1\over 8\pi g x } \left[ \e^{-4gi x} {\Phi(-x)\over\Phi(x)} + \e^{4gi x} {\Phi(x)\over\Phi(-x)} \right] +O(1/g^2). 
\end{align}
The analogous expression for $ \Gamma^{(1)}(2gx,2gx)$ scales as $O(1/g^2)$ and produces a subleading contribution
to \re{meq1}.

Compared to  \re{G0}, the two terms inside the brackets in \re{G0-h} have an opposite relative sign. As we will see in a moment, this has important consequences for the properties of $\mathcal F_\ell$.
Substituting \re{G0-h} into \re{meq1} we find that the first term on the right-hand side of \re{G0-h} yields $(-2g I_0)$, whereas the second one leads to the integral that is similar to \re{arc}. Because  
$ \partial_x\log \Phi(x)\sim -\beta/x$ at large $x$ (see \re{Phi-inf}),  this integral vanishes for $\beta=0$.

The contribution of the last term in \re{G0-h} to \re{meq1} can be written as
\begin{align}\label{pv-int}
   { (-1)^\ell\over 4\pi } \lr{ \int_{C_+} dx+\int_{C_-} dx}\,  \partial_x\log(1-\chi(ix))
 \e^{-4g x} {\Phi(ix)\over\Phi(-ix)} \equiv  \varphi(g-i0)+\varphi(g+i0) \,,
\end{align}
where the integration contours $C_+$ and $C_-$ start at the origin and go to $+\infty$ slightly above $(C_+)$ and below ($C_-$) the real axis. Here we took into account that the two terms inside the brackets on the last line of \re{G0-h} have poles 
located along the imaginary axis and rotated the integration contour as $x\to -ix$ and $x\to ix$, respectively. 

The integrals on the left-hand side of \re{pv-int} define two branches of the function $\varphi(g)$.
After the change of variable $x=\sigma /(4g)$, they take the form of the Borel transform \re{Bor}
\begin{align}\notag\label{bor-phi}
& \varphi(g) = \int_0^\infty d\sigma\, \e^{-\sigma} \mathcal B_\varphi(\sigma/g)\,,
\\
& \mathcal B_\varphi(\sigma) =  { (-1)^\ell\over 4\pi }   \partial_\sigma \log(1-\chi(i\sigma/4))
{\Phi(i\sigma/4)\over\Phi(-i\sigma/4)}\,.
\end{align}
Using  \re{Phi} we find that $\mathcal B_\varphi(\sigma)$ is a meromorphic function with poles located at real $\sigma$. 
As a consequence, the function $\varphi(g)$ has a cut running along the real axis in the complex $g$-plane.~\footnote{Formally expanding \re{bor-phi} in powers of $1/g$, one would obtain that  $\varphi(g)$ is given by a Borel non-summable series. } 
The discontinuity of this function across the cut is given by an  (infinite) sum over the residues of $\mathcal B_\varphi(\sigma/g)$
at the poles located at positive $\sigma$.  At large positive $g$ the leading contribution comes from the 
double pole at $\sigma=8\pi g x_1$ 
\begin{align}\notag\label{im-part}
\varphi(g+i0)-\varphi(g-i0) & =2\pi i \res_{\sigma=8\pi g x_1} \lr{\e^{-\sigma}\mathcal B_\varphi(\sigma)}
 \\[2mm]
& =  i (-1)^\ell 8g\pi x_1 \e^{-8\pi g x_1} \Lambda(x_1) (1+O(1/g))\,,
\end{align}
where $\Lambda(x_1)$ was  defined in \re{La}.~\footnote{At large negative $g$, the leading contribution to \re{im-part} comes from the 
double pole at $\sigma=-8\pi g y_1$.}

Combining together the above relations, we obtain from \re{meq1}  
\begin{align}\label{sum-phi}
g\partial_g  \mathcal F_\ell(g) &= -2gI_0+\varphi(g-i0)+\varphi(g+i0) +\dots\,,
\end{align}
where dots denote subleading corrections including those coming from $\Gamma^{(1)}(2gx,2gy)$.
The sum of two terms on the right-hand side of \re{sum-phi} is a real function of $g$, whereas each term separately develops an exponentially small imaginary part \re{im-part}. 
Applying \re{sum-phi} and \re{im-part} we can obtain two other representations of the same quantity
\begin{align}\notag\label{der-Borel}
g\partial_g  \mathcal F_\ell(g) &= -2gI_0+ 8i (-1)^\ell g\pi x_1 \e^{-8\pi g x_1} \Lambda(x_1)+2\varphi(g-i0)  +\dots
\\[2mm]
&= -2gI_0- 8 i (-1)^\ell g\pi x_1 \e^{-8\pi g x_1} \Lambda(x_1)+2\varphi(g+i0)  +\dots\,.
\end{align}
The three representations, Eqs.~\re{sum-phi} and \re{der-Borel}, are equivalent. This illustrates a universal feature of the strong-coupling expansion that has been mentioned previously. Namely, the definition of non-perturbative, exponentially small corrections to $g\partial_g  \mathcal F_\ell(g)$ depends on a  regularization that one employs to integrate through the Borel singularities in \re{bor-phi}. 
According to \re{bor-phi}, 
the functions $\varphi(g+ i0)$ and $\varphi(g- i0)$ are obtained by shifting the integration contour in the Borel plane slightly above and below the real axis, respectively. In the representation \re{sum-phi}, the Borel poles are integrated using the principal value prescription.

Comparing \re{der-Borel} with \re{der-Lambda} we notice that non-perturbative corrections to both expressions involve the same quantity $8(-1)^\ell g\pi x_1 \e^{-8\pi g x_1} \Lambda(x_1)$. For half-integer $\beta$, the perturbative series $f_\ell(g)$ is well-defined and the coefficient in front of the non-perturbative term in \re{der-Lambda} can be determined unambiguously. In contrast, for integer $\beta$ this coefficient depends on the regularization of the  Borel singularities of the perturbative function $\varphi(g)$.

Integrating the second relation in \re{der-Borel} and matching $ \mathcal F_\ell$ to \re{semi}, we can find the leading non-perturbative correction 
\begin{align}\label{np-scale1}
\Delta f_\ell(g)  =  i(-1)^{\ell+\beta}\Lambda(x_1) \e^{-8\pi g x_1}(1+O(1/g))\,.
\end{align}
We would like to emphasize that this relation should be supplemented with the analogous relation for the perturbative function
\begin{align}\label{f-phi}
g\partial_g f_\ell(g) = 2\varphi(g+i0)\,,
\end{align}
where the `$+i0$' prescription on the right-hand side specifies a deformation of the integration contour in the vicinity of the Borel singularities in \re{bor-phi}.  If we used `$-i0$' prescription, the expression on the right-hand side of \re{np-scale1} would have an opposite sign.
Note that the relation \re{np-scale1} and its complex conjugate coincide with \re{np-scale} evaluated at $\beta=0$ and $\sqrt{-1}=\pm i $.
 
Finally, we can find the difference $\Delta f_{\ell+2}-\Delta f_{\ell}$ by evaluating non-perturbative corrections to the ratio \re{feq}. To obtain  $D_\ell$ using \re{G-small}, we examine the leading asymptotic behaviour of the function $\Gamma(2gx,2gy)$ given by \re{Gamma-sum1} and \re{hard} for $x, y\to 0$
\begin{align}\notag\label{small-as1}
& \Gamma_{\ell=0}(2gx,2gy) =x g \left[1 - {1\over g}\lr{i\Phi'(0) - i{F(2\pi i x_1)\over  F(-2\pi i x_1)} \frac{ \e^{-8 \pi  g x_1}}{\pi x_1}}+O(1/g^2) \right],
\\[2mm]
& \Gamma_{\ell=1}(2gx,2gy) = \frac12 x^2 y g^3 \left[ 1- {2\over g}\lr{i\Phi'(0) + i{F(2\pi i x_1)\over  F(-2\pi i x_1)} \frac{ \e^{-8 \pi  g x_1}}{\pi x_1}}+O(1/g^2)\right].
\end{align}
The first term inside the parentheses in both relations is real whereas the second one is pure imaginary. The situation here is similar to the one we encountered in \re{der-Borel}. The non-perturbative correction to \re{small-as1} corresponds to a particular regularization of the perturbative function $f_\ell(g) \to f_\ell(g+i0)$, see Eq.~\re{f-phi}. It amounts to deforming the integration contour in the Borel transform \re{Bor} slightly below the real axis, thus avoiding the Borel poles. 

Matching \re{small-as1} to \re{G-small} we  find 
\begin{align}\label{rat-h} 
{D_\ell} = {1\over b}  \left[ 1- {1\over g}\left(\ell
   +1\right)  \lr{I_1-i(-1)^\ell  {\Lambda(x_1) \over \pi x_1}\e^{-8 \pi  g x_1}} + O(1/g^2) \right],
\end{align}
where  $\Lambda(x_1)$ is given by \re{La}.  This relation is valid for $\beta=0$ and $\ell=0,1$. As in the previous case of half-integer $\beta$, we repeated the calculation of ${D_\ell}$ for $\beta=1,2,3,4$ and found that the leading non-perturbative correction to $D_\ell$ can be obtained through the transformation \re{patt}. 
This leads to 
\begin{align}\label{fde}
\Delta f_{\ell+2}-\Delta f_{\ell} = i(-1)^{\ell+\beta}(\ell+\beta
   +1)\Lambda(x_1)  {\e^{-8 \pi  g x_1}\over g\pi x_1}\,,
\end{align}
up to corrections suppressed by $1/g$. Notice that the leading correction \re{np-scale1} cancels on the right-hand side of \re{fde}.
 
\subsection{Degenerate symbols} 

The obtained relations for the non-perturbative corrections \re{np-scale1} and \re{fde} are valid for the symbol of the form \re{Phi}. A distinguished feature of \re{Phi} is that $(1-\chi(x))$ has simple roots at $x=\pm 2\pi ix_n$.
Examining different expressions of the symbols in \re{chi's} we observe that this condition is satisfied for $1-\chi_{\rm BES}(x)$ whereas the two remaining symbols, $1-\chi_{\rm oct}(x)$ and $1-\chi_{\rm loc}(x)$ have double roots.

To define the symbols with double roots it is sufficient to take the limit of \re{Phi} as $x_2\to x_1$. We recall that 
deriving the non-perturbative corrections \re{np-scale1} and \re{fde}, we neglected subleading corrections to $\Delta f_\ell(g)$ suppressed by powers of $\exp(-8\pi x_n)$ with $n\ge 2$. For $x_2\to x_1$ we also have to retain corrections proportional to 
$\e^{-8\pi x_2}$. They are given by the same expressions \re{np-scale1} and \re{fde} with $x_1$ replaced by $x_2$. 
For instance, we get from \re{np-scale1}
\begin{align}\label{np-scale2}
\Delta f_\ell  =  i(-1)^{\ell+\beta}\left[ \Lambda(x_1) \e^{-8\pi g x_1}+\Lambda(x_2) \e^{-8\pi g x_2}\right](1+O(1/g))\,,
\end{align}
where $\Lambda(x_2)$ is obtained from \re{La} by replacing $x_1\leftrightarrow x_2$.

In particular, for $x_2\to x_1$ we have, up to $O(x_1-x_2)$ corrections,
\begin{align}\label{La1} 
\Lambda(x_2) =-\Lambda(x_1) &=  {2x_1\over x_1-x_2}\prod_{n\ge 3} {x_n+x_1\over x_n -x_1}
\prod_{n\ge 1} {y_n -x_1\over y_n +x_1}
= {2x_1\over x_1-x_2} \Lambda'(x_1) \,,
\end{align}
where the notation was introduced for
\begin{align}\label{F1}
\Lambda'(x_1) =  {F'(2i\pi x_1)\over F'(-2i\pi x_1)}\,,\qqqquad
F'(x)= {\Phi(x)\over (1-ix/(2\pi x_1))^2}\,.
\end{align}
Combining the above relations we get from \re{np-scale2} in the limit $x_2\to x_1$
\begin{align}\label{enh}
\Delta f_\ell  =  i(-1)^{\ell+\beta} 16\pi g x_1\Lambda'(x_1) \e^{-8\pi g x_1}(1+O(1/g))\,,
\end{align}
As compared with \re{np-scale1}, this expression contains the additional factor of $g$. In the similar manner, we get from 
\re{fde}
\begin{align}\label{fde1}
\Delta f_{\ell+2}-\Delta f_{\ell} = 16i(-1)^{\ell+\beta}(\ell+\beta
   +1)\Lambda'(x_1) \e^{-8 \pi  g x_1} \,.
\end{align}
The relations \re{enh} and \re{fde1} hold for integer $\beta$. They can be obtained from the analogous relations \re{np-scale1} and \re{fde} by replacing
\begin{align}
\Lambda(x_1) \to 16 \pi g x_1 \Lambda'(x_1)\,.
\end{align}
For half-integer $\beta$, we can apply the same transformation to \re{np-scale} and \re{fde0} to get the corresponding expressions for the function 
$\Delta f_\ell(g)$ for the symbol with double roots. 

\subsection{Non-perturbative corrections at strong coupling} \label{sect:bor}
 
Let us summarize the obtained results for the non-perturbative corrections to \re{semi} at strong coupling.
 
For half-integer $\beta$,  perturbative $f_\ell(g)$ and non-perturbative $\Delta f_\ell(g)$ terms in \re{semi} are well-defined functions of the coupling constant $g$.
The function $f_\ell(g)$ takes into account perturbative corrections in $1/g$ and it takes the form \re{f-well}. The function $\Delta f_\ell(g)$ describes non-perturbative, exponentially small corrections to \re{semi}. It  satisfies the relations
\re{np-scale}  and \re{fde0}. 

For integer $\beta$, the function $f_\ell(g)$ is given by a Borel non-summable series \re{f-PT} and requires a regularization.  
Combining together the relations \re{Bor}, \re{f-phi} and \re{bor-phi}, we find that its Borel transform satisfies
\begin{align}
\sigma \partial_\sigma \mathcal B_f(\sigma) = - 2 \mathcal B_\varphi(\sigma) \,.
\end{align}
According to \re{bor-phi}, the function $B_\varphi(\sigma)$ has double poles at $\sigma=8\pi x_n$ 
and $\sigma=-8\pi y_n$ (with $n\ge 1$). As a consequence, $\mathcal B_f(\sigma)$ has the following schematic form
\begin{align}\label{2poles}
\mathcal B_f(\sigma) \sim A {\sigma\over \sigma-8\pi x_1}+B {\sigma\over \sigma+8\pi y_1} +\dots \,,
\end{align}
where we displayed the contribution of the two poles closest to the origin.
Substituting \re{2poles} into \re{Bor} and expanding the integral in powers of $1/g$, we
find   
\begin{align}\label{f-large}
f_\ell(g) \sim -\sum_k \left[ {A\over (8\pi x_1)^{k+1}} + (-1)^k {B\over (8\pi y_1)^{k+1}}\right] k! g^{-k} +\dots
\end{align}
As expected, the expansion coefficients grow factorially. For $x_1< y_1$ 
and $x_1>y_1$ the large order behaviour is controlled, respectively, by the first and second term inside the brackets. In both cases, the series \re{f-large} suffers from Borel singularities.

Due to the presence of the pole in \re{2poles} at $x=8\pi x_1$, the integral \re{Bor} is not well-defined and requires a regularization. Different ways of deforming the integration contour in \re{Bor} in the vicinity of the pole lead to different results for the perturbative term $f_\ell(g)$. They differ from each other by exponentially small terms $O(\e^{-8\pi g x_1})$ which are proportional to the residue at the pole. The dependence on the regularization disappears in the sum of perturbative and non-perturbative terms \re{sum-f}. 
We demonstrated that this sum  can be written in three equivalent ways
\begin{align}\label{tot}
\Delta \mathcal F_\ell (g) \equiv
\frac12 \left[ f_\ell (g+i0)+ f_\ell (g-i0)\right]
 =   f_\ell (g+i0)+\Delta f_\ell(g)
 =   f_\ell (g-i0)-\Delta f_\ell(g) \,.
\end{align}
Here $f_\ell (g+ i0)$ and $f_\ell (g- i0)$  are given by the Borel transform \re{Bor} in which the integration contour is shifted, respectively, slightly below and above the Borel poles. The additional exponentially small term $\Delta f_\ell(g)=\ha(f_\ell (g-i0) - f_\ell (g+i0))$ satisfies the relations \re{np-scale1} and \re{fde}.  
 
\subsection{Physics applications}\label{sect:app}
 
Let us   now  specify the non-perturbative function $\Delta f_\ell(g)$ for
 three physically relevant  cases of the symbol $\chi(x)$  in \re{chi-BES}, \re{chi-oct} and \re{chi-loc}.

\subsection*{Easy case: BES}

Let us first consider the symbol \re{chi-BES} with $\beta=-1/2$. According to terminology adopted  in the previous subsection, this corresponds to an  `easy' case. 

The relations \re{np-scale} and \re{fde0} for the non-perturbative function $\Delta f_{\ell}(g)$ depend on $x_1$ and $\Lambda(x_1)$. 
We recall that $x_1$ is the solution to $\Phi_{_{\text{BES}}}(-2i\pi x_1)=0$ closest to the origin. We apply 
\re{Phi-BES}, \re{F}  and \re{La} to find that $x_1=\ha$ and 
\begin{align}
F_{_{\text{BES}}}(x) =    \sqrt{\pi}\frac{\Gamma \left(1-\frac{i x}{2\pi}\right)}{2\Gamma
   \left(\frac{3}{2}-\frac{i x}{2\pi}\right) }\,,\qqqquad
   \Lambda_{_{\text{BES}}} = {F_{_{\text{BES}}}(i\pi)\over F_{_{\text{BES}}}(-i\pi)} =  \frac12\,.
\end{align}
We then use \re{np-scale} and \re{fde0} to find the leading non-perturbative corrections as
\begin{align}\notag\label{easy-np}
& \Delta f_\ell^ {\text{BES}}  = {(-1)^{\ell}\over 2}\e^{-4\pi g}\big(1+O(1/g)\big)\,,
\\
& \Delta f_{\ell+2} ^{\text{BES}} -\Delta f_{\ell}^{\text{BES}} =   (-1)^{\ell}\left(2\ell
   +1\right){\e^{-4 \pi  g}\over 2\pi g}\,.
\end{align}
It follows from the first relation that the leading non-perturbative correction to $ \mathcal F^ {\text{BES}}_\ell$ is 
$ \ha (-1)^{\ell}\e^{-\sqrt \lambda}  $. It is easy to verify that for $\ell=0,1,2$ this result is in an agreement with the exact relations \re{B-guess}.

In a similar manner, we apply the second relation in \re{easy-np} together with \re{bes-ex} to obtain the ratio \re{feq} for $\ell=0$
\begin{align}\label{semi-rat}
D_0^{\text{BES}}=\e^{ \mathcal F_{\ell=2}- \mathcal F_{\ell=0}} = {1\over\pi g} \left[ 1-{\log 2 \over 2\pi g}+{1\over 2\pi g} \e^{-4\pi g} + O(\e^{-8\pi g})\right],
\end{align}
where the last term denotes subleading non-perturbative corrections. This relation should be compared with the exact expression of $D_0$ that follows from  \re{B-guess}
\begin{align}
D_0^{\text{BES}} =  {\log \cosh(2\pi g) \over 2\pi^2 g^2} = {1\over\pi g} \left[ 1-{\log 2 \over 2\pi g}+{1\over 2\pi g}\log(1+\e^{-4\pi g})\right].
\end{align}
We observe a perfect agreement.
 
\subsection*{Hard cases: octagon and localization}

Let us now consider the symbols \re{chi-oct} and \re{chi-loc}. As compared to the previous case, there are two important differences. 

First, because the strength of the Fisher-Hartwig singularity is integer in this case, $\beta_{\rm oct}=-\beta_{\rm loc}=1$ (see Eq.~\re{Phi-oct}), the perturbative correction to $\mathcal F_\ell$ has a more complicated form \re{f-large}. Matching \re{Phi-oct} to \re{Phi}, we identity the leading root ($x_1$) and pole ($y_1$) of the two symbols as
\begin{align}
x_1^{\text{oct}}=y_1^{\text{loc}}=1\,,\qqqquad  y_1^{\text{oct}}= x_1^{\text{loc}}=\frac12\,.
\end{align}
It follows from \re{f-large} that the perturbative series $f_\ell^{\text{oct}}(g)$ is sign alternating, \footnote{This does not mean however that $f_\ell^{\text{oct}}(g)$ does not suffer from Borel singularities. Due to $x_1^{\text{oct}}> y_1^{\text{oct}}$, the large order behaviour of $f_\ell^{\text{oct}}(g)$ is controlled by the Borel pole of \re{2poles} at $x=-4\pi$. The pole at $x=8\pi$ generates non-perturbative corrections.} whereas 
$f_\ell^{\text{loc}}(g)$ has the expansion coefficients of the same sign. This property is in agreement with the explicit expressions 
\re{f-oct-weak} and \re{-g}.

Secondly, the leading root of the symbols \re{Phi-oct} is double degenerate and, as a consequence, the non-perturbative corrections to $\mathcal F_\ell$ satisfy the relations \re{enh} and \re{fde1}. We apply \re{F1} and \re{Phi-oct} and identify the corresponding non-perturbative parameters as
\begin{align}
\Lambda_{\text{oct}}' = \frac1{64}\,,\qqqquad
\Lambda_{\text{loc}}' = \frac1{4}\,.
\end{align}
Then, it follows from \re{enh} that the leading non-perturbative correction satisfies 
\begin{align} \notag\label{rel1}
& \Delta f_\ell^{\text{oct}}(g) =  {(-1)^{\ell+1} \over 4}\, i \pi g \,  \e^{-8\pi g}(1+O(1/g))\,,
\\[2mm]
& \Delta f_\ell^{\text{loc}}(g) = (-1)^{\ell+1} \, 2i\pi g\,  \e^{-4\pi g}(1+O(1/g))\,.
\end{align}
Note that $\Delta f_\ell^{\text{oct}}$ is suppressed by the factor of $\e^{-4\pi g}$  compared to  $\Delta f_\ell^{\text{loc}} $. This property is expected because the Borel transform \re{2poles} of the perturbative functions $f_\ell^{\text{oct}}$ and $f_\ell^{\text{loc}}$ contains the leading pole located at $\sigma=8\pi$ and $\sigma=4\pi$, respectively. 

Similarly, the relation \re{fde1} takes the form
\begin{align}\notag\label{rel2}
& \Delta f^{\text{oct}} _{\ell+2}(g)-\Delta f^{\text{oct}} _{\ell} (g)= {i\over 4}(-1)^{\ell+1}(\ell 
   +2)\e^{-8 \pi g} \,,
\\[2mm]
& \Delta f^{\text{loc}} _{\ell+2}(g)-\Delta f^{\text{loc}} _{\ell}(g)  = 4i(-1)^{\ell+1} \ell \e^{-4 \pi g} \,.
\end{align}
Note that  the leading large $g$   terms  in \rf{rel1}  cancel  out in the difference.

It is important to emphasize that the relations \re{rel1} and \re{rel2} hold for large positive $g$. In fact, 
 the functions
$\Delta f^{\text{oct}} _{\ell} (g)$ and $\Delta f^{\text{loc}} _{\ell} (g)$ have different asymptotic behaviour 
at large positive and negative $g$ due to the Stokes phenomenon.

Recall that the perturbative series for $f^{\text{loc}} _{\ell}(g)$ and $f^{\text{oct}} _{\ell}(g)$ are related to each other as
\re{-g}. Due to the presence of Borel singularities in both series, this relation is formal. As discussed above, we can regularize these singularities 
by deforming the integration contour in \re{Bor} in the vicinity of the Borel poles. In this way, we get
\begin{align}\label{-g1}\notag
& f^{\text{oct}}_{\ell}(g+ i0)=f^{\text{loc}}_{\ell+2}(-g- i0)\,,
\\[2mm] 
& f^{\text{oct}}_{\ell}(g- i0)=f^{\text{loc}}_{\ell+2}(-g+ i0)\,.
\end{align}
This discontinuity  yields the non-perturbative correction $\Delta f_\ell(g) =\ha ( f_\ell(g-i0)-f_\ell(g+i0))$, i.e.   \re{-g1} leads to 
\begin{align}
\Delta f^{\text{oct}}_{\ell}(g) = - \Delta f^{\text{loc}}_{\ell+2}(-g)\,.
\end{align}
Combined together with \re{rel1} and \re{rel2}, this relation allows us to determine the asymptotic behaviour of the non-perturbative functions at large negative $g$.
 
Moreover, substituting \re{-g1} into the first relation in \re{tot} we find that the sum of the perturbative and non-perturbative contributions to $\mathcal F^{\text{oct}}_{\ell}(g)$   and   $\mathcal F^{\text{loc}}_{\ell}(g)$
 are related to each other as
\begin{align}
\Delta \mathcal F^{\text{oct}}_{\ell}(g)=\Delta  \mathcal F^{\text{loc}}_{\ell+2}(-g)\,.
\end{align}
Thus, the functions $\Delta \mathcal F^{\text{oct}}_{\ell}(g)$ and $\Delta \mathcal F^{\text{loc}}_{\ell+2}(g)$
can be identified as two branches of the same function defined for negative and positive $g$, respectively.

\section{Applications  to $\mathcal{N}=2$ superconformal models} \label{sect:N=2app}

Our aim in this section is to compute   the leading non-trivial corrections  to 
special observables in $\mathcal{N}=2$ four-dimensional superconformal models that are
  planar-equivalent to $\N=4$  SYM. These observables are controlled 
by the  localization matrix model and, as a result, 
 can be expressed  in terms of  the 
semi-infinite  matrix $K_{nm}$ defined in \re{K-def}
(see         \cite{Beccaria:2020hgy,Beccaria:2021ksw,Beccaria:2021vuc,Beccaria:2021ism}  and Appendix~A).

The relevant $\mathcal N=2$ models may be split into 
two classes I and II:
\begin{enumerate}
\item[I.] Models with gauge group $SU(N)$ or $Sp(2N)$ and matter  content summarized in Table~\ref{tab:1},
\begin{table}[h!]
\begin{equation}
\def\arraystretch{1.3}
\begin{array}[t]{ccccc}
\toprule
\text{model} & G & n_{\rm F} & n_{\rm S} & n_{\rm A} \\
\midrule
\sa & SU(N) & 0 & 1 & 1 \\
\fa & SU(N) & 4 & 0 & 2 \\
\midrule
\fatilde & Sp(2N) & 4 & 0 & 1 \\
\bottomrule
\end{array}\notag
\end{equation}
\caption{\label{tab:1}
Three $\mathcal N=2$ superconformal models of class I.}
\end{table}
where $n_{\rm F}$, $n_{\rm S}$, and $n_{\rm A}$ are the numbers of hypermultiplets in the 
fundamental, rank-2 symmetric and antisymmetric representations.
Guided by geometrical engineering, 
these models are  expected to be 
 holographically dual to  IIB superstring theory 
on orientifolds/orbifolds of  the type    AdS$_{5}\times S^{5}/\GG$  \cite{Park:1998zh,Ennes:2000fu}. 
They correspond to  a combination of  $2N$ D3-branes  together with   an
orientifold O7 plane
and, in   models with $n_{\rm F}\neq 0$,   also   with several 
 D7-branes.  
\item[II.]
Quiver theories (that we call $\mathsf{Q}_{L}$) 
with gauge group $SU(N)^{\otimes L}$, bi-fundamental matter,  and 
equal gauge couplings.
They   
are orbifold projections of the $\mathcal N=4$ SYM theory  
{and are dual to} 
 type  IIB superstring on AdS$_{5}\times S^{5}/\mathbb{Z}_{L}$   orbifold 
 \cite{Kachru:1998ys,Lawrence:1998ja,Oz:1998hr,Gukov:1998kk}.
\end{enumerate}

\subsection{Observables  in terms of  Bessel operator}

The  three simplest  observables in these  models   are  the 
 free energy on $S^4$,  half-BPS circular Wilson loop, and  some 
 correlators of  
chiral primary operators. They can   be computed  using localization  matrix model techniques.  It turns out  that the  leading non-trivial corrections to them 
can be expressed in terms of the Bessel operator  \re{K-def}. 

The interaction potential of the localization matrix model is  given by an (infinite) 
 sum of terms weighted   by  
powers of the  't Hooft   coupling. The evaluation of the 
observables is straightforward in  weak coupling  expansion, but 
the strong-coupling expansion poses  a  non-trivial problem. 
The latter limit is of main interest from the point of view of establishing correspondence with dual string theory, i.e. 
 for tests of the AdS/CFT correspondence. 

Explicitly,    semi-infinite matrix in \re{det-K}  appearing in the context of 
the above
$\mathcal N=2$ superconformal models  is given by~\cite{Beccaria:2020hgy}
\begin{align}
\label{K-loc}
\mathsf{X}_{nm}  &= -8\,(-1)^{n+m}\,\sqrt{(2n+\ell-1)(2m+\ell-1)}\  \notag \\
&\qquad  \times\int_{0}^{\infty}\frac{dt}{t}\frac{e^{2\pi t}}{(e^{2\pi t}-1)^{2}}\,J_{2n+\ell-1}(t\sqrt{\lambda})\,
J_{2m+\ell-1}(t\sqrt{\lambda})\,.
\end{align}
It is easy to see that,
upon change of the variable $t\to t/\sqrt{\lambda}$, the matrix $\mathsf{X}_{nm}$ coincides with the Bessel  matrix \re{K-def}  with the symbol 
  $\chi=\chi_{\rm loc}(x)$ given in \re{chi-loc}
\begin{align}\la{10}
\mathsf{X}_{nm} = K_{nm} \Big|_{\chi=\chi_{\rm loc}}\,.
\end{align}

\subsection*{Free energy}

The free energy is  defined as $F=-\log Z$, where $Z$ is the partition function of gauge  theory on $S^{4}$. It is a function of  $\lambda$ and $N$. Planar equivalence  
implies that at large $N$  the  free energy of  the above $\N=2$ models  is  equal   to the free energy of $\mathcal N=4$ SYM theory\foot{The free energy $F$   in general contains a UV divergence  proportional to the conformal a-anomaly  and thus is    scheme-dependent. 
Eq. \rf{free-N=4} is  the expression in a particular renormalization scheme
\ci{Pestun:2007rz,Russo:2012ay} (in \rf{free-N=4} we ignore a  $\lambda$ independent constant).}
\begin{align}\label{free-N=4}
F^{\mathcal N=4}(\lambda; N) = -\frac{1}{2}(N^{2}-1)\log \lambda \,.
\end{align}
The leading $O(N^2)$ term cancels in the difference $
\Delta F = F^{\mathcal N=2}-F^{\mathcal N=4} $
which is thus a {genuine} $\mathcal N=2$ quantity.

More precisely, let  us   define   
\begin{align}
\Delta F^{\M} &= \begin{cases}
F^{\mathsf M}(\lambda; N) -  F^{\mathcal N=4}(\lambda; N)\,, &\ \ \  \M = \sa\,, \fa\,, \fatilde\,, \\[2mm]
F^{\mathsf Q_{L}}(\lambda; N) - L\, F^{\mathcal N=4}(\lambda; N)\,, & \ \ \ \M = \mathsf{Q}_{L}\,. \la{11}
\end{cases}
\end{align}
We will  be interested in the leading $N\to \infty$ limit of $ \Delta F^{\M}$. 
This quantity has been studied previously in the $\sa$ \cite{Beccaria:2021vuc},  $\fa$ and $\fatilde$  \cite{Beccaria:2021ism}, and 
 $\mathsf{Q}_{2}$ models \cite{Beccaria:2021ksw}.

 In the $\sa$ and $\mathsf{Q}_{2}$ models, one finds the following explicit representations for the leading  correction to the free energy  in terms of the 
  Bessel operator  observable $\mathcal F_{\ell}$ defined as in 
  \rf{det-K}, \rf{K-def}, \rf{chi-loc} and \rf{10}
\begin{align}
\la{12}
& \Delta F^{\sa}(\lambda) = \frac{1}{2}\,\mathcal F^ {\text{loc}}_{\ell=2}\,, 
\\ \label{df-sa-q2}
& \Delta F^{\mathsf{Q}_{2}}(\lambda) = \frac{1}{2}\,\mathcal F^ {\text{loc}}_{\ell=2}+\frac{1}{2}\,\mathcal F^ {\text{loc}}_{\ell=1}\,.
\end{align} 
The relation \rf{12}  was  proved in \cite{Beccaria:2021vuc}, and    \re{df-sa-q2} is derived below  in Appendix~\ref{app:sc-orb}. 

From a string theory argument for the  non-planar correction to the 
half-BPS Wilson loop  \cite{Giombi:2020mhz} 
and  its relation to the free    energy  \cite{Beccaria:2021ksw}, one expects  that in both cases \re{12} and \re{df-sa-q2} the leading term at strong coupling should take the form 
\begin{equation}
\label{f-def}
\Delta F^{\mathsf M}(\lambda) = \f^{\M}\,\lambda^{1/2}+\dots\, , \qquad \qquad \l \gg 1 \ . 
\end{equation}
The constant $\f^{\mathsf{Q}_{2}}$  in the $L=2$   quiver  model 
was estimated  in  \cite{Beccaria:2021ksw} by a Monte Carlo numerical simulation 
(in a moderate range $\lambda < 450$) with the result  
\begin{equation}
\label{f-orb}
\f^{\mathsf{Q}_{2}} \simeq 0.08\, .
\end{equation}
In the $\sa$ model, an analytical determination of $\f^{\sa}$
was attempted in  \cite{Beccaria:2021vuc} by considering the leading-order (LO)  large $\lambda$ contribution
to the matrix elements of \re{K-loc} at $\ell=2$  before   computing the determinant.  
As a result, the  expected $\lambda^{1/2}$ scaling of $\Delta F^\sa $   was   reproduced   with the coefficient being  
\begin{equation}
\label{smallf-lo}
\f^{\sa}_{\rm LO} = \frac{1}{2\pi} = 0.159\dots\ .
\end{equation}
A numerical high-precision resummation of the weak-coupling expansion of $\Delta F^{\sa}$ was performed in \cite{Beccaria:2021vuc} 
using an improved conformal mapping  Pade' analysis.
 While the large $\l$ scaling exponent $1/2$  in \rf{f-def} was confirmed, the  Pade' estimate for its coefficient
 was substantially smaller $\f^{\sa}_{\rm Pade}\simeq 0.12$ than in \rf{smallf-lo},  
 so the final picture was not totally  satisfactory.

Turning to the 
 $\fa$ and $\fatilde$ models with fundamental
hypermultiplets,  one finds that the  large $N$ expansion of the free energy contains both odd and even powers of $1/N$.  
In the $\fa$ model with gauge group $SU(N)$ one gets \cite{Beccaria:2021ism} 
\begin{align}\notag  
& \Delta F^{\mathsf{FA}}= N\,F_{1}(\lambda)+F_{2}(\lambda)+\mathcal{O}(1/N)\,,  
\\[1.5mm] \notag
& F_{2}(\lambda) = \widetilde F_{2}(\lambda)+\Delta F^{\mathsf{SA}}(\lambda)\,, 
\\
& \partial_{\lambda}\widetilde F_{2}(\lambda) = -\frac{\lambda}{2}[\partial_{\lambda}^{2}(\lambda F_{1}(\lambda))]^{2}\,.\la{14}
\end{align}
Here the leading  $O(N)$ term $F_{1}(\lambda)$ can be found  in 
a closed form  \cite{Beccaria:2021ism}
\begin{equation}
F_1(\l)=\frac{2}{\sqrt{\lambda}} \int_0^\infty dt\,  \frac{e^{2\pi t}}{(e^{2\pi t}+1)^2}\Big[ \frac{J_1(2t\,\sqrt{\lambda})-t\, {\sqrt \lambda} +\frac{1}{2} (t \sqrt\lambda)^3}{t^2}\Big] \ . 
\la{eq:F1}
\end{equation}
It is    straightforward to   work out its   strong coupling  expansion  \cite{Beccaria:2021ism}
\begin{align}\notag
\la{6.2}
 F_1(\l)  =  {}&\frac{\log 2}{4\pi^{2}}\, \l -\frac{1}{4}   \log \l   -6\log\mathsf{A} + 
\frac{3}{4} +\frac{7}{6}\log 2 + \frac{1}{2}\log\pi
 \\
 &
-\frac{\pi^{2}}{4}  \,\lambda^{-1}+ {2^{5/2}\over \pi^{3/2}}\, \lambda^{1/4} \e^{-\sqrt\lambda}\big(1+O(\lambda^{-1/2})\big)\,,
\end{align}
Note that the second line of \rf{6.2}  contains only one perturbative term and that the coefficient of the non-perturbative correction\foot{We omit terms  with higher odd powers of $\e^{-\sqrt\lambda}$, see  \cite{Beccaria:2021ism}.}
  is real (cf. \rf{F-sa} and \rf{F-q2} below). 

 According to \rf{14}, the function 
$F_1(\lambda)$  determines the part $ \widetilde F_{2}(\lambda)$ of the subleading $O(N^0)$  correction. The remaining part of $F_2(\lambda)$   is the same  as in  free energy in the $\sa$ theory  given by  \re{df-sa-q2}.

In the $\fatilde$ model with the $Sp(2N)$ gauge group, the large $N$ expansion of the free energy  $\Delta F^{\fatilde}$ is much simpler (it is essentially determined by $F_1(\lambda)$) 
and thus does not involve
the Bessel operator.\footnote{The technical reason for this simplification is that 
the localization matrix model for $\fatilde$ model  has the interaction potential  which is  free from double-trace terms, \textit{cf.} \re{double-trace}. As a consequence,
 $\Delta F^{\fatilde}$ can be found analytically and an exact resummed expression is available for the leading strong coupling terms at each order in the $1/N$ expansion \cite{Beccaria:2021ism}.} 
 In this case one may directly  identify the  non-perturbative corrections to $\Delta F^{\fatilde}$ that are exponentially suppressed at large $\lambda$  and are partially controlled by 
resurgence properties \cite{Beccaria:2021ism}.

\subsection*{Non-planar correction to  half-BPS circular  Wilson loop}

In the  above $\N=2$ models  the localization yields a matrix model representation for the expectation value of the 
(suitably defined) half-BPS circular Wilson loop $\vev{\mathcal W}$.
The  planar equivalence implies that $\vev{\mathcal W}^{\mathsf{M}} $ in these models has a universal  large $N$ limit
\begin{equation}
\label{W-planar}
\lim_{N\to\infty} 
\vev{\mathcal W}^{\mathsf{M}} = \vev{\mathcal W}_{0} \equiv 
\frac{2}{\sqrt\lambda}\,I_{1}(\sqrt\lambda)\,,
\end{equation}
where $I_1$ is the modified Bessel function. 
The subscript `$0$' stands for the planar limit, \textit{i.e.} the leading term in the large $N$ expansion.
The leading non-planar correction to $\vev{\mathcal W}^{\mathsf{M}}$ defines the model-dependent function $q^{\mathsf M}(\lambda)$ \footnote{For a discussion of the Wilson loop 
  in the models $\fa$ and $\fatilde$ with the matter in the fundamental representation see \cite{Beccaria:2021ism}.}
\begin{equation}\la{414} 
\frac{\vev{\mathcal W}^{\M}}{\vev{\mathcal W}_{0}} = 1+\frac{1}{N^{2}}q^{\M}(\lambda) + \mathcal{O}(1/N^{4})\,, \qquad\qquad  \M = \sa, \mathsf{Q}_{L}\,.
\end{equation}
Remarkably, one can establish simple relations between $q^{\M}$ and
the leading correction $\Delta F^{\M}$ to the free  energy \cite{Beccaria:2021ksw,Beccaria:2021vuc}
\begin{align}\la{455}
\Delta q^{\mathsf{SA}}(\lambda) &= q^{\mathsf{SA}}(\lambda)-q^{\mathcal{N}=4}(\lambda) = -\frac{\lambda^{2}}{4}\frac{d}{d\lambda}\Delta F^{\mathsf{SA}}(\lambda)\,, \\
\label{dq-orb}
\Delta q^{\mathsf{Q}_{2}}(\lambda) & = q^{\mathsf{Q}_{2}}(\lambda)-q^{\mathcal{N}=4}(\lambda) = -\frac{\lambda^{2}}{8}\frac{d}{d\lambda}\Delta F^{\mathsf{Q}_{2}}(\lambda)\,,
\end{align}
where  $q^{\mathcal{N}=4}(\lambda) = {\l \ov 96} [ {\sqrt \lambda I_{2}(\sqrt\lambda)/ I_{1}(\sqrt\lambda)} - 12 ]=
{1 \ov 96} \lambda^{3/2}+\dots$ at strong coupling. 
Combining these relations together with \rf{12} and \re{df-sa-q2}, one can express the function $q^{\M}(\lambda)$ in terms of the Fredholm determinant $\mathcal F_\ell$ of the Bessel operator.

The string theory argument   suggests \cite{Giombi:2020mhz} that $q^{\M}(\lambda)$  should scale at strong coupling as $\l^{3/2}$.
The relations \rf{455} and \rf{dq-orb}   then 
 imply  that the coefficient of $\l^{3/2}$ should be proportional to $\rm C^\M$ in \rf{f-def}. 

\subsection*{Correlation  functions of chiral operators}
  
In contrast to the free energy and circular Wilson loop discussed above, the correlation functions of 
some chiral operators in $\mathcal N=2$ models of  class  I and II differ from their counterparts 
 in $\mathcal N=4$ SYM already at the leading large $N$ order. 

In the  class I  $\N=2$ models       this  happens for  correlators of chiral primary operators  $\mathcal O_{k}(x)= \text{tr}\, \varphi^{k}(x)$
involving odd power $k$ of the  complex scalar $\varphi(x)$ from  the $\mathcal N=2$ vector multiplet.
Defining the ratio of two-point  functions of anti-chiral/chiral operators in  
 $\mathcal N=2$  model and $\mathcal N=4$ SYM 
\begin{align}\label{ratio}
R_{k}^{\rm M }(\lambda)  &= \frac{\vev{\bar {\mathcal{O}}_{k}(x)\mathcal{O}_{k}(0)}^{\rm M  \phantom{aai}}}{\vev{\bar {\mathcal{O}}_{k}(x)\mathcal{O}_{k}(0)}^{\mathcal{N}=4}}\,, 
\end{align} 
one finds that, in the planar limit,  it is $1$ for even $k=2n$  but  a nontrivial function of $\lambda$ for odd $k=2n+1$. 

This property can be understood as follows.
Considering an  orbifold  projection of  a conformal  gauge  theory  with respect to some discrete subgroup 
of a global symmetry group  
 one can split  the observables into ``untwisted'' (even under the action of orbifold group) 
and ``twisted'' (odd under the orbifold group) \ci{Lawrence:1998ja,Bershadsky:1998mb,Bershadsky:1998cb,Zoubos:2010kh}. In the planar limit, the former are the same as in the original theory (but they differ at subleading order in $1/N$), while the latter deviate from the original theory ones already  at the leading large $N$  order. 
The  free energy, the circular Wilson loop and the ratio of the correlators $R_{2n}^{\rm M}$ are  examples of   ``untwisted''   observables whereas the ratio $R_{2n+1}^{\rm M }$
and analogous ratio of the correlators in class II models (see Eq.~\re{TT} below)  belong to the ``twisted'' sector.
 
In the $\mathsf Q_{2}$  orbifold  model with the  $SU(N) \times SU(N)$  gauge group the operators in the twisted sector are odd under interchanging 
the scalar fields $\varphi_{1}$ and $\varphi_{2}$  of the two  $\mathcal N=2$ 
$SU(N)$  vector multiplets. 
The $\mathsf{SA}$ model   is related to the $\mathsf Q_{2}$  model  by  an additional 
   orientifold projection (see, e.g., \cite{Billo:2021rdb}). 
In particular, 
the scalar fields $\varphi_{1}$ and $\varphi_{2}$  at the two nodes of  the $\mathsf Q_{2}$ quiver   are related to  the scalar field $\varphi$ belonging 
to the $\mathcal N=2$ vector multiplet of the $\mathsf{SA}$ theory
 as $\varphi_{1} =\varphi$ and  $\varphi_{2}=-\varphi$. This 
additional discrete modding  implies  that the chiral primary operators $\mathcal O_{2n+1}= \text{tr}\, \varphi^{2n+1}(x)$ and $\mathcal O_{2n}= \text{tr}\, \varphi^{2n}(x)$
belong, respectively, to the twisted  and untwisted sectors in the  SA  theory. This explains why the ratio of the correlators \re{ratio} 
is different from $1$ for the ``odd'' operators already at the planar level.
\footnote{For example,   the matrix model proof of planar equivalence is based on the assumption of an even distribution of eigenvalues in the large $N$ limit. This assumption is justified for  correlators of ``even'' chiral primaries $\mathcal O_{2n}(x)= \text{tr}\, \varphi^{2n}(x)$.
However, the correlators
of  ``odd'' chiral primaries $\mathcal O_{2n+1}(x)$ involve a deformed eigenvalue distribution % invalidating the derivation 
\cite{Beccaria:2020hgy}.} 

Explicitly,  in the $\sa$ model one finds for \rf{ratio} from the  localization matrix model representation  
\begin{align}
\label{CPO-ratio}
R_{2n+1}^{\sa}(\lambda)  
 = \left(\frac{1}{1- K_{[n]}}\right)_{11} \,,
\end{align}
where the semi-infinite matrix $K_{[n]}$ is obtained from the matrix $K$ in \re{K-def} with 
 $\ell=2$ by removing  its first $(n-1)$ rows and columns.
In \cite{Beccaria:2020hgy}, the relation in \re{CPO-ratio} was used to derive the  weak coupling expansion
 of $R^{\sa}_{2n+1}$.  At strong coupling, $R^{\sa}_{2n+1}$ can be determined in the leading-order  (LO) approximation as \cite{Beccaria:2021hvt}
\begin{equation}
\label{CPO-ratio-lo}
R_{2n+1}^{\sa}(\lambda) = \frac{8\pi^{2}}{\lambda}\,n(2n+1)+\dots \,.
\end{equation}
This result compares well to   direct Monte Carlo numerical evaluation of the matrix model  integral 
and to  Pade' resummation  of the perturbative series. 

For the correlators of even chiral operators in the $\sa$ model, the ratio \re{ratio} deviates from $1$  starting at order  $1/N^{2}$. Remarkably, like in the Wilson loop case \rf{455} and \rf{dq-orb},  
the $1/N^2$ term   
can be expressed in terms of $\l$-derivatives of the non-planar correction  in the 
free energy  $\Delta F^{\sa}(\lambda)$ defined in \re{12} \cite{Beccaria:2021vuc}
\begin{equation}\label{even-cor}
R^{\sa}_{2n}(\lambda) = 1-\frac{2n}{N^{2}}\Big[(2n^{2}-1)\,\lambda\,\partial_{\lambda}+(\lambda\,\partial_{\lambda})^{2}\Big]\,\Delta F^{\mathsf{SA}}(\lambda)+\mathcal{O}(1/N^{4})\,.
\end{equation}
According to \re{f-def}, the $1/N^2$  term  here 
 scales at strong coupling as $\lambda^{1/2}$ and is proportional to $\f^{\sa}$ in \rf{f-def}.
In an attempt to improve on the LO value of this coefficient  (\ref{smallf-lo}), a refined analysis 
was performed by  either  keeping the next to leading order  terms in the Bessel operator  matrix elements 
or  considering a sequence of its numerically exact finite dimensional truncations. 
The best estimates obtained by  these two  approximations   were  respectively \cite{Beccaria:2021hvt}
\begin{equation}
\label{smallf-nlo}
\f^{\sa}_{\rm NLO} = 0.113\,, \qqqquad \f^{\sa}_{\rm num} = 0.130\,.
\end{equation} 
The same methods were  also applied to   some  three-point functions of
 single-trace chiral primaries in the $\sa$ model \cite{Billo:2022xas}. The normalized extremal correlators
\begin{equation}
\label{3-point}
 R^{\sa}_{n_{1},n_{2}}(\lambda) = \frac{\vev{\mathcal{O}_{n_{1}}(x_1)\mathcal{O}_{n_{2}}(x_2)\, \overline{\mathcal{O}}_{n_{1}+n_{2}}(0)}^{\sa\phantom{aa}}}
{\vev{\mathcal{O}_{n_{1}}(x_1)\mathcal{O}_{n_{2}}(x_2)\, \overline{\mathcal{O}}_{n_{1}+n_{2}}(0)}^{\mathcal{N}=4}}\,, 
\end{equation}
can be expressed in terms of the resolvent $1/(1-K)_{nm}$. For even $n_1$ and $n_2$, the ratio tends, in the planar   limit, 
 to $1$  (as expected). 
In other cases the leading term in the strong-coupling expansion is given in the LO approximation by
\begin{align}\notag\label{3pt-LO}
& R^{\sa}_{2n,2m+1}(\lambda) = \frac{16\pi^{2}}{\lambda}\,m\,(n+m)+\dots\,, 
\\
& R^{\sa}_{2n+1,2m+1}(\lambda) = \frac{16\pi^{2}}{\lambda}\,n\,m+\dots\,.
\end{align}
In  class II models, i.e.  $\mathsf{Q}_{L}$ quiver theories, 
 a generalization of \re{ratio} and \re{CPO-ratio} has been worked out in \cite{Billo:2021rdb}. The operators in the ``twisted''   sector of the $\mathbb Z_L$ orbifold are defined as 
\begin{align}
 \label{twist}
 T_{\alpha, n} (x) &= \frac{1}{\sqrt L}\sum_{I=0}^{L-1}e^{-\frac{2\pi i\, I}{L}\alpha}\,  \mathcal{O}^{(I)}_{n}(x) \,, 
 \end{align}
where integer $\alpha$ satisfies $1\le \alpha \le L-1$ and
$ \mathcal{O}^{(I)}_{n} = \tr\,  \varphi_{I}^{n}$  corresponds to  the $I$-th node of the quiver. 
 The two-point  functions
 $\vev{\bar T_{\alpha, n} (x)T_{\alpha, n} (0)}$
of the twisted chiral operators \re{twist} can be computed using the localization technique. 
In  close analogy with \re{ratio}, one can define 
the ratio
\begin{align}\label{TT}
{\vev{\bar T_{\alpha, n} (x)T_{\alpha, n} (0)}^{\mathsf{Q}_{L}\phantom{a}} \over \vev{\bar {\mathcal{O}}_{n}(x)\mathcal{O}_{n}(0)}^{\mathcal{N}=4}}
\equiv 1+\Delta_{\alpha,n} (\lambda) \,.
\end{align}
In the planar limit, it admits the following representation in terms of the matrix \re{K-loc} for even and odd $n$  (see Eq.~(5.21) in \cite{Billo:2021rdb})
\begin{align}\notag\label{twist1}
&  1+\Delta_{\alpha,2k} (\lambda) =\Big({1\over 1-s_\alpha \mathsf{X}_{[k]}^{\rm even}}\Big)_{11}\ ,
\\
&  1+\Delta_{\alpha,2k+1} (\lambda) =\Big({1\over 1-s_\alpha \mathsf{X}_{[k]}^{\rm odd}}\Big)_{11},
\end{align}
where the matrices $\mathsf{X}^{\rm even}$ and $\mathsf{X}^{\rm odd}$ coincide with \re{K-loc} for $\ell=1$ and $\ell=2$, respectively. 
The matrix $X_{[n]}$ is again obtained from $X$ by removing its first $(n-1)$ rows and columns.
 The dependence on $\alpha$ enters \re{twist1} through the parameter\footnote{
 For $\alpha=0$ the relation \re{twist} defines untwisted chiral operator. In this case, $s_0=0$ and 
the corresponding ratio of the  two-point functions \re{twist1} is $1$ in the planar limit.}
 \begin{align}\label{s}
s_\alpha=\sin^2 \lr{\pi\alpha\over L}\,.
\end{align}
 At strong coupling, the expressions in  \re{twist1} can be evaluated in 
the LO approximation to give, for both even and odd $n$,  
\begin{equation}\label{2pt-tw}
1+\Delta_{\alpha,n} (\lambda)  {=} \frac{4\pi^{2}}{\lambda\,s_{\alpha}}\,n(n-1)+\dots\,.
 \end{equation}
Recently,  three-point functions  of  similar  BPS  twisted-sector operators 
in the orbifold $\mathsf{Q}_{L}$ theory  were  computed 
 in the   planar limit at LO  approximation in strong-coupling expansion 
and successfully matched to dual string  theory 
(which in this planar limit of correlators of BPS operators is represented by type IIB  supergravity) 
\cite{Billo:2022gmq,Billo:2022fnb}.

\subsection*{Comments}  

Let us   draw   some   conclusions  from   the above discussion.
\begin{enumerate}
\item 
The Bessel operator enters the $\N=2$  observables in two distinct ways:  the free energy and non-planar correction
 to  circular Wilson loop 
depend on its Fredholm determinant $\mathcal F^ {\text{loc}}_{\ell} = \log\det(1-K)$  (see Eqs.\re{df-sa-q2} and \re{dq-orb}). 
At the same time, the correlators
of chiral primary operators are expressed in terms of matrix elements of its resolvent $1/(1-K)$  (see Eqs.~\re{CPO-ratio} and \re{twist1}). 

\item The results for the strong-coupling expansion of    $\mathcal F^ {\text{loc}}_{\ell}$  available  so far in the literature 
 were  mostly  numerical  and inconclusive. This applies,  in particular, to  the  value of the 
 coefficient  of the leading $\lambda^{1/2}$ term in \rf{f-def}
({\em cf.} Eqs.~\re{smallf-lo} and \re{smallf-nlo}), not to mention   higher
subleading corrections  that  remained unknown.

\item The matrix elements of $1/(1-K)$ are more under  control. In particular, the leading $O(1/\lambda)$ term in the ratios of chiral correlation functions 
\re{CPO-ratio-lo}, \re{3pt-LO} and \re{2pt-tw}
 has been  computed 
numerically, showing a very good agreement with  analytic  expressions, and, in  some cases,  
was matched to dual supergravity results  \cite{Billo:2022gmq}. Still, 
subleading corrections at large $\lambda$ have not been computed yet,  as going beyond the LO
approximation appears to be  non-trivial.

\item  Apart from  ``perturbative''  $({1\ov \sqrt\lambda})^n$ strong-coupling corrections to the observables 
one expects also exponentially small non-perturbative  corrections  but   no  information about  them  is  available except in the case of much simpler  
$\fatilde$ model \cite{Beccaria:2021ism}.
\end{enumerate}

\subsection{Strong coupling expansion}\label{subsect}

Let us   now  apply the results obtained in  Sections 2 and 3   to address the open problems mentioned  above. 

\subsection*{Free energy and circular Wilson loop}

As was explained in the previous subsection, these observables can be expressed in terms of the Fredholm determinant of the Bessel operator \re{det-B} with the symbol \re{chi-loc}. 
From the results in Section \ref{sec:phys-app}
we have 
\begin{align}\label{ini}
\mathcal F^ {\text{loc}}_\ell= \pi g - \lr{\ell-\ft12}\log g +  B_\ell^ {\text{loc}}+ f^{\text{loc}}_{\ell}(g+i0) + \Delta f^{\text{loc}}_{\ell}(g) \,, \qquad \ \ \ 
g=\tfrac{\sqrt{\lambda}}{ 4\pi}  \ . \end{align}
Here the Widom-Dyson constant $ B_\ell^ {\text{loc}}$ is given by (see \re{Bs})
\begin{align}
B_\ell^ {\text{loc}} &= -6\log\mathsf{A}+\frac{1}{2}+\frac{1}{6}\log 2 -\ell\,\log 2+\log\Gamma(\ell)\,,
\end{align}
where $\mathsf{A}$ is Glaisher's constant.
Perturbative corrections in $1/g$ are described by the function $f^{\text{loc}}_{\ell}(g+i0)$, where `$+i0$' corresponds to a particular prescription
for integrating the Borel singularities in \re{Bor}. Its series expansion at large $g$ looks as (see \re{f-imp}, \re{In1}, \rf{3}  and \rf{4}) 
\begin{align} \la{111}\notag
 f^{\text{loc}}_{\ell}(g) =&\frac{1}{8} (2 \ell -3) (2 \ell -1) \log \left( {{g'}}/{g}\right)+(2 \ell -5) (2 \ell -3) (4 \ell^2 -1)\frac{\zeta (3)
   }{2048 \pi ^3 {g'}^3}
   \\
   & -(2 \ell -7) (2 \ell -5) (4 \ell -9) (4 \ell^2 -1)\frac{3
   \zeta (5) }{262144 \pi ^5 {g'}^5}+O({g'}^{-6}) 
    \ , 
\end{align}
where   $ g'= g- {\log 2\ov \pi} $. 
 Re-expanding this series in powers of $1/g$ one would generate terms proportional to powers of $\log 2\ov \pi$. 
Finally, the leading non-perturbative correction to $\Delta f^{\text{loc}}_{\ell}(g)$ is given by \re{rel1} and \re{rel2}.

Substituting \re{ini} into  the first relation in \re{df-sa-q2} we get the strong-coupling expansion of the free energy in the $\mathsf{SA}$ model~\footnote{Using the results of~\cite{Belitsky:2020qir} one can obtain the perturbative expansion up to order $O(1/g^{30})$. } 
\begin{align}\notag\label{F-sa}
\Delta F^{\mathsf{SA}}(\lambda) =
&{}\frac18 \lambda^{1/2}-\frac{3}{8}\log {\lambda}
-3 \log\mathsf{A} +\frac{1}{4}-\frac{11}{12}\log 2  + {3\ov 4} \log ( 4 \pi) 
\\\notag
& +\frac{3}{32} \log \left({\lambda'}/{\lambda}\right)-\frac{15 \zeta (3)}{64\,
    {\lambda'}^{3/2}}-\frac{945 \zeta (5)}{512\, {\lambda'}^{5/2}}-\frac{765 \zeta (3)^2}{128\,
   {\lambda'}^{3}}+O({\lambda'}^{-7/2})
\\[2.2mm]
& - {i\over 4} \lambda^{1/2} \e^{-\sqrt\lambda}\big(1+O(\lambda^{-1/2})\big)\,, \qquad \ \ \ \ \ \ \   \lambda'{}^{1/2} = \lambda^{1/2} -4\log 2 \ . 
\end{align} 
Let us make a few comments on this result.

The perturbative series on the second line of  \rf{F-sa}   has an interesting ``homogenous weight'' property. Namely, the coefficient in front of $1/\lambda'{}^n$ is proportional to the product of odd Riemann zeta values $\zeta(2n_i+1)$ with $\sum_i n_i=n$.

Note that both the leading $O(\lambda^{1/2})$ term  and non-perturbative correction on the last line of  \rf{F-sa} may also be expressed  in terms of $\l'$.\foot{  
The origin of this peculiar dependence on both $\l$ and $\l'$   may become  clear if one manages to find  an  exact 
finite $\l$ analog of the expansion  in \rf{F-sa},  by analogy with what happens in the
much simpler  BES  case (cf. \rf{bes-ex}  and \rf{B-guess}).}

Another comment is about the imaginary coefficient of the non-perturbative correction in \re{F-sa}. 
 It has the same origin as the coefficient in front of the first term in \re{rr}. The perturbative part
 of the strong-coupling expansion in  \re{F-sa}  is not Borel summable and the Borel singularities should be avoided by slightly moving $g$ in the complex plane. As a consequence, an imaginary part is produced 
 canceling  a similar contribution from the non-perturbative correction in \re{F-sa}.
 The  latter is  dressed by a  similar perturbative tail of  powers of $\lambda^{-1/2}$.

From \re{F-sa} we find, in particular, the exact value of the coefficient $\f^{\sa}$ defined in \re{f-def}
\begin{equation}
\f^{\sa}_{\rm exact} = \frac{1}{8}\,. \la{888}
\end{equation}
This  demonstrates that \rf{smallf-lo} is a rough estimate   and  that  numerical  results \re{smallf-nlo}
gives a relatively good approximation.
 
Similarly, for the orbifold model $\mathsf{Q}_{2}$, we obtain from the second relation in \re{df-sa-q2} 
\begin{align}\notag\label{F-q2}
\Delta F^{\mathsf Q_{2}}(\lambda) =&\frac14 \lambda^{1/2}-{1\ov 2} \log \l 
-6 \log\mathsf{A}+\frac{1}{2}-\frac{4}{3} \log 2  + \log (4 \pi) 
\\\notag
& +\frac{1}{16} \log \left({\lambda'}/{\lambda}\right)
-\frac{3 \zeta (3)}{32\,{\lambda'}^{3/2}} -\frac{135 \zeta (5)}{256 \,
   {\lambda'}^{5/2}}-\frac{99\zeta (3)^2}{64 \,{\lambda'}^3}+O({\lambda'}^{-7/2})
\\[2.5mm] 
& + i \c_1 \e^{-\sqrt\lambda} \big(1+O(\l^{-1/2} ) \big)  \,.
\end{align}
Notice that the leading non-perturbative correction in \rf{F-q2}  is suppressed by a factor of $\lambda^{1/2}$ as compared to the one in \re{F-sa}, i.e.  scales as $O(\e^{-\sqrt\lambda})$.
 This is a consequence of the fact  that while    both terms in  \re{df-sa-q2} receive the leading $O(\sqrt\lambda\e^{-\sqrt\lambda})$ nonperturbative corrections, 
   they cancel against each other in the sum.  
  The  normalization coefficient $\c_1$ of the subleading term remains to be determined: to find its value, one has to compute the subleading $O(1/g)$ correction to the functions \re{hard}. 

 The leading term of the expansion \re{F-q2} has the expected form \re{f-def}  with  the exact value of the coefficient  being 
\begin{align}
\f^{\mathsf Q_{2}}_{\rm exact} ={1 \ov 4} \ . \la{444}
\end{align}
This  may  be compared to  the    previous  numerical estimate \re{f-orb}. The disagreement is not surprising since the numerical analysis 
in  \cite{Beccaria:2021ksw} could only reach $\lambda \simeq 450$ where numerical fitting is not yet able to disentangle 
$\l^{1/2}$ from the slowly growing  $\log \lambda$ corrections. 

Interestingly, the  $\mathsf Q_{2}$ coefficient \rf{444}  is twice that of the  $\sa$ one  in \rf{888}.  
This may   be   related to the fact that
 the $\mathsf{SA}$ model   is  an   orientifold projection  of   the $\mathsf Q_{2}$  model 
  (see Appendix C of \cite{Billo:2021rdb} for details).
In particular, 
 the free-theory content of the $SU(N) \times SU(N)$ 
 $\mathsf Q_{2}$ quiver theory is twice that   of the $\sa$  theory (one free  bi-fundamental $SU(N)$ hypermultiplet  is the same as 
 the sum of   rank-2  symmetric  plus antisymmetric  hypers).\foot{In particular,  the conformal anomaly  coefficients of the two theories 
thus  also  differ  by   factor of 2:  a$^{\mathsf Q_{2}}= \ha N^2 - {5\ov 12},$ \  c$^{\mathsf Q_{2}}= \ha N^2 - {1\ov 3}$ compared to 
a$^{\sa}= {1\ov 4}  N^2 - {5\ov 24},$ \  c$^{\sa}= {1\ov 4}  N^2 - {1\ov 6}$.}
This    factor of 2  proportionality is, of course,  no longer true for the  subleading coefficients in \rf{F-sa} and \rf{F-q2}.

Substituting \re{F-sa} and \re{F-q2} into \rf{455} and \re{dq-orb} we obtain the strong-coupling expansion of the 
 leading non-planar correction \rf{414} to the circular Wilson loop 
\begin{align}\notag\label{dq-sa}
\Delta q^{\mathsf{SA}}(\lambda)= &
-\frac{\lambda ^{3/2}}{64}+\frac{3 \lambda }{32}-\frac{3 \lambda^{1/2}  \log 2}{32
   \big(1-{4 \log 2\over \sqrt{\lambda} }\big)}-\frac{45 \lambda ^{-1/2} \zeta (3)}{512
   \big(1-{4 \log 2\over \sqrt{\lambda} }\big)^4}+O(\lambda^{-3/2})
%\\   
% &  -\frac{4725 \lambda ^{-3/2} \zeta (5)}{4096
%  \big(1-{4 \log 2\over \sqrt{\lambda} }\big)^6}
%   -\frac{2295 \lambda ^{-2} \zeta (3)^2}{512
%   \big(1-{4 \log 2\over \sqrt{\lambda} }\big)^7} +O(\lambda^{-5/2})
  \\  
  & - {i\over 32} \lambda^{2} \e^{-\sqrt{\lambda}}\, (1+ O(\lambda^{-1/2}))\,,
\\[2.5mm]\notag
\Delta q^{\mathsf{Q}_{2}}(\lambda)  = & -\frac{\lambda
   ^{3/2}}{64}+\frac{\lambda }{16}- \frac{\lambda^{1/2}  \log 2}{32
   \big(1-{4 \log 2\over \sqrt{\lambda} }\big)}
 %  -\frac{\lambda  \log (2)}{32 \left(\sqrt{\lambda }
  % -4\log (2)\right)}
  -\frac{9 \lambda ^{-1/2} \zeta (3)}{512
   \big(1-{4 \log 2\over \sqrt{\lambda} }\big)^4}+O(\lambda^{-3/2})     
%\\   
%   -\frac{675 \lambda ^{3/2} \zeta (5)}{4096 \left(\sqrt{\lambda }-4 \log
%   (2)\right)^6}-\frac{297 \lambda ^{3/2} \zeta (3)^2}{512 \left(\sqrt{\lambda }-4 \log
%   (2)\right)^7} 
\\
\label{dq-q}
&  + \frac{i\,\c_1}{16}\,\lambda^{3/2}\,e^{-\sqrt{\lambda}}\, (1+O(\lambda^{-1/2}))\,.       
\end{align}
Here the first and the second line in each relation defines, respectively, the perturbative and non-perturbative corrections.
 The equality of the coefficients of the  leading $\l^{3/2}$   terms in \re{dq-sa} and \re{dq-q}  %is not an accident.
follows from  (i) the factor of 2 proportionality   between \rf{888} and \rf{444}  mentioned above  and  (ii) the fact 
 that the half-BPS 
Wilson loop in the  $\mathsf{Q}_{2}$ model 
is defined in terms of the fields associated with just one of the two $SU(N)$ factors of the gauge group resulting 
in   extra factor of 1/2   in \rf{dq-orb} as  compared to \rf{455}. 

\def \salpha {s_\alpha}

\subsection*{Two-point chiral correlators in $\sa$ and $\mathsf{Q}_{L}$ models}

We can  also  apply \re{ini} to derive the strong 
coupling expansion of the two-point correlation function of (anti) chiral operators in \re{ratio}.

To this aim, in the $\sa$ model, we write  (\ref{CPO-ratio}) in the form 
\begin{align}\label{CPO-ratio-im}
R^{\sa}_{2n+1}(\lambda) = {\det(\delta_{ij}- K_{ij})\Big|_{n+1\le i,j< \infty}\over \det(\delta_{ij}- K_{ij})\Big|_{n\le i,j< \infty}\phantom{aa}} = 
 \exp\big( {\mathcal F^{\rm loc}_{2n+2}-\mathcal F^{\rm loc}_{2n}}\big) \ ,
\end{align}
where we first   applied the Cramer's rule to (\ref{CPO-ratio})  and then  replaced the determinants by their expressions in terms of the functions ${\mathcal F_{\ell}}$ in   \re{det-K}. 

At weak coupling, we can use \re{F-weak} and \re{q-weak} to expand $R^{\sa}_{2n+1}(\lambda)$ in powers of 't~Hooft coupling. We checked that the resulting expressions are in agreement with the results of \cite{Beccaria:2020hgy} (see Eq.~(6.15)  there). At strong coupling, we find
from \re{CPO-ratio-im} and \re{ini} 
\begin{align}\notag\label{CPO-ratio-expanded}
R^{\sa}_{2n+1} =& {8 \pi ^2\over\lambda} n (2 n+1) 
\\\notag
& \times
(\lambda'/\lambda) ^{2 n} \Big[ 1+2   n \left(16 n^2-1\right) {\zeta
   (3)\over {\lambda'}^{3/2}}-  n \left(16 n^2-1\right) \left(16 n^2-9\right) {9\zeta (5)\over 8{\lambda'}^{5/2}}
+O(\lambda'{}^{-3})\Big]
\\
& \times \left[1-8in \e^{-\sqrt{\lambda}}\big(1+O(\lambda^{-1/2})\big) \right],
\end{align}
where $ (\lambda')^{1/2} = \lambda^{1/2} -4\log 2$. Here the expression on the second line yields perturbative corrections
in $\lambda^{-1/2}$. As above, using $\lambda'{}^{-1/2}$ as an expansion parameter is advantageous 
as  it automatically sums up all terms 
proportional to $\log 2$. The last line in \re{CPO-ratio-expanded} represents  the leading non-perturbative correction. It comes from the difference $\Delta f^{\rm loc}_{2n+2}-\Delta f^{\rm loc}_{2n}$ in the exponent of \re{CPO-ratio-im} and  is given by the second relation in \re{rel2} for $\ell=2n$.

Notice  that  the strong-coupling expansion  \re{CPO-ratio-expanded} involves half-integer powers of $\lambda^{-1}$, 
in agreement with the expectation  from the dual string theory side  (where $\lambda^{-1/2}$ is the inverse string tension).
 Such terms are not present in the large $\lambda$ asymptotic expansion of
individual matrix elements $K_{ij}$ in \re{CPO-ratio-im}. 
This illustrates (once again) non-trivial properties of the Fredholm determinant of the Bessel operator as well as  the power and efficiency of the techniques for computing it described in the first part of this
paper.

The expression on the first line of \re{CPO-ratio-expanded} defines the leading behaviour of  $R^{\sa}_{2n+1} $ for $\lambda\to \infty$. 
It agrees with (\ref{CPO-ratio-lo}), 
i.e. confirms the validity of the LO approximation used in \ci{Beccaria:2021hvt}.
 The subleading corrections in (\ref{CPO-ratio-expanded}) may be computed to  any desired order. 
 
We  remark that the relation \re{CPO-ratio-expanded} has an interesting behaviour in the double scaling limit $\lambda\to\infty$ and
$n\to\infty$ with the ratio $\bar n = 4\pi n\, {\lambda}^{-1/2}$ held fixed. It is equivalent to the so-called ``large bridge'' limit
previously studied in  \cite{Bargheer:2019exp,Belitsky:2020qir} in application to the octagon. 
From  \re{CPO-ratio-expanded} we  get in this limit 
\begin{align}\notag
\log R^{\sa}_{2n+1} =& 2\log \bar n  -{4 \log2\over \pi}\,\bar n+\frac{\zeta (3)}{2 \pi ^3}\bar n^3 -\frac{9 \zeta (5)}{32 \pi ^5}\bar n^5
+\frac{45\zeta (7)}{256 \pi ^7}\bar n^7 -\frac{2975 
   \zeta (9)}{24576 \pi ^9}\bar n^9
\\
&  +\frac{5859   \zeta (11)}{65536 \pi ^{11}}\bar n^{11} -\frac{72765   \zeta (13)}{1048576 \pi
   ^{13}}\bar n^{13}+\frac{2342769  \zeta (15)}{41943040 \pi ^{15}}\bar n^{15} + \dots \ .
\end{align}
This series admits a compact integral representation valid for arbitrary $\bar n$ \footnote{Its derivation can be found in 
Section 7.3 of \cite{Belitsky:2020qir} (\textit{cf.} Eq.~(7.14) there). 
}
\begin{align}
\log R^{\sa}_{2n+1} & =-{2\bar n\over \pi}
\int_{\bar n}^\infty dz
\frac{ \log \left(\coth ^2\left( {z}/{2}\right)\right)}{z \sqrt{z^2-\bar n^2}}\,.
\end{align}

For the correlators of even chiral operators in the $\sa$ model, we apply \re{even-cor} 
to find the ratio $R^{\sa}_{2n}$ in terms of the  non-planar correction  in the 
free energy  $\Delta F^{\sa}(\lambda)$ defined in \re{df-sa-q2}.

The same method can be used to compute  the two-point correlation functions \re{TT} of twisted-sector 
 operators in the $\mathsf{Q}_{L}$ orbifold  model. To start with, we notice that semi-infinite matrices in \re{twist1} coincide with the matrix \re{K-def} evaluated for special values of $\ell$ and the symbol replaced with $\chi(x)=s_\alpha \chi_{\rm loc}(x)$, schematically, 
\begin{align}
\mathsf X_{[k]}^{\rm odd} = K(s_\alpha \chi_{\rm loc})\Big|_{\ell=2k}\,, \qqqquad \mathsf X_{[k]}^{\rm even} = K(s_\alpha \chi_{\rm loc})\Big|_{\ell=2k-1}\,, \la{4420}
\end{align}
where $\chi_{\rm oct}(x)$ and $s_\alpha$ are defined in \re{chi-loc} and \re{s}, respectively.
In  close analogy with \re{CPO-ratio} and \re{CPO-ratio-im}, each  quantity in \re{twist1} can be expressed as  a ratio  of the  determinants
\re{det-K} and \re{det-B}
\begin{align}\label{rat-det-T}
 1+\Delta_{\alpha,k} (\lambda) =  {\det (1-\mathbf B_{k+1}(s_\alpha \chi_{\rm oct}))\over \det (1-\mathbf B_{k-1}(s_\alpha \chi_{\rm oct})) }= \exp \big({\mathcal F^\loc_{k+1}(g,\alpha) - \mathcal F^\loc_{k-1} (g,\alpha)}\big)\,,
\end{align}
where $\mathcal F^\loc _\ell(g,\alpha)$ is a logarithm of the Fredholm determinant of the Bessel operator with the symbol $\chi=s_\alpha \chi_{\rm loc}(x)$. The function $\mathcal F^\loc_\ell(g,\alpha)$ vanishes for $s_\alpha=0$ and 
 coincides with \re{ini} for $s_\alpha=1$. 

At weak coupling, it follows from \re{F-weak} and \re{rat-det-T} that  $\Delta_{\alpha,k} (\lambda)=O(\lambda^{k})$. 
Replacing   the constants $q_k$ in \re{F-weak} with their expressions $q_k=s_\alpha q_k^{\rm loc}=-4s_\alpha (2n)! \zeta(2n-1)$
(see \re{q-weak}) we immediately obtain the weak-coupling expansion of $\Delta_{\alpha, k}$, e.g., 
\begin{align}\notag
 \Delta_{\alpha,2}&  
  = -\frac{3s_\alpha \zeta (3) \lambda ^2}{32 \pi ^4}+\frac{5s_\alpha \zeta (5) \lambda ^3}{64 \pi
   ^6}+O\left(\lambda
   ^4\right),
\\  
  \Delta_{\alpha,3}&  
   =  -\frac{5s_\alpha \zeta (5) \lambda ^3}{256 \pi ^6}+\frac{105s_\alpha \zeta (7) \lambda ^4}{4096
   \pi ^8}+O\left(\lambda ^5\right).
\end{align}
These relations are 
in agreement with \cite{Billo:2021rdb}, see Eq.(5.23) there. 

At strong coupling, we apply \re{semi}--\re{f-PT} and \re{As} to obtain the following  
 expansion of the function $\mathcal F^\loc _\ell(g,\alpha)$
\begin{align}\label{F-alpha}\notag
\mathcal F^\loc_\ell(g,\alpha) =& -2  g I_0(\salpha) - (\ell-\frac{1}{2})\log g -\frac{\ell}{2}  \log
   \left(4s_{\alpha }\right) + \log \Gamma(\ell)+B(\alpha)
\\  \notag 
& +\frac{1}{8} (2 \ell-3) (2 \ell-1)\log(g'/g) - {(2
   \ell-5) (2 \ell-3) (4 \ell^2-1) } { I_2(\salpha) \over 3072
   {g'}^3} 
   \\
&   - {(2 \ell-7) (2 \ell-5) (4 \ell^2-9) (4 \ell^2-1)
    }{I_3(\salpha)\over 163840
   {g'}^5}+O(1/{g}^6)\,,
\end{align}
where $g'\equiv g-\ha I_1(\salpha)$. Here we replaced the expansion coefficients with \re{As} and denoted by $I_n(\salpha)$ the integrals \re{In}  
evaluated for $\chi=s_\alpha \chi_{\rm loc}(x)$. 
%\footnote{
%Notice that in this  notation the  dependence on $L$ is implicit. In particular,  
%v3-4
Note that 
$I_{n}(\salpha)$ is actually a function of $\alpha/L$ since this is the ratio that appears in $s_{\alpha}$ in \rf{s}. Also, 
in the $\mathsf{Q}_{2}$ model the only twisted sector has $\alpha=1$ and thus $s_{\alpha}=1$
in \rf{s}.
Hence, in this case, the integrals $I_{n}$ are given by  (\ref{In1}).
 For generic values of $\alpha/L$, the integrals $I_{n}(\salpha)$
can be expressed in terms of derivatives of the digamma function 
\be \la{iis}
I_{n}(\salpha) = \frac{1}{\Gamma(2n-1)}\frac{(-1)^{n}}{(2\pi)^{2n-1}}\Big[\psi^{(2(n-1))}\Big(\frac{\alpha}{L}\Big)+\psi^{(2(n-1))}\Big(1-\frac{\alpha}{L}\Big)-2\psi^{(2(n-1))}(1)\Big]
\ . \ee
The coefficient \re{A1} in front of $\log g$ in \re{F-alpha} does not depend on $\alpha$ because the 
 additional factor of $s_\alpha$  does not modify the strength $\beta_{\rm loc}=-1$ of the Fisher-Hartwig singularity \re{small-x} of the symbol
\begin{align}
1-s_\alpha \chi_{\rm loc}(x) \sim {4 s_\alpha \over x^2}\,.
\end{align}
It does affect, however, the normalization factor $b=4 s_\alpha$ in \re{small-x}. The last three terms on the first line of \re{F-alpha} come from 
the Widom-Dyson constant \re{B}. The first two terms on the right-hand side of \re{B} do not depend on $\ell$ and are denoted by $B(\alpha)$.
For the sake of simplicity, we did not display non-perturbative corrections to \re{F-alpha}. They are given by the general expressions \re{enh} and \re{fde1} with $x=\pm 2i \pi  x_1$ defined as the smallest root of $1-s_\alpha \chi_{\rm loc}(x)$. 
 For $s_\alpha=1$, the relation \re{F-alpha} coincides with \re{ini}. 

Substituting \re{F-alpha} into \re{rat-det-T} we obtain the strong-coupling expansion  
\begin{align}\notag\label{d-alpha}
 1+\Delta_{\alpha,k} (\lambda) =& {}  \frac{4 \pi ^2 (k-1) k}{\lambda s_{\alpha }} 
\\ & \notag 
\times 
(\lambda' /\lambda)^{k-1} \Big[ 1-  (k-1) (2 k-3) (2 k-1){2\pi ^3 I_2(\salpha) \over 3\lambda'{}^{3/2} } 
\\
& \qqqquad\quad \ \ \ 
 -  (k-1) (2 k-5) (2 k-3) (4 k^2-1){3\pi ^5 I_3(\salpha)\over 10\lambda'{}^{5/2}} +\dots \Big]\,,
\end{align}
where $ (\lambda')^{1/2} = \lambda^{1/2} -2\pi I_1(\salpha)$
and $I_n(\salpha)$   are given by \rf{iis}.
Here the first line defines the leading behaviour at strong coupling  and is in  agreement  with Eq.~(5.32) in \cite{Billo:2021rdb}. 
Expanding \re{d-alpha} in powers of $1/g=4\pi/\lambda^{1/2}$ would produce terms proportional to $I_1(\salpha)$, e.g.,
\begin{align}\notag
1+ \Delta_{\alpha,2}&={1\over s_\alpha} \Big[ \frac{1}{2 g^2}-\frac{I_1(\salpha)}{2 g^3}+\frac{I_1(\salpha)^2}{8
   g^4}-\frac{I_2(\salpha)}{64 g^5}+\dots\Big], 
\\    
1+ \Delta_{\alpha,3}&=  {1\over s_\alpha} \Big[  \frac{3}{2
   g^2}-\frac{3 I_1(\salpha)}{g^3}+\frac{9 I_1(\salpha)^2}{4 g^4}-\frac{\frac{3}{4}
   I_1(\salpha)^3+\frac{15
  }{32} I_2(\salpha)}{g^5}+\dots\Big]\,.
\end{align}
%v3-4
As was mentioned   above, in the $\mathsf{Q}_{2}$  theory where $s_\alpha=1$ 
 the values of $I_n$   are given by (\ref{In1}) and \rf{3}.
 We observe  that in this case $ 1+\Delta_{\alpha,k}$  in \rf{d-alpha} with $k=2n+1$   becomes the same as 
 the 2-point function \rf{CPO-ratio-expanded}  in the $\mathsf{SA}$ theory. This equality holds   to all orders  %in general 
 as follows from the second relation in \rf{twist1}, \rf{CPO-ratio}  and the first relation in \rf{4420}.
 %v4
 In general, this  is also a consequence of the fact   that the $\mathsf{SA}$ model is obtained 
    from   the $\mathsf{Q}_{2}$  one  by an extra projection that 
    effectively implies relation between these correlators (see discussion below \rf{ratio}).
    %does  not affect   this correlator.

\section{Conclusions}\la{s5}

In this paper, we have  elucidated the role of the (truncated) Bessel operator \re{K-def}
 in the description of  certain observables in  four-dimensional superconformal $\mathcal N=4$ and $\mathcal N=2$ theories.
The examples considered in this paper were 
 the special  four-point correlation function of half-BPS operators with infinite  $R$-charge in planar $\mathcal N=4$ SYM theory, 
as well as the  free energy, half-BPS circular Wilson loop, and various  two-point correlators of chiral primaries in $\mathcal N=2$ superconformal models % \GK
{that are planar equivalent to $\mathcal N=4$ SYM}.  

We demonstrated that these  quite  different  observables 
can be expressed in terms of logarithm of  Fredholm determinant   $\mathcal F_{\ell}(g)$, Eqs~\rf{det-K} and \re{det-B}, of the  Bessel operator  \re{def-B} and \re{ker-B}
after  choosing particular values of the non-negative integer $\ell$  and the ``symbol'' function  $\chi(x)$
 (see, e.g., \rf{12} and \rf{df-sa-q2}).\foot{
 Note that  while in the octagon   case the  
 relation  to Fredholm determinant follows  from a hexagon  representation   based 
    on  the integrability of  planar $\mathcal N=4$    SYM theory, in the case of the free energy 
 in the $\mathcal N=2$ models it arises  from  the  localization matrix model 
 expression for the leading non-planar correction and, thus, is not directly related to integrability in planar limit.}

The Fredholm determinant representation of the observables  is exact  in the 
't~Hooft coupling $\lambda$. While their small $\l$ expansion  is   straightforward, it is quite non-trivial 
to develop a  strong-coupling expansion. The latter  is of prime interest as it  
should be  equivalent  to the  inverse string tension expansion
according to the AdS/CFT  duality.
The advantage of the Fredholm determinant representation is that this difficult problem can be solved by applying the strong Szeg\H{o} limit theorem.
It requires  an application of special  techniques  partially available 
in  mathematical  literature. 

We found that, for the physically relevant cases, the perturbative expansion of  $\mathcal F_{\ell}(g)$  
in powers of $1/g ={4\pi/ \sqrt\lambda}$ reveals a number of remarkable properties that were  previously observed in the case of the octagon  correlator  in planar  $\N=4$  SYM  in \cite{Belitsky:2019fan,Belitsky:2020qrm,Belitsky:2020qir}. In particular, we demonstrated that $\mathcal F_{\ell}(g)$ receives a 
$\log g$ correction.  
It originates   from the  Fisher-Hartwig singularity of the symbol $\chi$ of the Bessel operator and is given by a simple expression $(\beta \ell+\ha \beta^2)\log g$, which only depends on
  $\ell$ and the strength of the singularity $\beta$ (cf. \rf{2}).  
The  structural simplicity of this term calls for its interpretation  on the dual string theory side. 

Examining the  expansion of $\mathcal F_\ell(g)$ in powers of $1/g$, we found that the resulting  series 
can be significantly simplified by changing the expansion parameter to $g'=g-\ha I_1$ where  a  transcendental  constant $I_1$
 is given by \re{In1} and \re{3}. This effectively performs a resummation of all terms containing  powers of $I_1$. 
A similar  phenomenon was  previously noticed  in  the strong-coupling expansion of the cusp anomalous dimension in planar $\mathcal N=4$ SYM
\cite{Basso:2007wd}  and,  thus,  may  be a universal feature  of strong-coupling expansion  of observables  in superconformal theories. 

The resulting perturbative expansion of  the determinant in \rf{det-K}  
takes the following factorized form (see Eqs.~\re{semi}, \rf{In}, \re{A1} and \re{f-imp})
\begin{align}\label{fact}
\exp \big({\mathcal F_\ell (g)} \big)
= \left[g^{-{1\ov 8}(4\ell^2-1)}\, \e^{-2g I_0+B}\right] \times    {g'}^{{{1\ov 8}(4\ell_\beta^2-1)}}\Big[1-\frac{(4 \ell_\beta^2-1)(4 \ell_\beta^2-9)}{3072 {g'}^3}I_2+O(g'{}^{-5})\Big],
\end{align}
where $\ell_\beta=\ell+\beta$. Here the two factors  depend separately on $g$ and $g'$. Notice that, due to our choice of the $g'{}^{-1}=(g-\ha I_1)^{-1}$ as the expansion parameter, the first few  powers of $g'{}^{-1}$  are absent inside the second brackets. 
The power  of $g$ in the first factor in  \re{fact} does not depend on the symbol of the Bessel operator, 
whereas that  of $g'$ in the second factor only depends on the strength of  the Fisher-Hartwig singularity $\beta$ of its symbol.
It would be interesting to understand the  meaning of the   $g$ vs $g'$  factorization  in \re{fact}  from the  dual string theory perspective.
  
Let us note that in the simplest $\ell=0$ octagon case we have at strong coupling $\mathbb O_0 = \exp \big[{\mathcal F^{\rm oct}_0 (g)} \big]  \sim   g^{1/8} g'{}^{3/8}\,\e^{-g A_0}\sim  g^{1/2}\,\e^{-g A_0}$, 
where  $  A_0= 2 I_0^{\rm oct} = \pi$  \cite{Bargheer:2019exp,Belitsky:2020qrm}.
 This   
implies  that the corresponding  planar correlator  $G_0$ of  four BPS operators  scales as $G_0=[\mathbb O_0]^2 \sim   g\,\e^{-2g A_0}\sim \sqrt \l
\,\e^{-{\sqrt{\l}\ov 2\pi } A_0 }$.   The constant  $A_0$  was  conjectured to   have  a  dual semiclassical   string theory   interpretation 
 as   minimal area of  a  world sheet surface  in AdS$_5 \times S^5$   \ci{Bargheer:2019exp}. 
 The prefactor 
 $\sqrt \l$  of the exponential  may be  given  the following heuristic string theory explanation. 
The dual  string   theory representation for $G_0$ is in terms of  a correlator  of 4  BPS vertex operators  on a plane or $S^2$. 
Each of them 
comes essentially from string action   and thus 
carries a normalization factor proportional to the string tension  $T = { \sqrt{\lambda}\ov 2\pi}$.
There is also  a  string tension dependence in the    M\"obius volume  for $S^2$   (including which  is like  dividing by a
3-point function);  that   gives a factor of $T^{-3}$. In total, we get  
$T^4 \times T^{-3} \sim \sqrt{\lambda}$,   in  agreement   with the  above strong-coupling scaling of $G_0$. 

In this paper  we extended the analysis of the Fredholm determinant of the Bessel operator to a wider class of symbols 
 that appear in   special  $\mathcal N=2$ superconformal models  that are planar equivalent to $\mathcal N=4$ SYM. We derived the strong-coupling expansion of the  relevant 
  observables   and  resolved some  issues  that existed in earlier work \cite{Beccaria:2021ksw,Beccaria:2021vuc,Beccaria:2021hvt}. 
 In particular, we obtained the strong-coupling expansion of the free energy \re{F-sa}
in the $\sa$ theory ($\mathcal N=2$ model with matter in rank-two symmetric  and  antisymmetric representations of  $SU(N)$)
 determining the 
exact value \rf{888}  of the coefficient of the leading $O(\sqrt\lambda)$ term.
We also computed systematically the strong coupling corrections to certain two-point functions of chiral primary operators, 
confirming the conjecture for the  coefficient of the leading term  \cite{Beccaria:2021hvt}. 

Apart from analytically  deriving the  coefficients  of 
the perturbative strong-coupling expansion,  one  main result of this paper is the identification of the non-perturbative (exponentially small at large $\l$) 
corrections \re{f-nonPT} to the Fredholm determinant  $\exp(\mathcal F_{\ell}(\lambda))$ 
and, thus, to  the  related observables in $\mathcal N=4$ and $\mathcal N=2$ theories. 
Such corrections are known to be
present, in particular, in the cusp anomalous dimension in planar $\mathcal N=4$ SYM where they drive the transition from strong to weak coupling \cite{Basso:2007wd,Basso:2009gh}. In that case,   the leading
non-perturbative correction  to the cusp anomalous dimension has 
a clear physical meaning  on the  dual string theory side \cite{Alday:2007mf}: it coincides with the  dynamical mass gap of the effective  two-dimensional bosonic O(6)
sigma model describing massless excitations of the Gubser-Klebanov-Polyakov  string.
We showed that for the Bessel operator with a general symbol $\chi(x)$, the leading non-perturbative  term  \re{f-nonPT} depends on the smallest root of 
$1-\chi(2i\pi x_1)=0$ and its multiplicity $n_1$.
 It would be interesting to understand the dual string theory origin  of this  correction
  in the $\mathcal N=2$ superconformal models and whether it also   admits some dynamical mass  scale  interpretation.\foot{Like in the cusp case, 
  non-perturbative corrections  may  be associated with certain semiclassical 
  string  configurations. Their stability under quantum fluctuations   may be related to  the Borel properties of the strong-coupling expansion    as suggested by the Wilson loop example  \cite{Drukker:2006ga}.}

In general,  the wealth of new results for the strong-coupling expansion of various observables obtained in this paper  calls for  a detailed  comparison  with  dual  string  theory.  The  required  computations on the string theory side  are, however,  appear 
to be  very non-trivial.
For example,  the  leading non-planar  corrections to free  energy in $\sa$ \rf{F-sa} or  $\mathsf Q_{2}$  \rf{F-q2} theories 
should  be reproduced by the torus  correction in type IIB  superstring theory on the corresponding 
 orientifold/orbifold of AdS$_5 \times S^5$.
It is unclear   if even the leading $\sqrt \l$ term \rf{F-sa} and \rf{F-q2}    can be reproduced    from  the one-loop 
   string effective action ${1\ov \alpha'} \int d^{10} x \sqrt g\,  RRRR+ \dots$ or 
 one needs to  compute the  torus   partition function exactly  before expanding in $\alpha'\sim {1\ov \sqrt \lambda} $
(for some related  discussion  see \ci{Beccaria:2021vuc,Beccaria:2021ism}).\foot{ 
 While there is no logical connection, it is still 
    curious to note that the $\zeta(n)$   coefficients of  the perturbative terms in \rf{F-sa} and \rf{F-q2} 
 are reminiscent of    those   in  the low-energy string effective action (that come out of the  expansion of the Shapiro-Virasoro amplitude). In particular,   a  similar  $\zeta(3)  \lambda^{-3/2}$  term originating from the 
 ${\alpha'}^{3} \zeta(3) \int d^{10} x \sqrt g\,  RRRR$  correction  in the tree-level  type IIB    string effective action
 appears  in the  strong-coupling expansion of the  finite-temperature free   energy of planar $\N=4$ SYM theory 
  \cite{Gubser:1998nz}. 
  }
 
\section*{Acknowledgments}

We are grateful to J. Russo  for  related discussions. GK would like to thank Alessandro Sfondrini and Dima Sorokin for interesting discussions and 
Theoretical Physics Group at the University of Padova for their hospitality. MB was supported by the INFN grant GSS (Gauge Theories, Strings and Supergravity).
The work of GK was supported by the French National Agency for Research grant ANR-17-CE31-0001-01.
AAT was supported by the STFC grant ST/T000791/1. 
 
\paragraph{Note added in v2:}
  The first few terms of the strong coupling expansion of the observables in the $\sa$ theory presented in Section~\ref{subsect} were reproduced in  a very recent paper \ci{Bobev:2022grf} 
 by  using a  high precision numerical calculation.

\appendix

\section{Matrix model representation}\label{app:mat}

The localization technique allows us to evaluate the partition function of $\mathcal N=2$ superconformal models on the 4-sphere as a matrix integral \cite{Pestun:2007rz}. For the models defined in Table~\ref{tab:1}, 
this integral (normalized to $\N=4$ SYM  expression) 
can be expressed  in the large $N$ limit  
as a Fredholm determinant of the Bessel operator with the symbol of the form \re{chi-loc}.
Below we review a derivation of this relation \cite{Beccaria:2021vuc} and, then, construct the matrix integral representation of the observable \re{det-K} and \re{det-B} for an arbitrary symbol $\chi(x)$.  

The partition function of $\mathcal N=2$ theory with the $SU(N)$ gauge group on  $S^4$ is given    by 
\begin{align}\label{Z-S4}
Z^{\mathcal N=2}  = \int da \e^{-{8\pi^2\over g_{\rm  YM}^2} \tr\, a^2 -S_{\text{int}}(a)}\,,
\end{align}
where integration goes over Hermitian traceless matrices $a$ of dimension $N$. 
Here we neglected the instanton contribution because we are interested in the large $N$ limit. 
The interaction potential is given by  (see also  \ci{Fiol:2015mrp})
\begin{align}\label{H}
S_{\text{int}}(a) = \tr_{\mathcal R} \log H(ia) - \tr_{\rm adj} \log H(ia) \equiv \tr_{\mathcal R'} \log H(ia)\,,
\end{align}
where the function $H(x)$ can be expressed in terms of the Barnes function 
\begin{align}\label{H1}
\log  H(x) =\log \Big(G(1+x)G(1-x)\Big) = -(1+\gamma_{\rm E})x^2 - \sum_{p=1}^\infty {\zeta(2p+1)\over p+1} x^{2p+2}\,.
\end{align}
The two terms in the first relation in \re{H} involve traces over the matter representation ${\mathcal R}$ and the adjoint representation of the $SU(N)$ group, respectively. Their difference is denoted as a trace over ${\mathcal R'}$.

In $\mathcal N=4$ theory,  ${\mathcal R}$ is the adjoint representation and the interaction term \re{H} vanishes. 
As a consequence, the partition function $Z^{\mathcal N=4}$ is given by a Gaussian integral whose evaluation gives the free energy \re{free-N=4}, $Z^{\mathcal N=4}=\exp(-F^{\mathcal N=4})$.
In  $\mathcal N=2$ superconformal models of type  $\mathsf{SA}$   (see Table~\ref{tab:1}),
 the trace \re{H} can be evaluated in the large $N$ limit   using the identity
\begin{align}
\label{double-trace}
\tr_{\mathcal R'}  a^{2k} =   \sum_{n=2}^{2k-2} \lr{2k\atop n} [1- (-1)^n] \,  \tr (a^n) \, \tr (a^{2k-n})  \,.
\end{align}
Note that the sum involves traces of odd powers of  matrix $a$. 
In particular, $\tr_{\mathcal R'}  a^{2}=0$ and, as a consequence, the $O(x^2)$ term on the right-hand side of \re{H1} does not contribute to the partition function \re{Z-S4}. Neglecting this term and introducing new variables $\omega_k(a)$ (with $k\ge 1$)  that satisfy 
\begin{align}
 \tr (a^{2n+1}) = g^{2n+1}  \sum_{k=1}^{n}  \omega_k(a)\sqrt{2k+1} \lr{2n+1\atop n-k} \, ,
 \end{align}
 the interaction term \re{H} can be rewritten in the planar limit as 
\begin{align} \label{S-X}
S_{\text{int}}(a) = - \frac12\sum_{n,m\ge 1} \omega_n (a) \, X_{nm} \, \omega_m(a)\,.
\end{align}
Here a semi-infinite matrix $X_{nm}$ is given by
\begin{align}\label{X}
X_{nm} = 2(-1)^{n+m}\sqrt{(2n+1)(2m+1)}\int_0^\infty {dx\over x}J_{2n+1}(x)J_{2m+1}(x)\ \chi_{\rm loc} (x/(2g))\,,
\end{align}
where $g^2= \lambda/(4\pi)^2 = g_{\rm YM}^2 N/(4\pi)^2$ and the function $\chi_{\rm loc}(x)$ is defined in \re{chi-loc}.
The equivalence of the two representations of the interaction term, Eqs.~\re{H} and \re{S-X}, relies on the relation satisfied by the function $\chi_{\rm loc}(x)$
\begin{align}\label{zeta}
 {\zeta(2p+1)\over p+1}=-{1\over 2(2p+2)! }\int_0^\infty {dt}\, t^{2p+1} \chi_{\rm loc}(t)\,.
\end{align}

In the absence of the interaction term  in \re{Z-S4}, i.e.  $S_{\rm int}(a)=0$, the $\omega_n$-variables  have diagonal two-point expectation values with respect to a Gaussian integration measure: $\vev{\omega_n(a)\omega_m(a)}=\delta_{nm}$ in the large $N$ limit. 
Adding the interaction term \re{S-X}, one obtains the following representation of the partition function \re{Z-S4} in the $\sa$ model
\begin{align}
Z^{\mathcal N=2}/Z^{\mathcal N=4}=\int \mathcal D\omega \e^{- \frac12 \omega_n  \lr{\delta_{nm}- X_{nm}} \omega_m} =  \Big[\det(1-X)\big|_{1\le n,m<\infty}\Big]^{-1/2}\,.
\end{align} 
Comparing \re{X} with the analogous matrix defined in \re{K-def} we observe that they coincide for $\ell=2$ and $\chi=\chi_{\rm loc}$
\begin{align}
X_{nm}= K_{nm}\Big|_{\ell=2,\, \chi=\chi_{\rm loc}}\,.
\end{align}
Combining together the above relations with \re{det-K} and \re{det-B}, we conclude that
\begin{align} 
Z^{\mathcal N=2}/Z^{\mathcal N=4}= {1\over \sqrt {\det(1 -\mathbf B_{\ell=2} (\chi_{\rm loc})   ) } } \,.
\end{align}
Thus, the matrix integral \re{Z-S4} can be expressed in terms of the Fredholm determinant of the Bessel kernel with the symbol \re{chi-loc}.

It is natural to ask whether a similar relation holds for an arbitrary symbol $\chi(x)$. Let us return to \re{H} and replace $\zeta(2p+1)/(p+1)$ in \re{H1} with its representation \re{zeta}. Then, substituting \re{zeta} into \re{H1} we obtain a general expression for the corresponding function $\log H_\chi(x)$
\begin{align}\notag
\log H_\chi(x) &= \sum_{p\ge 1}   {x^{2p+2}\over 2(2p+2)! }\int_0^\infty {dt}\, t^{2p+1} \chi(t)
\\
&= \int_0^\infty {dt\over 2t}\ \chi(t)\Big({\cosh (xt)-1-\frac12 x^2 t^2}\Big)\,,
\end{align}
where the sum of the three terms in the parenthesis scales  as $O(t^4)$ at small $t$. 
For $\chi$ given by \re{chi-loc}, $\log H_\chi(x)$ coincides with \re{H1} up to $O(x^2)$ term.
For the symbol \re{chi-oct} we find 
\begin{align}\label{H-oct}
\log H_{\rm oct}(x) = 4 \log H(x/2)-\log H(x)\,,
\end{align}
where $H$ is given by \re{H1}.
Repeating the above analysis we arrive at the following
identity (for $N\to \infty$) 
\begin{align}
 {\int da \e^{-{8\pi^2\over g_{\rm YM}^2} \tr\, a^2 -\tr_{\mathcal R'} \log H_\chi(ia)} 
 \over \int da \e^{-{8\pi^2\over g_{\rm YM}^2} \tr\, a^2} }
 = {1\over \sqrt {\det(1 -\mathbf B_{\ell=2} (\chi)   ) } } \,,
\end{align}
where the trace in the exponent is evaluated over the same representation ${\mathcal R'}$ as in the $\sa$ model, see \re{double-trace}.
In the special case \re{H-oct},  it yields a matrix model representation of the octagon in planar $\mathcal N = 4$ SYM.

\section{Bessel kernel}\label{app:B}

In this Appendix, we establish the relation between the semi-infinite matrix \re{K-def} and the integral Bessel operator defined in \re{def-B} and \re{ker-B}.

We show below that the matrix $K_{nm}$ represents the Bessel operator $\mathbf B_\ell$ on a space spanned by basis functions
\begin{align}\label{B-K}
\mathbf B_\ell \,\psi_m(t_1) \equiv  \int_0^{2g} dt_2\, B_\ell(t_1,t_2) \psi_m(t_2) =K_{nm} \psi_n(t_1)\,.
\end{align}
Here the kernel $B_\ell(t_1,t_2)$ is defined in \re{ker-B} and the functions $\psi_m(t)$ (with $m\ge 1$) have the following form
\begin{align}\label{psi1} 
\psi_m(t) =  (-1)^m \sqrt{2m+\ell-1} \sqrt{t} \int_0^\infty dy\, \chi(y) J_\ell(t y) J_{2m+\ell-1}(2gy)\,,
\end{align}
where $\chi(x)$ is the symbol of the Bessel operator.

To prove \re{B-K} it proves convenient to introduce an auxiliary function
\begin{align}\notag\label{K-rep}
K_\ell(x,y) &= {2\over xy} \sum_{n\ge 1}(2n+\ell-1) J_{2n+\ell-1}(x) J_{2n+\ell-1}(y)
\\\notag
&={x J_{\ell+1}(x) J_\ell(y) - y J_{\ell+1}(y) J_\ell(x)\over x^2-y^2}
\\
&=  \int_0^1 dt\, t J_\ell(t x) J_\ell(t y) \,.
\end{align}
Here the second relation is known as the Bessel kernel, see e.g. \cite{Tracy:1993xj}.
Defining an integral operator $\mathbf K_\ell$ with the kernel $K_\ell(x,y)$, 
it is straightforward to verify using the first relation in \re{K-rep} that  
\begin{align} \label{K-op}
 \mathbf K_\ell \, \phi_m(x) \equiv
 \int_0^\infty  dy\,y\, K_\ell(x,y) \chi \lr{y\over 2g} \phi_m(y) = K_{nm} \phi_n(x)\,,
\end{align}
where the functions $\phi_n(x)$ are given by
\begin{align}
 \phi_n(x) = (-1)^n \sqrt{2n+\ell-1}{J_{2n+\ell-1}(x)\over x}\,.
\end{align}
We can  use the well-known orthogonality property of the Bessel functions to check that the functions $\phi_n(x)$ are orthogonal with respect to the scalar product
\begin{align}
\vev{\phi_n|\phi_m} = 2\int_0^\infty dx \, x \, \phi_n(x) \phi_m(x) = \delta_{nm}\,.
\end{align}
The integral operator $\mathbf K_\ell$ has previously appeared in the study of the octagon in planar $\mathcal N=4$ SYM~\cite{Belitsky:2020qrm,Belitsky:2020qir}.
 
 Going back to \re{B-K}, we replace the kernel $B_\ell(t_1,t_2)$ with its expression \re{ker-B} to get
\begin{align}\label{K-K}
\mathbf B_\ell \,\psi_m(t_1) & = \sqrt{t_1} \int_0^\infty dx\,x \, J_\ell(t_1 x) \chi(x) \widetilde \psi_m(x)\,,
\end{align}
where the notation was introduced for
\begin{align}\notag\label{las}
\widetilde \psi_m(x) &= \int_0^{2g} dt_2 \sqrt{t_2} \, J_\ell(t_2 x) \psi_m(t_2)
\\
& = (-1)^m \sqrt{2m+\ell-1} \int_0^\infty dy\, \chi(y) J_{2m+\ell-1}(2gy) (2g)^2 K_\ell(2g x, 2g y)\,.
\end{align}
Here in the first relation $\psi_m(t_2)$ was replaced with its expression \re{psi1} and in the second relation
the integral over $t_2$ was evaluated using the last relation in \re{K-rep}. Changing variable $y\to y/(2g)$ in \re{las} and taking into account \re{K-op},
we arrive at
\begin{align}
 \widetilde \psi_m(x) ={1\over x} \sum_{n\ge 1}   (-1)^n \sqrt{2n+\ell-1}J_{2n+\ell-1}(2gx) K_{nm} \,.
\end{align}
Substituting this expression into \re{K-K}, we observe that the integral over $x$ is proportional to $\psi_n(t_1)$ with the proportionality coefficient
being $K_{nm}$. In this way, we reproduce \re{B-K}.

\section{General solution for half-integer $\beta$}\label{app:int-eq}

In this Appendix, we present some details of the  derivation of the relations \re{Ge} and \re{Go} for the
function $\Gamma(x,y)$. For the sake of simplicity we concentrate on the case of even $\ell$, for odd $\ell$ analysis goes along the same lines.

For even $\ell$, we use the relation $\Gamma(-x,y) = \Gamma(x,y)$ to simplify \re{Four}  as
\begin{align}\label{Four1}
\Gamma (x,y) = \int_{-\infty}^\infty dk\, \cos(kx) \, \tilde \Gamma(k,y)\,,
\end{align}
where $\tilde \Gamma(k,y)$ is an even function of $k$.
Substituting this relation into \re{eq-simp} and taking into account the identity
\begin{align} \label{Bes-int}
& \int_0^\infty {dx\over x} J_{2n+1}(x)\cos(k x) = (-1)^n {U_{2n}(k)\over 2n+1} \sqrt{1-k^2}\,\theta(1-k^2)\,,
\end{align}
where $U_k(x)$ is Chebyshev polynomial of the second kind, we arrive at an infinite system of equations for the function $\tilde \Gamma(k,y)$
\begin{align} \label{G-U}
& \int_{-1}^1 dk\,  \tilde \Gamma(k,y) U_{2m}(k)\sqrt{1-k^2} = (-1)^m(2m+1) {J_{2m+1}(y)\over y} \,,&& (m\ge \ell/2)\,.
\end{align}
Taking advantage of the fact that $U_m(k)=(-1)^k U_m(-k)$ are orthogonal polynomials 
\begin{align}
\int_{-1}^1 dk\,\sqrt{1-k^2} \, U_m(k) \, U_n(k) = {\pi\over 2} \delta_{mn}\,,
\end{align}
we can write a general solution to \re{G-U} as
\begin{align}\label{tildeG-sum} 
&\tilde \Gamma(k,y)={2\over\pi}\sum_{m\ge \ell/2} (-1)^m (2m+1) U_{2m}(k) {J_{2m+1}(y)\over y} + \sum_{n=0}^{\ell/2-1} a_n(y) U_{2n}(k) \,,
\end{align}
where $a_n(y)$ are arbitrary coefficient functions.
Since  $\tilde \Gamma(k,y)$ is even in $k$, the sum involves the Chebyshev polynomials with even indices.  The relation \re{tildeG-sum} only holds for $k^2<1$.

To find the function $\tilde \Gamma(k,y)$ for $k^2>1$, we invert \re{Four} and replace $\Gamma(x,y)$ with its expression \re{Gamma}
\begin{align}\label{tildeG-res}
\tilde \Gamma(k,y) =  {1\over y}\int_{-\infty}^\infty {dx\over 2\pi}\, \e^{-ikx}\gamma(x,y)  \Big[1-\chi(x/(2g))\Big]\,.
\end{align}
Here the function $\gamma(x,y)$ is given by a double Neumann series over  the Bessel functions \re{Gamma}.
For each term in this series, $\e^{-ikx}\gamma(x,y)$ vanishes as $x\to \infty$ for $k^2>1$. For $k>1$ and $k<-1$ this allows us to deform the integration contour in \re{tildeG-res} to the lower and upper half-plane, respectively, and pick up the residues at the poles of the function $1-\chi(x/(2g))$ defined in \re{zeros} and \re{Phi}. The poles are located at $x=\pm 4\pi i g y_n$ and their contribution to \re{tildeG-res} takes the form
\begin{align}\label{k>}
\tilde \Gamma(k,y) = i\sum_{n\ge 1} c_n(y) \Big[ \theta(k-1) \e^{-4\pi g y_n (k-1) }+\theta(-k-1) \e^{4\pi g y_n (k+1) }\Big],
\end{align}
where $c_n(y)\e^{4\pi g y_n }$ (with $n\ge 1$) are given by the residue of \re{tildeG-res} at the poles.

To find the function $\Gamma(x,y)$ we split the integral in \re{Four1} into $k^2<1$ and $k^2>1$ and replace 
$\tilde \Gamma(k,y)$ in each of the regions by the corresponding expression, Eqs.~\re{tildeG-sum} and \re{k>}, respectively.  
Integrating \re{tildeG-sum} over $-1<k<1$, we encounter the integrals
\begin{align}
\int_{-1}^1 dk\, \e^{ikx} U_\ell (k) = {\e^{ix}p_\ell (ix)  - \e^{-ix}p_\ell (-ix) \over (ix)^{\ell+1}}  \,,
\end{align}
where $p_\ell(x)$ is a polynomial in $x$ of degree $\ell$ defined in \re{p-pol}. Then, the two terms on the right-hand side of 
\re{tildeG-sum} give rise to the first two lines of \re{Ge}. Finally, integrating \re{k>} over $k^2>1$ we obtain the last line of \re{Ge}. Repeating the above calculation for odd $\ell$, one can derive \re{Go}.

\section{Resummation}\label{app:res}

In this Appendix, we present some details of derivation of the relations \re{rr2} and \re{gam2}. 

For $\ell=0$ and $\beta=-1/2$, 
the function $\Gamma(x,y)$ is given by \re{ini1}. 
The condition \re{zero-G} leads to the relations \re{c-exp} and \re{rc} for the coefficient functions $c_n(y)$. 
Inverting the Cauchy matrix $1/(x_n-y_j)$ that appears on the left-hand side of \re{rc}, we obtain  
\begin{align}\label{c0}\notag
& c_j^{(0)}(2gy) = {\e^{-2i g y}\over 2\pi}
 {x_j-y_j\over x_j-iy/(2 \pi)} \prod_{n\neq j} {x_n-y_j\over y_n-y_j}  { y_n-iy/(2 \pi )  \over  x_n-iy/(2 \pi) } +(y\to -y)\,,
\\
& c_j^{(1)}(2gy) = A(y)   {(x_1-y_1)(x_j-y_j)\over (x_1-y_j)} \prod_{n\ge 2} {y_n-x_1\over x_n-x_1}\prod_{n\neq j}{x_n-y_j\over y_n-y_j}\,,
\end{align}
where 
\begin{align}\label{A}
A(y)=-\frac{ \e^{-2ig y}}{2 \pi  \left(x_1+iy/(2 \pi)\right)}-\frac{\e^{2ig y}}{2 \pi 
   \left(x_1-iy/(2 \pi)\right)}+\sum_{j\ge 1}{c_j^{(0)} (2gy) \over x_1+y_j}  \,.
\end{align}
Substituting \re{c0} and \re{c-exp} into \re{ini1} we encounter the sum
\begin{align}\notag\label{inf-sums}
& \sum_{j\ge 1} {c_j^{(0)}(2gy)\over y_j-ix/(2\pi)} = {\e^{-2i g y}\over 2\pi} S_0(x,y) +{\e^{2i g y}\over 2\pi} S_0(x,-y) \,,
\\
& \sum_{j\ge 1} {c_j^{(1)}(2gy)\over y_j-ix/(2\pi)} = A(y) S_1(x)\,,
\end{align}
where we introduced the notation  
\begin{align}\notag\label{sums}
& S_0(x,y)=\sum_{j\ge 1} {x_j-y_j\over (x_j-iy/(2\pi))(y_j-ix/(2\pi))} \prod_{n\neq j} {x_n-y_j\over y_n-y_j}  { y_n-iy/(2\pi)  \over  x_n-iy/(2\pi) } \,,
\\
& S_1(x) = \sum_{j\ge 1} {(x_1-y_1)(x_j-y_j)\over (x_1-y_j)(y_j-ix/(2\pi))} \prod_{n\ge 2} {y_n-x_1\over x_n-x_1}\prod_{n\neq j}{x_n-y_j\over y_n-y_j}\ . 
\end{align}
Let us show  that $S_0$ and $S_1$   and thus $A(y)$  can be expressed in terms of the functions $\Phi(x)$ and $F(x)$ defined in \re{Phi} and \re{F}, respectively.
It is convenient to introduce an auxiliary function
\begin{align}\notag
f(x,z) &=   {1\over (z-iy/(2\pi))(z-ix/(2\pi))} \prod_{n\geq 1} {x_n-z\over y_n-z}  { y_n-iy/(2\pi)  \over  x_n-iy/(2\pi) }
\\
&= {1\over (z-iy/(2\pi))(z-ix/(2\pi))}{\Phi(-2\pi i z)\over \Phi(y)}\,,
\end{align}
where to get  the second relation we applied \re{Phi}. It is a meromorphic function of $z$ that decreases at infinity as
$f(x,z) \sim z^{-5/2}$. This property follows from \re{Phi-inf} for $\beta=-1/2$. As a consequence, the sum of the residues of $f(x,z)$ at all poles on the complex $z$-plane should vanish. It is easy to see that, up to a sign,  $S_0(x,y)$ is given by the residue of $f(x,z)$ at $z=y_n$ (with $n\ge 1$). Therefore, 
\begin{align}\notag\label{s0}
S_0(x,y) &= -\sum_{n\ge 1} \res_{z=y_n} f(x,z)  = \res_{z={ix\over 2\pi}} f(x,z)+\res_{z={iy\over 2\pi}} f(x,z)
\\ &
 =-{2\pi i\over x-y} \left[{\Phi(x)\over \Phi(y)} -1\right].
\end{align}
Repeating the same analysis for the second sum in \re{sums} we obtain 
\begin{align}\label{s1}
S_1(x) = - {F(x)\over F(-2\pi x_1)} \,.
\end{align}
Finally, we find from \re{A} and \re{c0}
\begin{align}\notag\label{a}
A(y) 
&= -\frac{ \e^{-2ig y}}{2 \pi  \left(x_1+iy/(2 \pi)\right)} + {\e^{-2i g y}\over 2\pi}S(2i\pi x_1,y) + (y\to -y)
\\\notag
&= -\frac{ \e^{-2ig y}}{2 \pi  \left(x_1+iy/(2 \pi)\right)} {\Phi(2i\pi x_1)\over \Phi(y)} + (y\to -y)
\\
&=-{x_1 F(2\pi i x_1)\over \pi(x_1^2+y^2/(2\pi )^2)}
\left[{\e^{2ig y}\over F(-y)}+{\e^{-2ig y}\over F(y)}\right],
\end{align}
where in the last relation we used \re{F}.
Taking into account \re{s0}, \re{s1} and \re{a} we can express the infinite sums \re{inf-sums} in terms of the functions \re{Phi} and \re{F}. Using the resulting expressions, we can reproduce \re{rr2} and \re{gam2} for $\ell=0$. For $\beta=-1/2$ and $\ell=1$ the calculation goes along the same lines.

Let us generalize the consideration to half-integer $\beta=p-1/2$ (with $p\ge 0$) and $\ell=0$. According to \re{Ge}, the function $\Gamma(x,y)$ is given by \re{ini1}. It follows from \re{G-small} that this function has to satisfy $\Gamma(2gx,2gy) \sim x^{2p}$ for $x\to 0$. Expanding the expression on the right-hand side of \re{ini1} at small $x$, we find that this leads to the additional $p$ relations for the coefficient functions $c_j(y)$. For instance, for $p=1$ we get
\begin{align}
\sum_{j\ge 1} {c_j(y)\over y_j} = -4 g {\sin y\over y}\,.
\end{align}
These relations should be supplemented with an infinite system of equations \re{inf-sys} that follow from \re{zero-G}. At strong coupling, its solution looks similar to \re{c-exp}
\begin{align}\label{c-p}
c_j(y)= g^p \left[ c_j^{(0)} (y)+ \e^{-8\pi g x_1}c_j^{(1)}(y)\right] +\dots \,.
\end{align}
It is straightforward to solve the resulting  equations for $c_j^{(0)} (y)$ and $c_j^{(1)} (y)$ for any given $p$.
To save the space, we do not present their  explicit expressions.

For $\ell\ge 2$ the relations \re{Ge} and \re{Go} also involve $ \left \lfloor{\ell/2}\right \rfloor $ functions $a_n(y)$. They can be expressed in terms of the functions $c_j(y)$ by imposing the condition  \re{G-small}. For instance, for $\beta=-1/2$ and $\ell=2$, the relation \re{Ge} combined with $\Gamma(2gx,2gy) \sim x^{2}$ as $x\to 0$ leads to
\begin{align}
 a_0(y) =-\frac{\sin (y)}{  y}- {1\over 4g}\sum_{j\ge 1} {c_j (y)\over y_j}  \,.
\end{align}
Replacing the functions $a_n(y)$ in \re{Ge} and \re{Go} with their expressions and imposing the relation \re{zero-G}, we 
obtain an infinite system of equations for the functions $c_j(y)$. As in the previous case, its solution takes the form
\re{c-p}. 

\section{Borel singularities}\label{app:bor}

The strong-coupling expansion \re{semi} involves perturbative Borel non-summable series \re{f-PT} and  non-perturbative exponentially small term, Eq.~\re{f-nonPT}. To define unambiguously
the corresponding functions $f(g)$ and $\Delta f(g)$, we have to specify a  prescription for integrating the Borel singularities of \re{f-PT}. 

To illustrate the procedure,  here we discuss  the strong-coupling expansion \re{rr}. In  close analogy with \re{r-def}, we define the ratio of the modified Bessel functions 
\begin{align}\label{r(x)}
r(x)={I_0(x)/ I_1(x)}\,.
\end{align}
The expression on the left-hand side of \re{rr} involves $r_n=r(4\pi gx_n)$. 

Recall that for integer index $n$,  $I_n(x)$ are entire functions of $x$ with 
\begin{align}
I_n(-x) = (-1)^n I_n(x)\ , 
\end{align}
that  lead to $r(-x)=-r(x)$. At large $x$, they have the following asymptotic expansion
\begin{align}\label{In-im}
I_n(x) = {\e^{x}\over \sqrt{2\pi x}} \sum_{k\ge 0} {a_k\over x^k} + i {\e^{-x}\over \sqrt{2\pi x}} \sum_{k\ge 0} (-1)^{k+n} {a_k\over x^k} \,,
\end{align}
where $a_k={\lr{\ft12-n}_k \lr{\ft12+n}_k/ (2^kk!)}$. Note that for $x>0$ the second, exponentially small term on the right-hand side of \re{In-im} is pure imaginary. It is this term that gives rise to non-perturbative correction in \re{rr}.
Its appearance is closely related to the fact that the series which accompanies the first term in \re{In-im} is not Borel summable. The problem  here is similar to that of  separation of  the perturbative $f(g)$ and non-perturbative $\Delta f(g)$ terms in \re{semi} mentioned above.

To define the two terms on the right-hand side of \re{In-im}, we apply the identities between the modified Bessel functions
of the first and second kind
\begin{align}\notag
I_n(x) =& {i\over\pi} \left[K_n(-x+i0)-(-1)^n K_n(x-i0) \right]\\
= &{i\over\pi} \left[-K_n(-x-i0)+(-1)^n K_n(x+i0) \right] \label{I=K}
\,.
\end{align}
In contrast to $I_n(z)$,  the function $K_n(z)$ has a cut on the complex $z$-plane running along the negative real axis. The argument of the Bessel functions on the right-hand sides of \re{I=K} is shifted slightly away from the real axis to avoid the cut.  

The function $K_n(z)$ admits an integral representation
\begin{align}\label{Kn-int}\notag
K_n(z) &=\lr{\pi \over 2 z}^{1/2} \e^{-z} \int_0^\infty d\sigma\, \e^{-\sigma} {}_2F_1\left(\ft{1}{2}-n ,\ft{1}{2}+n;1;-\frac{\sigma }{2z}\right)
\\
&=
\lr{\pi \over 2 z}^{1/2} \e^{-z}\Big[ 1
-  {\ft{1}{4}-n^2 \over 2z} \int_0^\infty d\sigma \e^{-\sigma}
 {{}
   _2F_1\left(\ft{3}{2}-n ,\ft{3}{2}+n;2;-\frac{\sigma }{2 z}\right)} 
 \Big]  
\,,
\end{align}
where in the second relation we integrated by parts.
At large $z$, we can expand the hypergeometric function in powers of $1/z$ and integrate term-by-term to obtain a power series representation of $K_n(z)$. In this way, we verify that the second term on the right-hand side of \re{In-im} arises from $K_n(x-i0)$ in \re{I=K}. The first term in \re{In-im} comes from the large $x$ expansion of $K_n(-x+i0)$ in \re{I=K}. The Borel singularities arise in this term because the function $K_n(-x)$ has a discontinuity across the cut $x>0$.  

The simplest way to see this is to notice that the integral in \re{Kn-int} takes the form of the Borel transform \re{Bor}.
The hypergeometric function in \re{Kn-int} has a pole at $\sigma=-2z$
\begin{align}
\, _2F_1\left(\ft{3}{2}-n ,\ft{3}{2}+n;2;-\frac{\sigma }{2 z}\right) = \frac{2z}{(\sigma +2z)\,  \Gamma \left(\frac{3}{2}-n \right) \Gamma \left( \frac{3}{2}+n\right)}+\dots
\end{align}
For $z<0$ this pole is located on the integration contour in  \re{Kn-int} making $K_n(z)$ ill defined. Replacing $z\to z=-x\pm i0$ in  \re{Kn-int} amounts to deforming the integration contour in the vicinity of the pole as $\sigma\to\sigma\pm i0$, schematically, 
\begin{align}\label{K-pole}
K_n(-x\pm i0) \sim \pm  {\e^{x}\over \sqrt{x}}\int_0^\infty {d\sigma \e^{-\sigma}\over \sigma -2x \pm i0}\,.
\end{align}
At large positive $x$, we can identify the first term inside the brackets on both lines of \re{I=K} as defining a perturbative correction to $I_n(x)$. Notice that this correction is different for the first and the second relation in \re{I=K}. The difference  
is proportional to the sum of the functions $K_n(-x+ i0)+K_n(-x- i0)$. It is given by the
residue at the pole $\sigma=2x$ and yields an exponentially small correction $\sim \e^{-x}/\sqrt{x}$
that matches the second term in \re{In-im}. Due to the equivalence of the two representations \re{I=K}, it is compensated by the analogous exponentially small correction coming from $K_n(x)=K_n(x-i0)=K_n(x+i0)$. 

Taking an average of the two relations in \re{I=K}, we can get another equivalent representation  
\begin{align}\label{I-av}
I_n(x) &= {i\over 2\pi} \left[K_n(-x+i0)-K_n(-x-i0)\right],
\end{align}
which is valid for $x>0$.
It corresponds to the principal value prescription for integrating the Borel pole \re{K-pole}. 

We conclude from above analysis that the relations \re{I=K} and \re{I-av} 
correspond to three different prescriptions of regularizing Borel singularities in the asymptotic expansion \re{In-im}.

Let us return to \re{r(x)} and apply the first relation in \re{I=K}
\begin{align}\label{rn1}\notag
r(x) &= {K_0(-x+i0)- K_0(x)\over K_1(-x+i0)+K_1(x)} 
\\
&={K_0(-x+i0)\over K_1(-x+i0)} +{i\pi \over x[K_1(-x+i0)]^2} \lr{1+ {K_1(x)\over K_1(-x+i0)}}^{-1}\,,
\end{align}
where in the last equality  we used the identity for the Bessel function. Applying \re{Kn-int} we find that for large positive $x$ the first term 
in \re{rn1} defines a perturbative correction to $r(x)$. The second term in \re{rn1} is exponentially suppressed
\begin{align}
{i\pi \over x[K_1(-x+i0)]^2}=-2i \e^{-2x}+\dots\,,
\end{align}
together with $K_1(x)/K_1(-x+i0)=O(\e^{-2x})$. This allows us to simplify \re{rn1} as 
\begin{align}\notag\label{sign-flip}
 r(x) {}&= {K_0(-x+i0)\over K_1(-x+i0)} +{i\pi \over x[K_1(-x+i0)]^2}+O(\e^{-4x})
\\
&= \Big({ 1+\frac{1}{2 x}+\frac{3}{8 x^2}+O(1/x^3)}\Big) -i \e^{-2 x} \Big(2+\frac{3 }{2 x}+\frac{21
   }{16 x^2}+O(1/x^3)\Big)+O(\e^{-4x}),
\end{align}
where in the second line  we replaced the Bessel functions by their leading large $x$ expansion. The series inside the brackets suffer from Borel singularities. As explained above, the prescription $-x+i0$ in the argument of the Bessel functions is equivalent to deforming the integration contour in their Borel transform slightly above the real axis. Had we applied the second relation in \re{I=K} we 
would get a similar relation for $r(x)$ with the sign in front of the second term in \re{sign-flip} reversed. 

Applying \re{sign-flip} we can obtain the strong-coupling expansion of $r_n=r(4\pi gx_n)$ and, then, use it to reproduce the relation \re{rr}.
  
\section{Free energy in $\mathsf Q_{2}$ model}
 
\label{app:sc-orb}
The weak-coupling expansion of non-planar correction (\ref{dq-orb}) to the circular Wilson loop $\Delta q^{\mathsf Q_{2}}$ looks as
\begin{equation}
\Delta q^{\mathsf Q_{2}}(\lambda) = \lambda\,\sum_{n=2}^{\infty}d_{n}\,\Big(\frac{\lambda}{8\pi^{2}}\Big)^{n},
\end{equation}
where the expansion coefficients $d_{n}$ are listed in Eqs.~(3.24) and (3.29) of \cite{Beccaria:2021ksw}. Integrating the relation (\ref{dq-orb}) we can get the free energy difference  as
\begin{equation}
\Delta F^{\mathsf Q_{2}}(\lambda) = -8\,\sum_{n=2}^{\infty}\frac{d_{n}}{n}\,\Big(\frac{\lambda}{8\pi^{2}}\Big)^{n}\,.
\end{equation}
Its  weak coupling expansion reads 
\begin{align}\notag
\label{4.3}
\Delta  F^{\mathsf Q_{2}}(\lambda) = &
3 \zeta (3) \Big(\frac{\lambda}{8\pi^2}\Big)^2
% -----------------------------------------------------------------------------------------------------------------
-15 \zeta (5) \Big(\frac{\lambda}{8\pi^2}\Big)^3
% -----------------------------------------------------------------------------------------------------------------
%\\ &
-\Big(9 \zeta (3)^2-\tfrac{315}{4} \zeta (7)\Big) \Big(\frac{\lambda}{8\pi^2}\Big)^4
% -----------------------------------------------------------------------------------------------------------------
 \\ &\notag
+\Big(120 \zeta (3) \zeta (5) 
-441 \zeta (9)\Big) \Big(\frac{\lambda}{8\pi^2}\Big)^5
% -----------------------------------------------------------------------------------------------------------------
 \\ &  
+\Big(36 \zeta (3)^3-450 \zeta (5)^2-735 \zeta (3) \zeta (7)+\tfrac{10395}{4} \zeta (11)\Big) \Big(\frac{\lambda}{8\pi^2}\Big)^6
+\dots \,.
%\\ &
% -----------------------------------------------------------------------------------------------------------------
%+\bigg(-720 \zeta (3)^2 \zeta (5)+5985 \zeta (5) \zeta 
%(7)+4536 \zeta (3) \zeta (9)-\frac{127413}{8} \zeta (13)\bigg) \left(\frac{\lambda}{8\pi^2}\right)^7 +\dots.
%% -----------------------------------------------------------------------------------------------------------------
%+\bigg(-162 
%\zeta (3)^4+4950 \zeta (3) \zeta (5)^2+4410 \zeta (3)^2 \zeta 
%(7)-\frac{337365}{16} \zeta (7)^2-\frac{78435}{2} \zeta (5) \zeta 
%(9)-\frac{114345}{4} \zeta (3) \zeta (11)\lp
%+\frac{6441435}{64} \zeta 
%(15)\bigg) \left(\frac{\l}{8\pi^2}\right)^8
%% -----------------------------------------------------------------------------------------------------------------
%+\bigg(4320 \zeta (3)^3 \zeta (5)-11500 \zeta (5)^3-62160 
%\zeta (3) \zeta (5) \zeta (7)-27216 \zeta (3)^2 \zeta 
%(9)\lp
%+\frac{576765}{2} \zeta (7) \zeta (9)+\frac{517275}{2} \zeta (5) \zeta 
%(11)+184041 \zeta (3) \zeta (13)-\frac{5214495}{8} \zeta (17)) 
%\left(\frac{\l}{8\pi^2}\right)^9+\dots.
\end{align}
To find $\Delta F^{\mathsf Q_{2}}(\lambda)$ at arbitrary coupling, we use the  determinant representation of the partition function in the quiver model $\mathsf{Q}_{L}$ derived in  \cite{Galvagno:2020cgq,Billo:2021rdb}. Specialising the results of these papers to $L=2$, we get
\begin{equation}
\label{4.4}
\Delta F^{\mathsf Q_{2}}(\lambda) = \frac{1}{2}\log\det\Big[(1-\mathsf{X}^{\rm even})(1-\mathsf{X}^{\rm odd})\Big] =  \frac{1}{2}\,\mathcal F^\loc_{\ell=1}+\frac{1}{2}\,\mathcal F^\loc_{\ell=2}\, , 
\end{equation}
where the matrices $\mathsf{X}^{\rm even}$ and $\mathsf{X}^{\rm odd}$ coincide with \re{K-loc} for $\ell=1$ and $\ell=2$, respectively, 
\begin{align}
\mathsf{X}^{\rm even}_{nm} &= -8\,(-1)^{n+m}\sqrt{2n}\sqrt{2m}\int_{0}^{\infty}\frac{dt}{t}\frac{e^{2\pi t}}{(e^{2\pi t}-1)^{2}}\,J_{2n}(t\sqrt{\lambda})\,J_{2m}(t\sqrt{\lambda})\,, \notag \\
\mathsf{X}^{\rm odd}_{nm} &= -8\,(-1)^{n+m}\sqrt{2n+1}\sqrt{2m+1}\int_{0}^{\infty}\frac{dt}{t}\frac{e^{2\pi t}}{(e^{2\pi t}-1)^{2}}\,J_{2n+1}(t\sqrt{\lambda})\,J_{2m+1}(t\sqrt{\lambda}) \,.
\end{align}
The product $(1-\mathsf{X}^{\rm even})(1-\mathsf{X}^{\rm odd})$ arises in \re{4.4} due to the even-odd block factorization of the quadratic form associated with the underlying matrix model at large $N$, 
({\em cf.}  Eq.~(5.9)
of \cite{Billo:2021rdb}). The second relation in \re{4.4} follows from \re{det-K} by noticing that $\mathsf{X}^{\rm even}$ and $\mathsf{X}^{\rm odd}$ coincide with the matrix \re{K-def} evaluated for $\ell=1$ and $\ell=2$, respectively, and for the symbol given by \re{chi-loc}.
 
Notice that $\ha \mathcal F^\loc_{\ell=2}$ in  \rf{4.4}  
  gives  the free energy \re{12} in the  $\mathsf{SA}$ model.
  This is   a consequence  of the fact   that the   $\mathsf{SA}$ model may be viewed as  an   
 orientifold projection of the 
$L=2$ quiver \cite{Billo:2021rdb}.
After the projection, the even-even double trace terms in the matrix model interaction potential \rf{H}
 proportional to $\mathsf{X}^{\rm even}$ are removed and the  matrix $\mathsf{X}$   reduces to 
 $\mathsf{X}^{\rm odd}$. 

At weak coupling, we can apply \re{F-weak} and \re{q-weak} to expand the determinants in \re{4.4} in powers of $g^2={\lambda\ov (4\pi)^2}$
\begin{align}\notag
  \mathcal F_{\ell=1} =& 24 \zeta (3) g^4-320 \zeta (5) g^6 
-\Big(288 \zeta (3)^2-4200 \zeta (7)\Big) 
g^8
+\Big(7680 \zeta (3) \zeta (5)-56448 \zeta (9)\Big) g^{10} 
\\  &
+\Big(4608 \zeta (3)^3-54400 \zeta (5)^2-94080 \zeta (3) \zeta (7)+776160 \zeta (11)\Big) 
g^{12}+\dots\ , 
\notag 
\\
%+(-184320 \zeta (3)^2 \zeta (5)+1397760 \zeta (5) \zeta 
%(7)+1161216 \zeta (3) \zeta (9)-10872576 \zeta (13)) g^{14}\notag \\ &
%+(-82944 
%\zeta (3)^4+2534400 \zeta (3) \zeta (5)^2+2257920 \zeta (3)^2 \zeta 
%(7)-9337440 \zeta (7)^2-17902080 \zeta (5) \zeta (9)\notag \\ &
%-14636160 \zeta 
%(3) \zeta (11)+154594440 \zeta (15)) g^{16}+\dots, \\
%%%
 \mathcal F_{\ell=2} =& 80 \zeta (5) g^6-1680 \zeta (7) g^8+28224 \zeta (9) g^{10}-\Big(3200 
\zeta (5)^2+443520 \zeta (11)\Big) g^{12}
+\dots
%\notag \\ &
%+(134400 \zeta (5) \zeta (7)+6795360 \zeta (13)) g^{14}\notag \\ &
%+(-1458240 \zeta (7)^2-2177280 \zeta(5) \zeta (9)-103062960 \zeta (15)) g^{16}+\dots\ .
%%%%
%\label{2.13}
%\log & \det(1-k^{(3)}) = 280 \zeta (7) g^8-8064 \zeta (9) g^{10}+166320 \zeta (11) 
%g^{12}-3020160 \zeta (13) g^{14}\notag \\ &
%+(-39200 \zeta (7)^2+51531480 \zeta (15)) g^{16}+\dots\ .
\end{align}
 Substituting these relations into \re{4.4} we reproduce \re{4.3}.

 \newpage

 \bibliography{BT-Biblio}
 
\bibliographystyle{JHEP}
\end{document}